\documentclass{aa}
\usepackage{graphicx}

\def\ltsim{\raise 2pt \hbox {$<$} \kern-1.1em \lower 4pt \hbox {$\sim$}}
\def\ltapprox{\raise 2pt \hbox {$<$} \kern-1.1em \lower 5pt \hbox {$\approx
$}}
\def\gtsim{\raise 2pt \hbox {$>$} \kern-1.1em \lower 4pt \hbox {$\sim$}}
\def\gtapprox{\raise 2pt \hbox {$>$} \kern-1.1em \lower 5pt \hbox {$\approx
$}}

\begin{document}

%\title{High sensitivity GMRT observations of cD galaxies in rich and 
%poor galaxy clusters}
%\subtitle{I. Morphologies and spectral analysis}
\title{Radio morphology and spectral analysis of cD galaxies in rich and 
poor galaxy clusters} 
\author{S.~Giacintucci\inst{1,}\inst{2,}\inst{3} \and 
T.~Venturi\inst{1} \and
M. Murgia\inst{1,}\inst{4}
D.~Dallacasa\inst{1,} \inst{3} \and
R.~Athreya\inst{5}\and
S.~Bardelli\inst{2} \and
P.~Mazzotta \inst{6,}\inst{7}\and
D.J.~Saikia\inst{5}
}

\institute
{
INAF -- Istituto di Radioastronomia, via Gobetti 101, I-40129, Bologna, Italy 
\and
INAF -- Osservatorio Astronomico di Bologna,
via Ranzani 1, I--40126 Bologna, Italy
\and
Dipartimento di Astronomia, Universit\`a di Bologna,
via Ranzani 1, I--40126, Bologna, Italy
\and
INAF -- Osservatorio Astronomico di Cagliari,
Loc. Poggio dei Pini, Strada 54, 09012 Capoterra, Italy
\and
Tata Institute of Fundamental Research,
National Centre for Radio Astrophysics,
Ganeshkhind, Pune 411 007, India
\and
Dipartimento di Fisica,  Universit\'a di Roma Tor Vergata,
via della Ricerca Scientifica 1, I--00133 Roma,  Italy
\and
Harvard--Smithsonian Centre for Astrophysics, 
60 Garden Street, Cambridge, MA 02138, USA                        
}

\date{}
%\date{Received 00 - 00 - 0000; accepted 00 - 00 - 0000}
%
\titlerunning{cD galaxies in rich and poor galaxy clusters}
%\authorrunning{}
%
\abstract
{}
{We present a radio morphological study and spectral analysis for a sample 
of 13 cD galaxies in rich and poor clusters of galaxies.}
{Our study is based on new high sensitivity Giant Metrewave Radio Telescope 
(GMRT) observations at 1.28 GHz, 610 MHz and 235 MHz, and on archival data.
From a statistical sample of cluster cD galaxies we selected those sources
with little information available in the literature and promising for
the detection of aged radio emission. Beyond the high sensitivity images for 
all 13 radio galaxies, we present also a detailed spectral analysis for 7 of
them.}
{We found a variety of morphologies and linear sizes, as typical for radio
galaxies in the radio power range sampled here (low to intermediate power radio
galaxies). The spectral analysis shows that 10/13 radio galaxies have steep 
radio spectrum, with spectral index $\alpha \ge 1$.
In general, the radiative ages and growth velocities are consistent with
previous findings that the evolution of radio galaxies at the cluster
centres is affected by the dense external medium (i.e. low growth velocities
and old ages).   
We suggest that the dominant galaxies in A\,2622 and MKW\,03s are dying radio 
sources, which at
present are not fed by nuclear activity. On the other hand, the spectacular 
source at the centre of A\,2372 might be a very interesting example of 
restarted radio galaxy. For this source we estimated a life cycle of the 
order of 10$^6$ yr.}
{}


\maketitle
\section{Introduction}\label{sec:intro}

Nuclear radio emission associated with elliptical
galaxies, i.e. the {\it radio galaxy phenomenon}, is
one of the most studied in radio astronomy. 
Among ellipticals, cluster dominant cD galaxies (Mathews et al.
\cite{mathews64}) are the most
extreme and interesting population, since they are among
the most luminous and massive galaxies known. They are 
found at the centre of the potential 
wells of both rich and poor clusters of galaxies, and they are
usually located at the peak of the cluster X--ray emission
(Jones et al. \cite{jones79}).  

Despite the large amount of data available for this class of
radio sources, some basic questions related to their birth, evolution 
and death have no answer yet.
For example, it is still unknown what triggers radio
emission in elliptical galaxies, 
%what is the phenomenology underlying the radio luminosity function
%(RLF) for these objects, 
and what is the role of the large-- and small--scale environment. 
The presence of a central massive black hole accreting material is 
crucial, but it is not the only requirement, since it is 
widely accepted that massive and supermassive black holes 
are hosted in radio quiet sources as well
(Merritt \& Ferrarese \cite{merritt01}). 

The accumulating evidence for recurrent radio activity in
elliptical galaxies suggests that all ellipticals  may 
alternate phases of radio activity and quiescence (i.e.  Lara 
et al. \cite{lara99}; Giovannini  et al. \cite{giovannini98}; 
Schoenmakers et al. \cite{schoen00}), however the cause of
such intermittent activity and the duty cycle of the radio
emission are still basically unknown.
\\
In recent years, thanks to the high--resolution X--ray images from
the satellites {\it Chandra} and XMM--Newton, it has become clear that
major feedback processes take place in the central regions of galaxy 
clusters and groups, between the AGN activity of the dominant galaxy 
and the intracluster gas (for recent literature on the subject refer
to B\"ohringer et al. \cite{boehring07}).
The relativistic plasma of the central radio galaxies may
interact with the ICM, displacing the X--ray emitting gas and creating 
depressions in  the X--ray surface brightness, referred to as 
{\it holes}, {\it cavities} or {\it bubbles} 
(see for instance the review by Blanton \cite{blanton04}). 
Such cavities may be filled with radio emission
from the lobes of the central AGN ({\it radio--filled cavities}, 
e.g. Hydra A, McNamara et al. 
\cite{mcnamara00}; Perseus, Fabian et al. \cite{fabian00};
A\,2052, Blanton et al. \cite{blanton01}; MKW\,03s, Mazzotta et al. 
\cite{mazzotta04}). 
In other cases an offset is found between radio lobes and cavities 
({\it ghost cavities}, e.g. Perseus, Fabian et al. \cite{fabian00}; A\,2597, 
McNamara et al. \cite{mcnamara01}). Both types of 
cavities may be present in the same cluster (like Perseus), suggesting
that repeated outbursts of radio 
activity of the central galaxy could play a role in the formation of 
the holes, being the ghost cavities the result of a previous AGN burst
(McNamara et al. \cite{mcnamara01}) moved outwards by buoyancy.
The central regions of galaxy clusters
and groups are thus expected to be very promising for the study of aged
radio galaxies, for recurrent activity, and at a more general level for
our understanding of the birth and evolution of the radio galaxy phenomenon.
\\
In order to study the morphological and spectral properties of dominant
cluster galaxies in the radio band, and to address the question of their age 
and evolution, in this paper we present new radio observations carried out 
with the Giant Metrewave Radio Telescope (GMRT, Pune, India) at 1.28 GHz, 
610 MHz and 235 MHz for a sample of 13 cD galaxies located at
the centre of rich clusters and poor groups.  
The sources belong to a large statistical sample of 132 cD galaxies 
we selected in order to investigate the properties of this class 
of radio sources and to understand how the  external environment may 
influence them.
%, we undertook a 
%study of such objects in rich and poor galaxy clusters, 
%and selected a statistical sample of 132 cD galaxies.
\\
\\
The paper is organised as follows: 
in Section \ref{sec:sample} we report on the sample selection and on the 
list of clusters observed with the GMRT; in Section \ref{sec:obs} we 
describe the observations and the data reduction; the radio images and the 
spectral analysis are given in Section \ref{sec:results} and 
\ref{sec:spectra} respectively; 
our results are discussed in Section \ref{sec:disc} 
and a brief summary is reported in Section \ref{sec:summary}.
In the Appendix we present the 235 MHz image of the cD galaxy 
in A\,2634, which  is located within the GMRT field of view of 
one of the cluster  presented in this paper. We point out that A\,2634 
belongs to our complete sample of cD galaxies, but it was not 
selected as target for the GMRT observations, given the 
amount of radio data  available for this cluster in the literature.
\\
\\
In this paper we adopt the $\Lambda$CDM cosmology,
with H$_0$=70 km s$^{-1}$ Mpc$^{-1}$, $\Omega_m=0.3$ and
$\Omega_{\Lambda}=0.7$. The spectral index $\alpha$ is defined
according to S$\propto \nu^{-\alpha}$. 

\section{The sample of cluster cD galaxies}\label{sec:sample}

For completeness, we report here the most relevant information on
the compilation of our statistical sample of 132 cD galaxies in rich 
and poor clusters, selected in the light of a comprenhensive study 
of their radio properties (work in progress).
% and to investigate the role of 
% the large scale environment on the cD radio activity, we compiled a 
% statistical sample of 132 galaxies. 
The sample includes the 109 Abell clusters
%extracted from the morphological catalog by Struble \& Rood (\cite{struble87})
from the Struble \& Rood (\cite{struble87}) morphological catalogue
and from the cD galaxy sample in
Valentjin \& Bijleveld (\cite{val83}) with
DEC$_{J2000} > - 30^{\circ}$ and redshift 
z$\le$0.1 \footnote{We used the NASA/IPAC Extragalactic Database NED to obtain 
updated (although inhomogenous) redshift information for all the clusters.}, 
optically  classified as cD clusters.
From Valentjin \& Bijleveld (\cite{val83}) we further selected 23 poor 
clusters hosting a cD, with the same declination and redshift constraints.
The poor cluster cDs belong to the MKW sample (Morgan, Kayser \& White 
(\cite{mkw75}) and to the AWM sample (Albert, White \& Morgan \cite{awm77}).

We point out that our selection was carried out on the basis of the optical 
information only, 
therefore no {\it a priori} knowledge of the radio properties of the cDs 
and the X--ray properties of the hosting clusters was considered.  

We searched for information on the radio emission for all the 
cD galaxies in our sample by inspection of the 1.4 GHz 
NVSS (Condon et al. \cite{condon98}) and FIRST 
(Becker et al. \cite{becker95}) surveys, and from the literature.
We found that $\sim$ 60 \% of the cDs in the sample are radio 
loud above the NVSS sensitivity limit, at levels going from few 
mJy to $\sim$Jy. 
On the basis of the radio information available in the literature, 
we selected all clusters still lacking high sensitivity and
high resolution data at 1.4 GHz (beyond the NVSS and FIRST
images), and a number of clusters which appeared promising for 
the search of aged radio emission related to a previous burst 
of activity of the central galaxy.  Our final list includes
a total of 13 objects (in both rich and poor clusters)
given in Table 1.
The columns in the table provide respectively the following information:
cluster name; cluster richness (available for the Abell clusters only); 
cD galaxy name; J2000 optical coordinates of the cD; 
cluster redshift; 1.4 GHz flux density measured on the NVSS image, and 
corresponding radio power at the redshift of the optical galaxy. 
%
%------------- Table 1-------------------------------------------------------------
%
\begin{table*}[h!]
\label{tab:sample1}
\caption[]{Rich Abell and poor cluster cDs}
\begin{center}
\begin{tabular}{lcllccrc}
\hline\noalign{\smallskip}
Cluster name  & R$^{(\diamond)}$ & Galaxy name & RA$_{J2000}$ &  DEC$_{J2000}$ & z & S$_{\rm 1.4\,GHz}$ & logP$_{\rm 1.4\,GHz}$ \\
  & &  h,m,s            &  $^{\circ}$, $^{\prime}$, $^{\prime \prime}$ & & (mJy) & (W Hz$^{-1}$) \\
\noalign{\smallskip}
\hline\noalign{\smallskip}
\multicolumn{1}{l}{Rich Abell clusters} &&&&&&&\\
\noalign{\smallskip}
\hspace{0.3cm}A\,1663 & 1  & 2MASX J1302$-$0230   & 13 02 52.6 & $-$02 30 59 & 0.0847 &  39 & 23.84 \\
\hspace{0.2cm} A\,1775$^{(*)}$ & 2  & UGC 08669 (1) & 13 41 49.1 &   +26 22 25 & 0.0757 &  24 & 23.52 \\
                             & & UGC 08669 (2) & 13 41 50.4 &   +26 22 13 & 0.0734 & 298 & 24.71 \\
\hspace{0.3cm}A\,2162  & 0     & NGC 6086      & 16 12 35.6 &   +29 29 04 & 0.0319 & 111 & 23.40 \\
\hspace{0.3cm}A\,2372  & 0     & 2MASX J2145$-$1959   & 21 45 15.5 & $-$19 59 41 & 0.0587 & 417 & 24.53 \\
\hspace{0.3cm}A\,2480  & 1     & 2MASX J2245$-$1737   & 22 45 59.0 & $-$17 37 33 & 0.0684 & 406 & 24.66 \\
\hspace{0.3cm}A\,2622  & 0  &  2MASX J2335+2722    & 23 35 01.5 &   +27 22 20 & 0.0613 &  78 & 23.84 \\
\noalign{\bigskip}
\multicolumn{1}{l}{Poor clusters} &&&&&&\\
\noalign{\smallskip}
\hspace{0.3cm}MKW\,01s   &&  UGC 04956          & 09 20 02.2 &   +01 02 18 & 0.0172 &  11 & 21.86 \\
\hspace{0.3cm}MKW\,01    &&  NGC 3090           & 10 00 30.2 & $-$02 58 06 & 0.0203 &  25 & 22.37 \\
\hspace{0.3cm}MKW\,02    &&  CGCG 009$-$062     & 10 30 10.6 & $-$03 09 49 & 0.0380 & 385 & 24.11 \\  
\hspace{0.3cm}MKW\,06    &&  CGCG 018$-$111     & 14 17 36.5 &   +02 03 18 & 0.0530 &  42 & 23.44 \\
\hspace{0.3cm}MKW\,07    &&  UGC 09371          & 14 33 59.1 &   +03 46 41 & 0.0302 &  94 & 23.28 \\
\hspace{0.3cm}MKW\,03s   &&  NGC 5920           & 15 21 51.8 &   +07 42 32 & 0.0453 & 105 & 23.72 \\
\hspace{0.3cm}AWM\,05    && NGC 6269            & 16 57 58.1 &   +27 51 16 & 0.0348 &  51 & 23.15 \\
\noalign{\smallskip}
\hline
\end{tabular}
\end{center}
$^{(*)}$ this cluster hosts a Dumbell system at its centre; the 1.4 GHz flux densities given for the two radio 
galaxies are from the FIRST survey.\\
$^{(\diamond)}$ Richness class R, defined by Abell (e.g. 
ACO catalogue of rich galaxy clusters, Abell, Corwin \& Olowin \cite{aco89}), 
on the basis of the
number of galaxies in the magnitude range [m$_3$; m$_3$+2].

\end{table*}
%
%------------------- end of table 1------------------------------------
%

\section{Radio observations and data reduction}\label{sec:obs}

The observations were carried out with the GMRT at 1.28 GHz, 610 and
235 MHz. In particular, 1.28 GHz observations were carried out for 8
of the 13 galaxies in our sample still lacking high sensitivity 
and high resolution observations 
at this frequency, while the observations at 610 MHz and 
235 MHz were carried out for those sources (7/13) candidate for recurrent 
radio activity. For this reason, not all galaxies listed in Table 1 were 
observed at all three frequencies. The observing logs are reported in Table 
\ref{tab:obs}, where the following information is provided: 
cluster name; observing date; observed frequency;
total frequency bandwidth (MHz); total time on source (minutes); 
half power bandwidth (HPWB) of the full array 
(arcsec $\times$ arcsec, position angle in degrees); rms level 
(1$\sigma$) in the full resolution image (mJy b$^{-1}$).

The 1.28 GHz observations were performed using both 
the upper and lower side band, for a total observing bandwidth of 
32 MHz; the August 2003 610 MHz observations were carried out using the 
dual receiver 235/610 MHz, and have a bandwidth of 16 MHz. Since the 
235 MHz band of those observations was affected by major interference 
which jeopardized the whole data reduction, those sources were reobserved at 
235 MHz in July 2005 with a bandwidth of 8 MHz. 

The data at all frequencies were collected in spectral--line mode 
(128 channels at 1.28 GHz 
and 610 MHz, and 64 channels at 235 MHz), with a spectral resolution 
of 125 kHz/channel. The data calibration and reduction were performed using 
the NRAO Astronomical Image Processing System (AIPS) package. An accurate
editing was needed to identify and remove the data affected by radio frequency 
interference (RFI) at 235 MHz. In order to find a compromise between the 
size of the dataset and the need to minimize bandwidth smearing effects 
within the primary beam, after bandpass calibration the central channels in 
each individual dataset were averaged to 6 channels of $\sim$1 MHz each at 
235 MHz and $\sim$2 MHz each at 610 MHz. At 1.28 GHz the central 
band was averaged to 1 single channel of $\sim$ 11 MHz.

In each step of the data reduction we used the wide--field imaging technique, 
and after a number of phase self--calibration cycles, we produced the final
images for each cluster. The upper and lower side band datasets at each 
observing frequency were self--calibrated separately and then combined 
to produce the final images. 
The rms noise level (1$\sigma$) achieved in the final full resolution images 
is in the range 13--30 $\mu$Jy at 1.28 GHz, 50--150 $\mu$Jy at 610 MHz
and 0.65--2 mJy at 235 MHz (see Tab.~\ref{tab:obs}). 
The spread in the noise level depends mostly on the total time on source and 
residual RFI, especially at 235 MHz.
Unless specified otherwise, the average residual amplitude errors in our data 
are of the order of $\ltsim$ 5\% at all frequencies.

%%%%%%%%%%%%%%%%%%%%%% Table 2 &&&&&&&&&&&&&&&&&&&&&&&&&&&&&&&&&&&&&&&&&&&&&

\begin{table*}[h!]
\caption[]{GMRT observations.}
\begin{center}
\begin{tabular}{lcccccl}
\hline\noalign{\smallskip}
Cluster  & Obs. date & Obs. freq.  &  $\Delta \nu$ & Obs. time & Beam, PA  &   rms      \\ 
            &                 &   MHz       &      MHz         &   min        &
	     $^{\prime \prime} \times^{\prime \prime}$, $^{\circ}$& mJy b$^{-1}$ \\
\noalign{\smallskip}
\hline\noalign{\smallskip}
A\,1663    & Aug 2003 &1280 & 32 &  60 &  3.0 $\times$ 2.1, 17   & 0.015 \\
&&&&&& \\
A\,1775    & Aug 2003 & 610 & 16 &  45 &  5.1 $\times$ 3.9, 49   & 0.130 \\
           & Jul 2005 & 235 & 8 & 100 & 12.7 $\times$ 9.1, 63   & 1.500 \\
&&&&&& \\
A\,2162    & Aug 2003 & 610 & 16 &  60 &  7.1 $\times$ 4.9, --83 & 0.150 \\
           & Jul 2005 & 235 &  8 &  50 & 14.7 $\times$10.8, --85 & 1.000 \\
&&&&&& \\
A\,2372    & Jul 2005 & 610 & 32 & 160 &  7.5 $\times$ 5.7, 38   & 0.065 \\
           & Jul 2005 & 235 & 8 & 160 & 19.7 $\times$13.8, --8  & 1.500 \\
&&&&&& \\
A\,2480    & Aug 2003 &1280 & 32 &  45 &  2.9 $\times$ 2.0, 13   & 0.030 \\
&&&&&& \\
A\,2622    & Jul 2005 & 610 & 32 & 220 &  5.5 $\times$ 4.9, 89   & 0.050 \\
           & Jul 2005 & 235 & 8 & 220 & 17.1 $\times$11.4, 53   & 0.800 \\
&&&&&& \\
MKW\,01s   & Aug 2003 &1280 & 32 &  60 &  3.5 $\times$ 2.3, 2    & 0.015 \\
&&&&&& \\
MKW\,01    & Aug 2003 &1280 & 32 &  60 &  3.1 $\times$ 2.1, 28   & 0.020 \\
&&&&&& \\
MKW\,02    & Aug 2003 & 610 & 16 & 100 &  5.1 $\times$ 4.6, 42   & 0.130 \\
           & Jul 2005 & 235 &  8 & 100 & 12.1 $\times$ 9.7, 65   & 0.650 \\
&&&&&& \\
MKW\,06    & Aug 2003 &1280 & 32 &  60 &  3.3 $\times$ 2.3, 45   & 0.020 \\
&&&&&& \\
MKW\,07    & Aug 2003 &1280 & 32 &  60 &  3.5 $\times$ 2.4, 49   & 0.030 \\
           & Aug 2003 & 610 & 16 & 120 &  5.8 $\times$ 4.4, 57   & 0.100 \\ 
           & Jul 2005 & 235 & 8  & 100 & 14.0 $\times$ 9.1, 50   & 1.500 \\
&&&&&& \\	       
MKW\,03s   & Aug 2003 &1280 & 32 &  60 &  3.0 $\times$ 2.5, --38 & 0.030 \\
           & Aug 2003 & 610 & 16 &  90 &  5.4 $\times$ 4.5, 61   & 0.150 \\
	   & Jul 2005 & 235 &  8 &  90 & 12.1 $\times$ 9.1, 52   & 2.000 \\
&&&&&& \\
AWM\,05    & Aug 2003 &1280 & 32 &  60 &  3.2 $\times$ 2.6, --72 & 0.025 \\
\hline{\smallskip}
\end{tabular}
\end{center}
\label{tab:obs}
\end{table*}
%%%%%%%%%%%%%%%%%%%%%%%% end tab. 2 %%%%%%%%%%%%%%%%%%%%%%%%%%%%%%%%%%%%%%%%%%%

\section{The radio images}\label{sec:results}

In this section we present our new GMRT radio images of the 
clusters listed in Tables 1 and \ref{tab:obs}. 
In the presentation of the images we chose to group
the sources according to the frequency of the observations, for a 
clearer reference to the information reported in the tables.
We used the AIPS task JMFIT to measure the total flux density and
angular size of the unresolved components, while the task TVSTAT
was used to determine the total flux density of extended radio sources.
Given the high signal--to--noise ratio in all our images, 
the error associated with the flux density measurement is
dominated by the uncertainty in the residual amplitude calibration
errors.
The conversion factor from angular to linear scale is reported
in the caption of each figure.

Where possible, we carried out a detailed study of the radio  
integrated spectrum, and/or spectral index images. Such analysis is 
presented in Section \ref{sec:spectra}.

\subsection{Clusters observed at 1.28 GHz}\label{sec:1.28}

We present the radio images of the clusters observed at 1.28 GHz only 
(Tab.~\ref{tab:obs}). In Tab.~\ref{tab:flux1} we summarize the most relevant
observational properties of each source, i.e.  
the 1.28 GHz total flux density and largest linear size (LLS). 
In those cases where the source structure can be unambiguosly 
separated into a number of individual components, 
information is listed for each of them. A brief note for
all sources is reported below.
\\
\\
%%%%%%%%%%%%%%%%%%%%%%%%%%%%%% Tab.3 %%%%%%%%%%%%%%%%%%%%%%%%%%%%%%%%%%%%%%%%%%%
\begin{table*}[h!]
\caption[]{Radio source data at 1.28 GHz}
\begin{center}
\begin{tabular}{rccc}
\noalign{\smallskip}
\hline\noalign{\smallskip}
Cluster & component & S$_{\rm 1280\,MHz}$  &  LLS \\ 
        &           & (mJy)                &  kpc \\
\hline\noalign{\smallskip}			      
A\,1663  &  tot.  &  40 & $\sim40\times$16 \\
         &  A     &  26 & $\sim$16 \\
         &  B     &  10 & $\sim$16 \\ 
         &  C     &   4 &  $<$3.5  \\ 
&&&\\
A\,2480  & tot.   & 472 & $\sim35\times$14 \\
&&&\\
MKW\,01s & tot.   &  12 & $<$0.02 \\
&&&\\
MKW\,01  &  tot.  &  30 & $\sim 21$ \\ 
         &   C    &  23 & $<0.7  $  \\
         &   H    &   7 & $\sim 21$ \\
&&&\\
MKW\,06  &  tot.  &  34 & $\sim 165\times$46 \\ 
         & A      &  14 & $\sim  23\times$10 \\
         & B      &  16 & $\sim  57\times$46 \\
	 & C      &   4 & $\sim  52\times$33 \\
&&&\\
AWM\,05  & tot.   &  50 & $\sim  35\times$9 \\
         & A      &  30 & $\sim   8\times$5 \\
         & B      &  10 & $\sim  10\times$9 \\
	 & C      &  10 & $\sim  12\times$8 \\
\noalign{\smallskip}
\hline\noalign{\smallskip}
\end{tabular}
\end{center}
\label{tab:flux1}
\end{table*}
%%%%%%%%%%%%%%%%%%%%%%%%%%%%%% end tab.3 %%%%%%%%%%%%%%%%%%%%%%%%%%%%%%%%%%%%
%
{\it A\,1663} -- In Fig.~\ref{fig:a1663} we present the GMRT 1.28 GHz 
full resolution 
%($3.0^{\prime\prime} \times 2.1^{\prime\prime}$, p.a. $17^{\circ}$)
contours of the source, superposed on the  Sloan Digital Sky Survey (SDSS)
optical image of the cD galaxy. At this resolution 
the source can be classified as triple, and
extends on the galactic scale with 
% an angular size of $\approx 25^{\prime \prime} \times 10^{\prime \prime}$,
%corresponding to a 
linear size of $\approx$ 40$\times$16 kpc. Between the two extended radio 
lobes  (labelled A and B in Fig.~\ref{fig:a1663}), 
a third compact component  (C) is detected
at the location of the optical nucleus of the cD galaxy.
No radio emission  from the cD was detected with the VLA in the B 
array configuration at 4.7 GHz by Ball et al. (\cite{ball93}) at a 
sensitivity limit of 0.7 mJy.
%
%
%%%%%%%%%%%%%%%%%% fig. 1 - A1663 %%%%%%%%%%%%%%%%%%%%%%%%%
\begin{figure}
\centering
\caption{ A\,1663 -- GMRT 1.28 GHz radio contours of the central 
radio galaxy overlaid on the SDSS optical frame. 
Different source components are labelled as A, B and C.
The 1$\sigma$ level in the image is 15 $\mu$Jy b$^{-1}$.
Logarithmic contours are reported, starting from $\pm$0.045 
mJy b$^{-1}$. The contour peak flux is 13.53 mJy b$^{-1}$.
The HPWB is $3.0^{\prime\prime} \times 2.1^{\prime\prime}$, 
p.a. $17^{\circ}$. 
For this source 1$^{\prime\prime}$ = 1.59 kpc.}
\label{fig:a1663}
\end{figure}
%%%%%%%%%%%%%%%%%%%%%%%%%%%%%%%%%%%%%%%%%%%%%%%%%%%%%%%%%

The GMRT 1.28 GHz total flux density of the source (Tab.~\ref{tab:flux1}) 
is in good agreement with the 1.4 GHz value measured on the NVSS image 
(Tab. 1), despite the difference in angular resolution.
The source radio power 
%log P$_{\rm 1.28\,GHz}$ (W Hz$^{-1}$)= 23.85  
is typical of FR\,I sources (Fanaroff \& Riley \cite{fr74}), and consistent 
with its radio morphology.  
\\
\\
{\it A\,2480} -- The GMRT 1.28 GHz image is presented in Fig.~\ref{fig:a2480}.
The cD galaxy in A\,2480 hosts a double radio galaxy of galactic size
($\sim 35 \times 14$ kpc),  with two jets and Z--shaped lobes.
Our image is consistent with the VLA--B 4.7 GHz image published 
in Ball et al. (\cite{ball93}).
Both the radio morhology
and power are typical of a FR\,I/FR\,II transition source. It is  
the most powerful radio source among those presented in this paper. 
%
%%%%%%%%%%%%%%%%%% fig. 2  A2480 %%%%%%%%%%%%%%%%%%%%%%%%%%%%%%
\begin{figure}
\centering
\caption{A\,2480 -- GMRT 1.28 GHz radio contours of the central radio 
galaxy, on the POSS--2 optical image. The 1$\sigma$ level in the 
image is 30 $\mu$Jy b$^{-1}$. Logarithmic contours are 
reported, starting from $\pm$0.09 mJy b$^{-1}$. The contour peak 
flux is 51.26 mJy $b^{-1}$. The HPWB is 
$2.9^{\prime\prime} \times 2.0^{\prime\prime}$, p.a. $13^{\circ}$.
For this source 1$^{\prime\prime}$ =1.31 kpc.}
\label{fig:a2480}
\end{figure}
%%%%%%%%%%%%%%%%%%%%%%%%%%%%%%%%%%%%%%%%%%%%%%%%%%%%%%%%%%%%%%%%%%%%
\\
A detailed study of the spectral properties for this object is
given in \ref{sec:spec_a2480}.
\\
\\
{\it MKW\,01s} -- The GMRT 1.28 GHz full resolution image of the radio 
emission from the dominant
galaxy in MKW\,01s is shown in Fig.~\ref{fig:mkw01s}, superposed to the
optical SDSS frame. At this frequency and resolution we detected a compact 
radio source with a flux density of 12 mJy (Tab.~\ref{tab:flux1}), in agreement
with the value from the NVSS (Tab. 1) and FIRST images 
(S$_{\rm 1.4\,GHz}$=11 mJy). 
This is source is the closest to us and the least powerful among those 
presented in this paper. 
A two dimensional Gaussian fit of the source provides 
a deconvolved linear size of  $\sim$170 $\times$140 pc.

A 9 mJy unresolved radio source was also detected with the VLA at 1.45 GHz, 
with a resolution of $\sim13^{\prime \prime}$, by Burns et al. 
(\cite{burns87}). A total flux density of 10$\pm$3 mJy was measured at 2.38 GHz
with the Arecibo telescope by Dressel \& Condon (\cite{dressel78}). 
Using this value a spectral index of $\alpha=0.3\pm0.6$ is obtained between
1.28 GHz and 2.38 GHz.

%
%%%%%%%%%%%%%%%%%% fig. 3 - MKW01s %%%%%%%%%%%%%%%%%%%%%%%%%%%%%%%%
\begin{figure}
\centering
\caption{MKW01s -- GMRT 1.28 GHz radio contours of the central radio galaxy
overlaid on the SDSS image. The 1$\sigma$ level in the image is 15 
$\mu$Jy b$^{-1}$. Logarithmic contours are reported, starting 
from $\pm$0.045 mJy b$^{-1}$. The contour peak flux is 30.83 mJy b$^{-1}$.
The HPWB is $3.5^{\prime\prime} \times 2.3^{\prime\prime}$, 
p.a. $2^{\circ}$. For this source 1$^{\prime\prime}$ = 0.35 kpc.}
\label{fig:mkw01s}
\end{figure}
%%%%%%%%%%%%%%%%%%%%%%%%%%%%%%%%%%%%%%%%%%%%%%%%%%%%%%%%%%%%%%%%%%%%%%

The nature of this radio source is unclear. Its radio morphology, 
total power and unknown spectral shape make it difficult to 
discriminate between a compact low power AGN and a radio emission of 
nuclear starburst origin. 
\\
\\
{\it MKW\,01} -- The radio emission of the cD galaxy is reported in 
Fig. \ref{fig:mkw01},
where a  full resolution and a tapered image are given in the left 
and right panel respectively.
In both images the source is characterised by a compact bright component 
(labelled C in Fig.~\ref{fig:mkw01}) surrounded by an amorphous low surface 
brightness feature (H in the figure). 
This morphology is reminiscent of the core--halo radio sources found in a 
number of cool core clusters, as for example 3C\,317 (Venturi et al. 
\cite{venturi04}), 2A\,0335+096 (Sarazin et al. \cite{sarazin95}), and
PKS\,0745--191 (Baum \& O'Dea \cite{baum91}).  
In Tab.~\ref{tab:flux1} we report the individual flux density values of 
components C and H. The flux density of C was estimated through a two 
dimensional Gaussian fit on the highest resolution image (Fig.~\ref{fig:mkw01},
left panel).
The fit provides a deconvolved size of 
$\sim$0.7$\times$0.4 kpc. The halo emission extends over a scale of 
$\sim$ 21 kpc in the low resolution image, and it is completely 
embedded in the optical galaxy (Fig.~\ref{fig:mkw01}, right panel).

Previous 1.45 GHz VLA observations (Burns et al. \cite{burns87})
show only a compact component coincident with the cD galaxy in MKW\,01.
The flux density in that image ($\sim 13^{\prime \prime}$ resolution) is 
S$_{\rm 1.45\,GHz}$=21.4 mJy. This value is consistent with the
1.4 GHz NVSS flux density reported in Tab. 1, and they are
both $\sim 20$\% lower than our measurement at 1.28 GHz. 
Such difference is significant, even if we account for spectral 
index effects, and might suggest variability of the central component 
associated with the nucleus of the cD galaxy.

No information is available in the literature at other frequencies, 
therefore no spectral study is possible for this source.
\\
\\
%%%%%%%%%%%%%%%%%%%%%%% fig. 4  MKW01 %%%%%%%%%%%%%%%%%%%%%%%%%%%%%%%%%%%
\begin{figure*}
\centering
\hspace{0.8truecm}
\caption{MKW01 -- GMRT 1.28 GHz full resolution (on the POSS--2 optical image, 
{\it left panel}) and u--v tapered low resolution ({\it right panel}) 
contours of the central radio galaxy. The 1$\sigma$ level in the image is 
20 $\mu$Jy b$^{-1}$ and 30 $\mu$Jy b$^{-1}$ respectively. Logarithmic contours 
are reported, starting from $\pm$3$\sigma$. The contour peak flux is 22.9 
mJy b$^{-1}$ in the right panel and 24.4 mJy b$^{-1}$ in the left panel. 
The HPWB is $3.1^{\prime\prime} \times 2.1^{\prime\prime}$, p.a. $28^{\circ}$ 
and $10.5^{\prime\prime} \times 9.6^{\prime\prime}$, p.a. $-6^{\circ}$ 
respectively. In both panels letters C and H indicate the core and 
halo components of the source. For this source 1$^{\prime\prime}$ = 0.41 kpc.}
\label{fig:mkw01}
\end{figure*}
%%%%%%%%%%%%%%%%%%%%%%%%%%%%%%%%%%%%%%%%%%%%%%%%%%%%%%%%%%%%%%%%%%%%%%%%%%%%%%
%
{\it MKW\,06} -- The 1.28 GHz image of the source is presented in 
Fig.~\ref{fig:mkw06}.
The left panel shows the full resolution image overlaid on the SDSS
optical frame: a radio galaxy of galactic size
%($\sim 22^{\prime \prime} \times 10^{\prime \prime}$, 
($\sim 23 \times 10$ kpc) is located at the 
position of the cD galaxy and is characterised by a double component 
(labelled as A in the figure). 
A region of low surface brightness emission, indicated as B in the 
figure, is detected South of A.
In the low  resolution image (right panel of Fig.~\ref{fig:mkw06})
this region may be interpreted as the southern lobe of a double radio source,
extending on a scale of $\sim 165$ kpc, larger 
than the structure detected at high resolution.  
The northern lobe (labelled as C in the right panel of Fig.~\ref{fig:mkw06}) 
has a very low surface brightness and is much fainter than the southern one.
An image of this radio source was reported also in Burns et al. 
(\cite{burns87}), who classified it as a possible head--tail, but they did 
not exclude a double morphology, possibly lost due the limited sensitivity of 
their observations. 

The 1.28 GHz total flux density of the source reported in Tab.~\ref{tab:flux1} 
was measured on  the low resolution image. It is in agreement with the Burns 
et al. ({\cite{burns87}) measurement (S$_{\rm 1.5\,GHz}$=33.9 mJy), 
and it is slightly lower than the NVSS value (Tab.~\ref{tab:sample1}), 
probably due to the different angular resolution. 
Table \ref{tab:flux1} summarises the flux density and size of each component. 
\\
\\
%%%%%%%%%%%%%%%%%%%% fig. 5  MKW06 %%%%%%%%%%%%%%%%%%%%%%%%%%%%%%%%%%% 
\begin{figure*}
\centering
\hspace{0.7truecm}
\caption{MKW06 -- GMRT 1.28 GHz full resolution (on SDSS optical
image, {\it left panel}) and low
resolution ({\it right panel}) of the central radio galaxy. The 1$\sigma$ level in the image is 
20 $\mu$Jy b$^{-1}$ and 25 $\mu$Jy b$^{-1}$ respectively. Logarithmic contours 
are reported, starting from $\pm$4$\sigma$. The contour peak flux is 
2.81 mJy b$^{-1}$ and 7.00 mJy b$^{-1}$ in left and right panels respectively. 
The HPWB is $5.4^{\prime\prime} \times 2.8^{\prime\prime}$, p.a. $47^{\circ}$ 
in the left panel, and $10.9^{\prime\prime} \times 8.9^{\prime\prime}$, 
p.a. $62^{\circ}$ in the right panel. Letters A, B and C indicate the source 
components. For this source 1$^{\prime\prime}$ = 1.03 kpc.}
\label{fig:mkw06}
\end{figure*}
%%%%%%%%%%%%%%%%%%%%%%%%%%%%%%%%%%%%%%%%%%%%%%%%%%%%%%%%%%%%%%%%%%%%%%%%%%
%
{\it AWM\,05} -- In Fig.~\ref{fig:awm05} we show the GMRT 1.28 GHz 
full resolution contours of  the source, overlaid on the POSS--2 optical frame.
The source is resolved into a triple. The three components 
are labelled as A, B and C. The whole radio source is located within the 
optical galaxy %(NGC\,6269), 
and has a largest linear  
size of $\sim$35 kpc along the North--South axis.
The axis of the central component A is misaligned with respect 
to the whole source elongation. 

No other images at 1.4 GHz are available in the literature for this source,
except for the NVSS and FIRST surveys.
The total flux density of the source is given in Tab.~\ref{tab:flux1}, 
and is in good agreement with the 1.4 GHz NVSS value (Tab. 1), 
despite the different angular resolution. 
The amount of total flux density measurements for this source allowed
us to study its integrated spectrum. Our analysis is reported in 
Section \ref{sec:spec_awm05}.
\\
\\
%%%%%%%%%%%%%%%%%%%%% fig. 6  AWM05%%%%%%%%%%%%%%%%%%%%%%%%%%%%%%%%%%
\begin{figure}
\centering
\caption{AMW05 -- GMRT 1.28 GHz radio contours of the central radio galaxy
overlaid on the optical POSS--2 frame. The 1$\sigma$ level in the image is 
25 $\mu$Jy b$^{-1}$. Logarithmic contours are reported, starting from 
$\pm$0.075 mJy b$^{-1}$. The contour peak flux is 8.29 mJy b$^{-1}$. 
The HPWB is $3.2^{\prime\prime} \times 2.0^{\prime\prime}$, p.a. $-72^{\circ}$.
The source components are labelled as A, B and C. For this source 
1$^{\prime\prime}$=0.69 kpc.}
\label{fig:awm05}
\end{figure}
%%%%%%%%%%%%%%%%%%%%%%%%%%%%%%%%%%%%%%%%%%%%%%%%%%%%%%%%%%%%%%%%%%%%%%

\subsection{Clusters observed at 235 \& 610 MHz}\label{sec:235-610}

In this section we present the radio images of the clusters observed both 
at 235 MHz and 610 MHz (Tab.~\ref{tab:obs}). For all sources the radio 
information 
at each frequency is summarised in Tab.~\ref{tab:flux2}, where we give  
the flux density, the 235 MHz--610 MHz spectral index, and the largest
linear size as estimated from the 235 MHz image (unless specified otherwise). 
Individual information is listed for those sources whose structure can
be unambiguously separated into more than one component.
A brief morphological description of the sources is given below.
\\
\\
%%%%%%%%%%%%%%%%%%%%%%%%%%%%% Tab.4 %%%%%%%%%%%%%%%%%%%%%%%%%%%%%%%%%%%%%%%%%%%
\begin{table*}[h!]
\caption[]{Radio source data at 235 and 610 MHz}
\begin{center}
\begin{tabular}{rccccc}
\noalign{\smallskip}
\hline\noalign{\smallskip}
Cluster & component & S$_{\rm 235\,MHz}$ & S$_{\rm 610\,MHz}$ & $\alpha_{\rm 235\,MHz}^{\rm 610\,MHz}$ $\pm 0.07$ & 
 LLS \\ 
        &           &  (mJy) & (mJy)  &  &  kpc  \\
\hline\noalign{\smallskip}			      
A\,1775  & tot. & 2035 &  896 & 0.85 &         --            \\
         &  D   &  135 &   62 & 0.82 & $\sim  56 \times 21$  \\
	 &  HT  & 1900 &  834 & 0.86 & $\sim 381 \times 56$  \\
&&&&&\\	     
A\,2162  & tot. &  218$^{(*)}$  & 240& 0.10$^{(*)}$ & $\sim 90\times38$ \\
&&&&&\\
A\,2372  & tot  & 2657 & 1106 & 0.92 & $\sim 842 \times 117$ \\
         & A    &    8 &   14 &--0.59& $\sim  28 \times  15^{\diamond}$ \\ 
         & B    & 1310 &  616 & 0.79 & $\sim 260 \times 115$ \\
         & C    & 1410 &  476 & 0.92 & $\sim 380 \times 100$ \\
&&&&&\\
A\,2622  & tot. & 1184 &  310 & 1.40 & $\sim 180 \times 60$ \\
         & A    &  --  &  193 &  --  & $\sim  32 \times 24^{\diamond}$ \\
         & B    &  --  &   59 &  --  & $\sim  47 \times 47^{\diamond}$ \\ 
         & C    &  --  &   54 &  --  & $\sim  47 \times 35^{\diamond}$ \\ 
	       &&&&&\\
MKW\,02  & tot. & 1700 &  245 & 2.03 & $\sim$670            \\
         &  A   &  571 &   60 & 2.36 & $\sim 135 \times 90$\\
	 &  B   &  940 &  150 & 1.92 & $\sim 180 \times 90$\\
         &  C   &   50 &   31 & 0.50 &     $<2.1$           \\
	 &  D   &   15 &    3 & 1.69 &     $\sim 23$        \\
\noalign{\smallskip}
\hline\noalign{\smallskip}
\end{tabular}
\end{center}
$^{\star}$ this value has large uncertainties due to problems with the secondary 
calibrator at 235 MHz \\
$^{\diamond}$ values estimated from the full resolution 610 MHz image
%{\it(a)}: the error is $\pm 0.07$;
\label{tab:flux2}
\end{table*}
%%%%%%%%%%%%%%%%%%%%%%%%%%%%%%%%%% End Tab. 4 %%%%%%%%%%%%%%%%%%%%%%%%%%%%%%%%%
%
%
{\it A\,1775} -- The rich Abell cluster A\,1775 is known to host a binary 
galaxy system  (dumb--bell galaxy) at its centre, with two optical components 
of similar luminosity, separated by a projected distance of 32 kpc 
(Parma et al. \cite{parma91}). 
In Fig.~\ref{fig:a1775} we show the GMRT full resolution radio images at 
610 MHz (left panel) and 235 MHz (right panel) of A\,1775.
At 610 MHz (left panel, superposed to the optical image), we can clearly 
recognise the different radio 
structures associated with the two galaxies: the North--Western galaxy
hosts a $\sim$30$^{\prime \prime}$ ($\sim 42$ kpc)
double radio source, labelled as D; a head--tail radio source, 
labelled as HT  and extending on a scale of $\sim 270 ^{\prime \prime}$ 
($\sim 380$ kpc), is associated with the companion galaxy.
As clear from the 610 MHz image shown in left panel of Fig.~\ref{fig:a1775}, 
the radio tail shows a wiggling structure.
The morphology of the dumb--bell system is consistent with the 
1.4 GHz VLA image presented in Owen \& Ledlow (\cite{owen97}).

%%%%%%%%%%%%%%%%%%%%%%%% fig.7 -  A1775 %%%%%%%%%%%%%%%%%%%%%%%%%%%%%%% 
\begin{figure*}
\centering
\hspace{1.5truecm}
\caption{A\,1775 -- GMRT radio contours at 610 MHz ({\it left panel}),
overlaid on the POSS--2 optical frame, and 235 MHz ({\it right panel}) 
of the central cluster region. The 1$\sigma$ 
level in the image is 130 $\mu$Jy b$^{-1}$ and 1.5 mJy b$^{-1}$  respectively.
Logarithmic contours are reported, starting from $\pm$3 $\sigma$. The contour 
peak flux is 118.3 mJy b$^{-1}$ and 480.8 mJy b$^{-1}$ in left and right 
panels respectively. The HPWB is $5.4^{\prime\prime} \times 
4.2^{\prime\prime}$, p.a. $49^{\circ}$ in the 610 MHz image and
$12.7^{\prime\prime} \times 9.1^{\prime\prime}$, p.a. $63^{\circ}$ in the
235 MHz image. D indicates the double radio source associated with the central 
cD galaxy, while HT stands for the head--tail radio galaxy associated with 
the companion galaxy. For this system 1$^{\prime\prime}$=1.41 kpc.}
\label{fig:a1775}
\end{figure*}
%%%%%%%%%%%%%%%%%%%%%%%%%%%%%%%%%%%%%%%%%%%%%%%%%%%%%%%%%%%%%%%%%%%%%%%

In Table \ref{tab:flux2} we give the flux density at 235 MHz and 610 MHz and
the spectral index for the whole system (D+HT) as well as for the individual
radio galaxies.

For this source we carried out a spectral study, reported in 
\ref{sec:spec_a1775}.
\\
\\
{\it A\,2162} -- The GMRT full resolution images at 235 MHz and 610 MHz of 
the source are shown in the right and left panels of Fig.~\ref{fig:a2162} 
respectively.  The source is characterised
by two opposite lobes, with lack of a central compact component at
both frequencies. The radio emission is weak and the surface brightness
low. A similar double morphology without central component was found also at 
1.4 GHz by Owen \& Ledlow (\cite{owen97}).
The source has a linear size of $\sim 90 \times 38$ kpc. 
The total flux density of the source at 610 MHz and 235 MHz, and the spectral 
index value between these two frequencies are given in Table \ref{tab:flux2}.  
We point out that the 235 MHz measurement is most likely underestimated, due to
problems with the flux density scale of the secondary calibrator (see also
note to Table \ref{tab:flux2}), therefore the reported value for the
spectral index should be taken with care. A detailed spectral study
for this source is reported in Sect. \ref{sec:spec_a2162}.
\\
\\
%%%%%%%%%%%%%%%%%%%%%% fig. 8 - A2162%%%%%%%%%%%%%%%%%%%%%%%%%%%%%%%%%%%%% 
\begin{figure*}
\centering
\hspace{0.7cm}
\caption{A\,2162 -- GMRT radio contours at 610 MHz, overlaid on the  SDSS 
optical frame ({\it left panel}), and 235 MHz ({\it right panel}) of the 
central 
radio galaxy. The 1$\sigma$ level in the image is 150 $\mu$Jy b$^{-1}$ at 610 
MHz and 1 mJy b$^{-1}$ at 235 MHz. Logarithmic contours are reported, starting 
from $\pm$3$\sigma$. 
The HPWB is $7.1^{\prime\prime} \times 4.9^{\prime\prime}$, p.a. $-83^{\circ}$
in the left panel and  $14.7^{\prime\prime} \times 10.8^{\prime\prime}$, 
p.a. $-85^{\circ}$ in the right panel. 
For this source 1$^{\prime\prime}$ = 0.64 kpc.}
\label{fig:a2162}
\end{figure*}
%%%%%%%%%%%%%%%%%%%%%%%%%%%%%%%%%%%%%%%%%%%%%%%%%%%%%%%%%%%%%%%%
%
{\it A\,2372} -- The cD galaxy in A\,2372 hosts a spectacular large radio 
source with a 
wide--angle--tail (WAT) morphology. Our GMRT full resolution images at 610 MHz 
and 235 MHz are shown in the upper and lower panels of Fig.~\ref{fig:a2372} 
respectively. The total angular extent of the WAT (end to end) is 
$\sim 730^{\prime \prime}$, corresponding to a linear size of $\sim$ 850 kpc.
Our images are in very good agreement with the morphologies observed
with the VLA at 1.4 GHz and 4.9 GHz and reported in 
Owen \& Ledlow (\cite{owen97}) and Gregorini et al. (\cite{gregorini94})
respectively.

%%%%%%%%%%%%%%%%%%%%%%%%% fig. 9 - A2372%%%%%%%%%%%%%%%%%%%%%%%%%%%%%%%%%%%%%
\begin{figure*}
\centering
\caption{A\,2372 -- GMRT 610 MHz, on POSS--2 optical image, 
({\it upper panel}) and 235 MHz ({\it lower panel}) radio contours of the 
central radio galaxy overlaid on 
the POSS--2 optical frame. The 1$\sigma$ level in the image is
65$\mu$Jy b$^{-1}$ and 1.5 mJy b$^{-1}$ respectively. 
In both panels logarithmic contours are reported, starting from $\pm$3$\sigma$.
The HPWB is $7.5^{\prime\prime} \times 5.7^{\prime\prime}$, p.a. 
$38^{\circ}$ in the 610 MHz image and 
$19.7^{\prime\prime} \times 13.8^{\prime\prime}$, p.a. $-8^{\circ}$ 
at 235 MHz. In both panels letters A, B, and C indicate the different source
components. For this source 1$^{\prime\prime}$ = 1.17 kpc.}
\label{fig:a2372}
\end{figure*}
%%%%%%%%%%%%%%%%%%%%%%%%%%%%%%%%%%%%%%%%%%%%%%%%%%%%%%%%%%%%%%%%%%%%%%%%%%

The central component, coincident with the optical cD galaxy (labelled as
A in Fig.~\ref{fig:a2372}), shows a double morphology, as clear from the 
insert in the upper panel of the figure, which zooms into the nuclear emission 
at 610 MHz. Such features are most likely the inner jets. On a larger
scale, the jets show symmetric wiggles until they lose collimation and
form the lobes (B and C in Fig.~\ref{fig:a2372}). A sharp edge in the western 
lobe suggests strong interaction with the intracluster medium, but
unfortunatley no high resolution X--ray observations are available for 
this cluster to investigate the origin of this feature.
\\
The total flux densities at 610 MHz and 235 MHz and the spectral index between 
these two frequencies are given in Tab.~\ref{tab:flux2}.  
A detailed spectral study is reported in Section \ref{sec:spec_a2372}.
\\
\\
{\it A\,2622} -- Fig.~\ref{fig:a2622} presents the GMRT full resolution 
images of the source at 610 MHz (left panel) and 235 MHz (right panel). 
The whole source extends over a total size of 
$\sim$ 120 kpc. North and South of the bright central component A, two 
diffuse 
radio lobes are detected, labelled respectively as B and C in the left panel
of the figure. Comparison of the images suggests that very low frequency 
emission is present in this source.
A\,2622 was observed with the VLA at 1.4 GHz at high
resolution (A and B array) by Owen \& Ledlow (\cite{owen97}). They could
detect only component A in our images.
At 4.7 GHz Ball et al. (\cite{ball93}) did not detect 
any radio emission from this galaxy at a sensitivity limit 
of 0.69 mJy. 

The total flux density of the source at 235 MHz and 610 MHz and the resulting 
spectral index are given in Tab.~\ref{tab:flux2}, where we also give 
the 610 MHz flux density of the individual source components.

A detailed spectral analysis for this source is given in Sec. 
\ref{sec:spec_a2622}.
\\
\\
%%%%%%%%%%%%%%%%%%%%%%%%%%%%% fig.10  A2622%%%%%%%%%%%%%%%%%%%%%%%%%%%%%%%%%%
\begin{figure*}
\centering
\hspace{1.5truecm}
\caption{A\,2622 -- GMRT radio contours at 610 MHz, overlaid on the
POSS--2 optical image ({\it left panel}), and 235 MHz ({\it right panel}) of 
the central radio galaxy. The 
1$\sigma$ level in the image is 50 $\mu$Jy b$^{-1}$ and 800 $\mu$Jy b$^{-1}$ 
respectively. Logarithmic contours are reported starting from $\pm$3$\sigma$ 
and $\pm$4$\sigma$ in the left and right panel respectively. The HPWB is 
$4.9^{\prime\prime} \times 4.2^{\prime\prime}$, p.a. $52^{\circ}$ in the 610 
MHz image and  $17.1^{\prime\prime} \times 11.4^{\prime\prime}$, p.a. 
$53^{\circ}$ in the 235 MHz image. In the left panel the letters A, B, and C
indicate the different source components. 
For this source 1$^{\prime\prime}$ = 1.18 kpc.}
\label{fig:a2622}
\end{figure*}
%%%%%%%%%%%%%%%%%%%%%%%%%%%%%%%%%%%%%%%%%%%%%%%%%%%%%%%%%%%%%%%%%%%%%%%%%%%%%%
%
{\it MKW\,02} -- The GMRT full resolution images at 235 MHz and 610 MHz are 
presented in  Fig.~\ref{fig:mkw02}.
The central cD galaxy is associated with a very large radio 
galaxy ($\sim 15^{\prime}$, $\sim$ 670 kpc), whose radio power is typical 
of FR\,I radio galaxies (see Table 1), but whose radio morphology is similar 
to that of FR\,II radio sources, with two jets, asymmetric in size and 
brightness, lobes and possibly hot spots.

At the resolution and frequency of our images only the central component 
associated with the cD galaxy (labelled as C in Fig.~\ref{fig:mkw02}) and 
the outer regions of the lobes (labelled as A and B) are visible. The
southern jet is completely resolved out at both frequencies, and hints of
the northern one (D in Fig.~\ref{fig:mkw02}) are visible only at 235 MHz. 
Our images are fully consistent with a 1.4 GHz VLA image reported in 
Burns et al. (\cite{burns87}).

%%%%%%%%%%%%%%%%%%%%%%%%% fig. 11 - MKW02 %%%%%%%%%%%%%%%%%%%%%%%%%%%%%%
\begin{figure*}
\centering
\hspace{1.5truecm}
\caption{MKW\,02 -- GMRT radio contours at 610 MHz, overlaid on the
POSS--2 optical image ({\it left panel}), and 235 MHz ({\it right panel}) of 
the central radio galaxy. The 1$\sigma$ level 
in the image is 130 $\mu$Jy b$^{-1}$ and 650 $\mu$Jy b$^{-1}$ respectively. 
Logarithmic contours are reported, starting from $\pm$3$\sigma$. 
The HPWB is $9.7^{\prime\prime} \times 8.6^{\prime\prime}$, p.a. $-24^{\circ}$
in the 610 MHz image and  $12.1^{\prime\prime} \times 9.7^{\prime\prime}$, 
p.a. $65^{\circ}$ in the 235 MHz image. In both panels letters A, B, C, and D 
indicate the different source components. 
For this source 1$^{\prime\prime}$ = 0.75 kpc.}
\label{fig:mkw02}
\end{figure*}
%%%%%%%%%%%%%%%%%%%%%%%%%%%%%%%%%%%%%%%%%%%%%%%%%%%%%%%%%%%%%%%%%%%%%%

The total flux density of the source at 235 MHz and 610 MHz and the derived 
spectral index are given in Tab.~\ref{tab:flux2}, where we also report 
the 610 MHz flux density and size of the source components.

The total flux density value at 610 MHz is most likely understimated, even
though no apparent calibration problem was encountered during the data 
reduction. We measured S$_{\rm 610~MHz}$ = 245 mJy, to be compared with the 
value from the NVSS, i.e. S$_{\rm 1.4 GHz}$=385 mJy. The 1.4 GHz 
flux density measured by Burns et al. (\cite{burns87}) is different from
the NVSS 
one, being S$_{\rm 1.4 GHz}$=668 mJy (resolution $\sim 13^{\prime \prime}$).
These values
seem to disagree even if we allow for some variability of the central
component, since it contains only a small fraction of the total flux 
(12\% at 610 MHz). Such disagreements suggest that the u--v coverage 
is critical for a source large and complicated as this one.
For this reason we think that the spectral index
values reported in Tab.~\ref{tab:flux2} should be taken with caution, and
considered as upper limits to the value of $\alpha$.
Due to the uncertain values of the flux density we did not undertake any 
study of the radio spectrum.

%%%%%%%%%%%%%%%%%%%%%%%%%%%%%%%%% Tab.4 %%%%%%%%%%%%%%%%%%%%%%%%%%%%%%%%%%%%%%
\begin{table*}[h!]
\caption[]{Radio source data at 235 MHz, 610 MHz and 1.28 GHz}
\begin{center}
\begin{tabular}{rcccccc}
\noalign{\smallskip}
\hline\noalign{\smallskip}
Cluster & component &  S$_{\rm 235\,MHz}$  & S$_{\rm 610\,MHz}$ &S$_{\rm 1.28\,GHz}$ & 
$\alpha_{\rm 235\,MHz}^{\rm 1.28\,GHz}\pm 0.04$ & LLS \\ 
        &           &  (mJy) & (mJy)  &  (mJy) &  &  kpc \\
\hline\noalign{\smallskip}			      
MKW\,07 & tot. & 700 & 244 &  95 & 1.18 & $\sim 98 \times 43$ \\
        &  A   & 294 & 110 &  44 & 1.13 & $\sim 43 \times 24$ \\
	&  B   & 392 & 131 &  50 & 1.22 & $\sim 43 \times 37$ \\
        &  C   &  14 &   3 &   1 & 1.67 & $\sim 1.6 \times 0.4^{\diamond} $ \\ 	
&&&&&&\\
MKW\,03s & tot. & 8436 & 1010 & 139 & 2.42 & $\sim 170 \times 80$ \\
         & A  & 7810 &  952 & 119 & 2.47 & $\sim  98 \times 90$ \\
         & B  &  586 &   40 &  10 & 2.40 & $\sim  58 \times 53$ \\
         & C  &   40 &   18 &  10 & 0.82 & $< 9$                \\
\noalign{\smallskip}
\hline\noalign{\smallskip}
\end{tabular}
\end{center}
$^{\diamond}$ this value is estimated from the 1.28 GHz image
%{\it(a)}: the error is $\pm 0.07$;
\label{tab:flux3}
\end{table*}
%%%%%%%%%%%%%%%%%%%%%%%%%%%%%%% End Tab. 4 %%%%%%%%%%%%%%%%%%%%%%%%%%%%%%%%

\subsection{Clusters observed at 235 MHz, 610 MHz and 1.28 GHz}\label{sec:235-610-1.28}

In this section we present the radio images of the poor clusters MKW\,07 and 
MKW\,03s, observed with the GMRT at 235 MHz, 610 MHz and 1.28 GHz 
(Tab.~\ref{tab:obs}). 
For these sources the radio information at each frequency is summarised in 
Tab.~\ref{tab:flux3}, where we give the flux density, the 235 MHz--1.28 GHz 
spectral index, and the largest linear size LLS, estimated from the 235 MHz 
image (unless specified otherwise). The radio information is given also for
each source component. Individual comments are reported below.
\\
\\
{\it MKW\,07} --  The cD galaxy at the centre of the poor cluster MKW\,07 
is embedded  in an asymmetric envelope, including also a compact fainter 
companion 
(Van den Bergh \cite{vandenberg77}). 
The radio emission from this galaxy is given in Fig.~\ref{fig:mkw07}, where 
we present the full resolution images at 1.28 GHz, overlaid on the
optical frame (left panel), 610 MHz (central panel) and  235 MHz (right panel).
Note that the bright object west of the central cD galaxy
is a star. 
At all frequencies the morphology of the source is dominated by two radio 
lobes, labelled as A and B in the figure. The source has a largest linear 
size of $\sim$ 98 kpc in the 235 MHz image.

%%%%%%%%%%%%%%%%%%%%%%%%%%%%%% fig.12  MKW07 %%%%%%%%%%%%%%%%%%%%%%%%%%%%%%% 
\begin{figure*}
\centering
\caption{MKW\,07 -- GMRT radio contours at 1.28 GHz, overlaid on the SDSS
optical frame ({\it left panel}), 
610 MHz ({\it central panel}) and 235 MHz ({\it right panel}) of the central 
radio galaxy. The 1$\sigma$ level in the 
image is 30 $\mu$Jy b$^{-1}$, 100 $\mu$Jy b$^{-1}$ and 1.5 mJy b$^{-1}$ 
respectively. Logarithmic contours are reported, starting from $\pm$3$\sigma$.
The contour peak flux is 0.96 mJy b$^{-1}$ in the left panel, 
3.52 mJy b$^{-1}$ in the central panel, and 33.51 mJy b$^{-1}$ in the right 
panel. The HPWB is $3.5^{\prime\prime} \times 2.4^{\prime\prime}$, p.a. 
$49^{\circ}$ at 1.28 GHz, $5.8^{\prime\prime} \times 4.4^{\prime\prime}$, 
p.a. $57^{\circ}$ at 610 MHz and 
$14.0^{\prime\prime} \times 9.1^{\prime\prime}$, p.a. $50^{\circ}$ at 235 MHz. 
In all panels letters A, B and C indicate the different components of the 
source. For this source 1$^{\prime\prime}$ = 0.61 kpc.}
\label{fig:mkw07}
\end{figure*}
%%%%%%%%%%%%%%%%%%%%%%%%%%%%%%%%%%%%%%%%%%%%%%%%%%%%%%%%%%%%%%%%%%%%%%%%%%%%%%%

A compact component, labelled as C and centered on the nucleus of the brightest
galaxy, is detected clearly at 1.28 GHz (left panel of Fig.~\ref{fig:mkw07}) 
and is visible also at the other two frequencies. No jets or other features 
connect the nuclear component C with the two lobes. On the other hand, the
two lobes are connected by a bridge of emission, visible at 1.28 GHz and 
610 MHz. This feature is reminiscent of the radio galaxy 3C\,338,
associated with the multiple nuclei galaxy at the centre of A\,2199,
whose morphology is interpreted in terms of an old radio galaxy with 
restarted nuclear activity (Giovannini et al. \cite{giovannini98}).

The total flux density at each frequency and the spectral index between 
235 MHz and 1.28 GHz are given in Tab.~\ref{tab:flux3}. 
The GMRT 1.28 GHz flux density measurement is in good agreement with the 
1.4 GHz NVSS value (Table 1), despite the different angular resolution.

The total spectrum and a spectral index image for this source are 
discussed in Section \ref{sec:spec_mkw07}.
\\
\\
{\it MKW\,03s} -- In Fig. ~\ref{fig:mkw03s} we present the new GMRT 235 MHz 
image of the source and refer to Mazzotta et al. (\cite{mazzotta04}) 
and Giacintucci et al. (\cite{giacintucci06}) for the GMRT radio images 
at 1.28 GHz and 610 MHz.
As observed at 1.28 GHz, 610 MHz, and at 327 MHz (VLA data, 
de Young \cite{young04}, also reported in Mazzotta et al. \cite{mazzotta04}), 
the source shows two opposite radio lobes. The southern one (labelled as A) 
is brighter than the northern one (labelled as B). 
A compact component (labelled as C) is detected in coincidence with the 
nucleus of the cD galaxy.
The flux density of the various source components at 235 MHz are reported in 
Tab.~\ref{tab:flux3}, where we summarise also the values at 610 MHz and 
1.28 GHz, taken from Mazzotta et al. (\cite{mazzotta04}). 
The radio emission from the cD in MKW\,03s is dominated at all the frequencies 
by the southern lobe A, which accounts for most of the total flux density of 
the source. 
A study of the total spectrum of this source is reported in Section
\ref{sec:spec_mkw03s}.

This cluster is one of the most amazing presented here. 
A combined radio (GMRT 1.28 GHz and 610 MHz) and X--ray analysis of the cD 
radio source in MKW\,03s  has already been published in Mazzotta et al. 
(\cite{mazzotta04}) and Giacintucci et al. (\cite{giacintucci06}). 
In those papers we discussed the 
connection between the radio activity of the central AGN and the features 
observed in the X--ray surface brightness and temperature structure of the 
cluster. 
MKW\,03s shows two opposite cavities in the X--ray surface brightness, filled 
by the radio emission from the lobes of the central radio galaxy.
The lobes appear confined by extended gas regions with temperature 
significantly higher than the radially averaged gas temperature at any radius. 
A possibile interpretation of this complex X--ray structure is that the radio 
lobes are expanding into 
the external ICM and heating the gas. Hence, MKW\,03s may be an example of
gas heated by the central AGN. A preliminary analysis of the synchrotron 
spectrum of the lobes provides radiative ages in reasonable agreement 
with this scenario (Mazzotta et al. \cite{mazzotta04}). A study 
of the implications of the possible connection between the morphological
and spectral properties of the source and the cluster thermal structure
will be discussed in details in a forthcoming paper. 

%%%%%%%%%%%%%%%%%%%%%%%%%% fig. 13   MKW03s%%%%%%%%%%%%%%%%%%%%%%%%%%%%%%%%%%
\begin{figure}
\centering
\caption{MKW\,03s -- GMRT 235 MHz radio contours of the central radio galaxy.,
The 1$\sigma$ level in the image is 2 mJy b$^{-1}$. 
Logarithmic contours are reported, starting from $\pm$6 mJy b$^{-1}$. 
The HPWB is $12.1^{\prime\prime} \times 9.1^{\prime\prime}$, p.a. $52^{\circ}$.
Letters A, B and C the different source component.
For this source 1$^{\prime\prime}$ = 0.89 kpc.}
\label{fig:mkw03s}
\end{figure}
%%%%%%%%%%%%%%%%%%%%%%%%%%%%%%%%%%%%%%%%%%%%%%%%%%%%%%%%%%%%%%%%%%%%%%%%%%%%%%%

\subsection{Linear sizes and the cluster environment}

A natural question in the study of radio galaxies located at the
cluster centres is whether the external dense intracluster medium
confines the radio structure, affecting their morphology and limiting their 
total size.

Ledlow et al. ({\cite{ledlow02}) compared the size distribution of
radio galaxies as function of the radio power for cluster and non--cluster 
sources (including both FR\,I and FR\,II types), and found no significant 
difference between the two samples. Using the combined cluster--non cluster 
sample, they found that the linear size of FR\,I sources tends to increase 
with the increasing radio power,
while they did not find any 
dependence between radio power and size for FR\,II sources.

Radio galaxies at the cluster centres represent the extreme population 
of cluster radio galaxies.
Using the radio power in Table 1 and the size from our 1.28 GHz (or 610 MHz) 
images (see Tables 3, 4 \& 5) we plotted the logP -- LLS diagram for the
cD galaxies under discussion here, which is reported in Fig.~\ref{fig:p-las}. 
The sources located at the centre of the rich Abell clusters are indicated as 
filled triangles, while empty triangles represent the sources in poor 
clusters. 
Our sample covers only a part of the wide radio power range explored by 
Ledlow et al. ({\cite{ledlow02}), since our objects have radio powers from 
P$_{\rm 1.4\,GHz} \sim 10^{22}$ W Hz$^{-1}$ to  
P$_{\rm 1.4\,GHz} < 10^{25}$ W Hz$^{-1}$. 
Within this interval of radio power the linear size distribution of our 
sources mainly extends over two orders of magnitude, being contained in the 
LLS range 10--1000 kpc (Fig.~\ref{fig:p-las}). The only exception 
is the unresolved radio galaxy in MKW\,01s ($\S$ \ref{sec:1.28}), which is 
the least powerful source in our sample and has a size upper limit of only 
0.2 kpc. We note that the observed range of source size is consistent with 
what expected in this radio power range from the (P$_{\rm 1.4\,GHz}$, LLS) 
diagram in Ledlow et al. ({\cite{ledlow02}). 
The fit obtained using the whole sample is 
LLS$\propto$P$_{\rm 1.4\,GHz}^{1.11\pm0.19}$ (solid line in 
Fig.~\ref{fig:p-las}), while, excluding the MKW\,01s data point, we find  
LLS$\propto$P$_{\rm 1.4\,GHz}^{0.83\pm0.13}$ (dashed line).  
\\
The fit obtained using only the Abell clusters,  
LLS$\propto$P$_{\rm 1.4\,GHz}^{1.01\pm0.19}$, is consistent within the errors 
with the fit obtained using only the poor clusters, i.e. 
LLS$\propto$P$_{\rm 1.4\,GHz}^{1.36\pm0.21}$ 
(LLS$\propto$P$_{\rm 1.4\,GHz}^{0.86\pm0.16}$ if we exclude MKW\,01s).
This consistency, however, may be affected by the large errors in
the fit, due to the limited statistics of our sample.
A similar analysis on a larger sample may allow us to draw a final
conclusion on this issue.
\\
Finally, we note that all the most powerful radio sources are 
located in rich clusters, while the least powerful is located in a poor group.
\\
%%%%%%%%%%%%%%%%%%%%%%%%%%%%% fig. 14 power-size plot %%%%%%%%%%%%%%%%%%%%%%
\begin{figure}
\centering
\caption{1.4 GHz radio power--radio size diagram for the combined rich 
(filled triangles) and poor (empty triangles) cD cluster sample.}
\label{fig:p-las}
\end{figure}
%%%%%%%%%%%%%%%%%%%%%%%%%%%%%%%%%%%%%%%%%%%%%%%%%%%%%%%%%%%%%%%%%%%%%%%%%%%%

\section{Spectral Analysis}\label{sec:spectra}

For a number of sources, the amount of information presented in this paper, 
coupled with literature and archival data, allowed us to carry out a radio 
spectral study, to address the questions of the source radiative age and the
kinematics.
In this section we perform two types of analysis: {\it (a)} fit of the
integrated radio spectrum, and {\it (b)} fit of the trend of the 
two--point spectral index along the source structure.
\\
\begin{itemize}
\item [{\it (a)}] For eight sources we obtained the integrated source spectra 
making use of the database CATS (Verkhodanov et al. 1997), and fitted them 
with the Synage++ package (Murgia 2001). 
A continuous injection model (CI, i.e. the source is fuelled at a constant
rate, Kardashev \cite{kardashev62})
was adopted for the spectral fitting. The derived 
injection spectral index $\alpha_{\rm inj}$ and the break frequency 
$\nu_{\rm b}$ were then used to estimate the equipartition magnetic field  
B$_{\rm eq}$, and the the radiative age $\rm t_{rad}$ of the source, 
using the following equation:

\begin{equation}
$$ {\rm t_{\rm rad} = 1590  \frac{B_{\rm eq}^{0.5}}{(B_{\rm eq}^2 + B_{\rm CMB}^2)} [(1+z) \nu_{\rm b}]^{-0.5}} $$
\label{trad1}
\end{equation}

\noindent where $\rm t_{rad}$ is expressed in Myr, $\nu_{\rm b}$ in GHz, 
B$_{\rm eq}$ and B$_{\rm CMB}$ in $\mu$G, with 
B$_{\rm CMB} = 3.2(1+z)^2$ is the magnetic field strength with energy 
density equal to that of the cosmic microwave background (CMB) at the 
redshift z.

For the ultra steep spectrum source in MKW\,03s, a diffusion 
MJP model (Slee et al. \cite{slee01}) provided the best fit to the total 
spectrum (see Sect. \ref{sec:spec_mkw03s} for details). The model accounts
for the diffusion of the synchrotron electrons in a magnetic field with 
a Gaussian distribution of field strengths (Murgia, in prep.).
In this case the radiative lifetime of the electrons was calculated 
according to the equation

\begin{equation}
$$ {\rm t_{rad} = 1590  \frac{B_{\rm rms}^{0.5}}{(B_{\rm rms}^2 + B_{\rm CMB}^2)} [(1+z) \nu_{\rm b}]^{-0.5}}. $$
\end{equation}

where B$_{rms}$ is a mean field value over the distribution of field
strenghts.

\item[{\it (b)}] For five sources we performed a point--to--point spectral 
analysis, which was carried out on two sets of images produced using the 
same u--v range and same cellsize, and restored with the same beam. 
The individual images were corrected for the primary beam attenuation, 
aligned and clipped at the 3$\sigma$ level, and finally combined to obtain the 
spectral index map. 
For each source, details on the images and frequencies used for the analysis
are given below in the appropriate sections.
In each case we determined the average spectral index in circular regions,
starting from the inner edge of each lobe and going outwards along the
major axis, and derived the spectral index distribution as a function 
of distance from the core. The size of the regions selected for our
analysis was chosen large enough to ensure sufficient signal--to--noise--ratio
and independent data points.
\\
We fitted the trend following Parma et al. 
(\cite{parma99} and \cite{parma02}). In particular, we used the relation 
$\nu_{\rm b} \propto x^{-2}$, $x$ being the distance from the core. Such
relation is expected if a constant expansion velocity is assumed 
(i.e. $x \propto t$), and reflects the fact that the radiating electrons
become older as they move away from the nucleus, which is appropriate
for the type of sources discussed here (FR\,I radio galaxies).
\\ 
As detailed in the individual sessions, the five sources studied here differ
in their properties, and different models were used to derive the best fit.
\end{itemize}

For all sources, B$_{\rm eq}$ was computed assuming a cut--off in the minimum 
energy for the radiating electrons corresponding to a Lorentz factor 
$\gamma$=50 (Brunetti, Setti \& Comastri \cite{brunetti97}). 
We further note that the equipartition magnetic field was 
derived making use of the injection index of the radiative electrons,
scaling the radio flux density at the value expected for $\alpha_{inj}$.

Our analysis is presented in detail in the following subsections for each 
source individally, and the results are summarised in Table 6, where
we list the observed spectral index $\alpha_{obs}$, the injection spectral 
index $\alpha_{inj}$, the frequency break $\nu_{\rm b}$, the equipartition
magnetic field B$_{\rm eq}$, the radiative age t$_{\rm rad}$ and a crude
estimate of the source velocity growth v$_{\rm growth}$/c, obtained assuming 
a constant velocity and using the linear sizes reported for each source in 
Tables 3, 4 and 5, i.e. v$_{\rm growth}$ = LLS/t$_{\rm rad}$.

Where possible, we report the values derived from the point--to--point
analysis, in all other cases the results from the fit of the integrated
spectrum are given. 

\subsection{A\,2480. Integrated spectrum and spectral trend along the jets}\label{sec:spec_a2480}

We used our 1.28 GHz flux density in Tab.~ \ref{tab:flux1},
the flux densities reported in the literature and the 74 MHz VLSS measurement 
to derive the total synchrotron spectrum of the source, shown in 
Fig.~\ref{fig:a2480_sp}. The spectrum can be described as a power law with 
$\alpha=0.6$ in the 160 MHz--11.2 GHz frequency range. Below 160 MHz the 
spectrum has a negative slope with $\alpha=-0.2$.

As clear from Fig.~\ref{fig:a2480_sp}, the integrated spectrum is scattered,
however it is still consistent with a single power law. The fit, shown
as solid line in the Figure, provides $\alpha_{inj}=0.54\pm0.03$.
%
%%%%%%%%%%%%%%%%%%% fig.15 total spectrum  A2480 %%%%%%%%%%%%%%%%%%%%
\begin{figure}[h!]
\centering
\caption{A\,2480 -- Radio spectrum of the central radio galaxy between
74 MHz and 11.2 GHz. The empy circles are literature data, while the 
filled circle is the GMRT 1.28 GHz value. The solid line is the best fit
of the CI model.}
\label{fig:a2480_sp}
\end{figure}
%%%%%%%%%%%%%%%%%%%%%%%%%%%%%%%%%%%%%%%%%%%%%%%%%%%%%%%%%%%%%%%%%%%%

In Figure \ref{fig:spix_a2480} (left panel) we show the spectral index image 
of the source (colours), obtained by comparison of the GMRT
1.28 GHz image and the 4.87 GHz image (contours) from VLA archival data 
which we re--analyzed (Obs. Id. AB398). 
The figure shows the presence of a central flat spectrum region with 
$\alpha = 0.2 \pm 0.1$. Both jets and lobes in the source have an average 
spectral index $\alpha \sim 0.7 \pm 0.1$. These features suggest 
that the radio source is currently fed by an active nucleus.

Using the spectral index image, we derived the average spectral index of 
the source along the two jets. The distribution of $\alpha$ is given in
the right panel of Fig. \ref{fig:spix_a2480}, and the insert shows
the circular regions used during the analysis (red open circles 
overlaid on the contour image). 
The spectrum steepens along
the jet from $\sim 0.6$ at the jet basis to $\sim$ 1 in the outer regions.
Following well--known arguments, we interpreted the spectral trend along 
the lobes in terms of radiative ageing of the relativistic electrons by 
synchrotron and inverse Compton processes (e.g. Parma et al. \cite{parma99} 
and references therein). It is assumed that the 
radiative losses dominate over expansion losses and re--acceleration processes.
The spectral trend was fitted with a simple JP model (Jaffe \& Perola
\cite{jp74}). This model assumes an isotropic distribution of the pitch
angles of the radiating electrons with respect to the local direction of the
magnetic field. This condition is likely to be satisfied in most sources of 
our sample since the inverse Compton energy losses due to the microwave 
background radiation are as important as the synchrotron losses, and in the 
former random orientations are expected between electrons and photons. 
The JP model is characterized by three free parameters: 
i) the injection spectral index $\alpha_{inj}$,
ii) the break frequency $\nu_{b}$, and 
iii) the flux normalization. 
By assuming a constant magnetic field across the source and a constant speed
for the plasma outflow, we can obtain the injection spectral index and the 
minimum break frequency from a global fit of the observed spectral index 
profiles.
The model fit, which is represented by the blue lines in Figure 
\ref{fig:spix_a2480}, provided $\alpha_{inj}=0.55 \pm 0.03$, 
$\nu_{\rm b}=8.4^{+2.5}_{-1.6}$ GHz, 
the magnetic field obtained with this analysis is again B$_{\rm eq} = 10~\mu$G
and the derived radiative age is t$_{rad} = 1.5 \times10^7$ yr. 
It is important to mention that part of the observed spectral steepening 
could be due to a decrease of the magnetic field strength with the increasing 
distance from the core. If this is the case, the radiative age reported above 
should be considered as an upper limit to the source age.

%%%%%%%%%%%%%%%%%%%%%% fig.16 -  A2480 spectral index map %%%%%%%%%%%%%%%%%%
\begin{figure*}
\centering
\caption{A2480 -- Left: Colour scale image of the spectral index distribution 
between 1.28 GHz and 4.87 GHz over the central radio galaxy. It
has been computed from images with a restoring beam
of 2.9$^{\prime \prime} \times 2.0^{\prime \prime}$. Overlaid
are the VLA 4.87 GHz contours at levels 0.15, 0.30, 0.60, 1.20,
2.40, 4.80, 9.60 and 19.20 mJy. Right: 1.28--4.87 GHz spectral index
distribtion along the jets as function of distance from the core. The
solid lines represent the best fit of the radiative model described in
the text.}
\label{fig:spix_a2480}
\end{figure*}
%%%%%%%%%%%%%%%%%%%%%%%%%%%%%%%%%%%%%%%%%%%%%%%%%%%%%%%%%%%%%%%%%%%%%%%

\subsection{AWM\,05. Study of the integrated spectrum}\label{sec:spec_awm05}

Beyond the flux density data reported here (1.28 GHz, GMRT, 
Tab.~\ref{tab:flux1} and 1.4 GHz, NVSS, Tab. 1), 74 MHz and 2.38 GHz flux 
density measurements are also available in the literature 
(VLSS and Dressel \& Condon \cite{dressel78} respectively).
The resulting integrated spectrum is reported in Fig.~\ref{fig:awm05_sp}, 
and the spectral index in the range 74 MHz -- 2.38 GHz is $\alpha=0.8$. 

We assume that the integrated radio source spectrum is described by a 
continuous injection model (CI), where the source is continuously replenished 
by a constant flow of fresh relativistic particles with a power law energy 
distribution. 
Under these assumptions, it is well--known that the radio spectrum has a 
standard shape (Kardashev 1962), with spectral index $\alpha_{inj}$ below a 
critical frequency $\nu_{\rm b}$ and $\alpha_{\rm h}=\alpha_{inj}+0.5$ 
above $\nu_{\rm b}$. The CI spectral fit allows us to determine the non--aged 
spectral index $\alpha_{inj}$ and the break frequency $\nu_{\rm b}$, which
are the free parameters characterizing the model together with the flux 
normalization.
The CI model applies to those sources whose total spectrum is dominated by
the emission from the lobes, which accumulated most of the electrons produced 
over the entire source life. 
If the radiative losses are dominant over the expansion losses 
and the magnetic field is constant, the break frequency may be used to 
estimate the time elapsed since the source formation using 
Equation \ref{trad1}.

The best fit of the CI model to the integrated spectrum of AWM\,05 is reported 
in Fig.~\ref{fig:awm05_sp} (solid line), and provides a break frequency 
$\nu_{\rm b} = $ 1.1 GHz, and $\alpha_{inj}=0.58^{+0.25}_{-0.18}$.
The equipartition magnetic field and radiative age for the relativistic 
electrons are respectively B$_{\rm eq} \sim 7\, \mu$G and
t$_{\rm rad}=6.5\times10^7$ yr (see Section \ref{sec:ages} and Table 6).
\\
We note that also for this source the radio spectrum is poorly
constrained, and the parameters obtained with the fit should be considered
as indicative.

%%%%%%%%%%%%%%%%%%% fig. 17 AWM05 spectrum  %%%%%%%%%%%%%%%%%%%%%%%%%%%%
\begin{figure}[h!]
\centering
\caption{AWM05 -- Radio spectrum of the central radio galaxy between 74 MHz and
2.38 GHz. The empy circles are literature data (see text), while the filled 
circle is the GMRT 1.28 GHz value. The solid line is the best fit of the 
CI model.}
\label{fig:awm05_sp}
\end{figure}
%%%%%%%%%%%%%%%%%%%%%%%%%%%%%%%%%%%%%%%%%%%%%%%%%%%%%%%%%%%%%%%%%%%%

\subsection{A\,1775. Spectral analysis along the tail}\label{sec:spec_a1775}

We used the flux densities in Table \ref{tab:flux2} and the literature values 
(which include both the head--tail and the double source) to derive the 
integrated synchrotron spectrum of the whole radio emission associated with 
the dumb--bell system. The resulting spectrum is shown in 
Fig.~\ref{fig:sp_a1775_2}. It can be fitted with a single power law with 
$\alpha=1.0$ between 74 MHz and 10.7 GHz. 
Fig.~\ref{fig:sp_a1775_2} clearly shows the reliability of the GMRT flux 
density measurements at 610 MHz and  235 MHz presented here.  

From the flux density values given in Tab.~\ref{tab:flux2} it is clear that 
the long head--tail radio galaxy is the dominant component in the total 
spectrum.
The 235 MHz--1.4 GHz radio spectra of the head-tail and the double 
separately are shown in Fig.~\ref{fig:sp_a1775}, where we used the 1.4 GHz 
flux densities from the FIRST survey (Tab. 1).
The two spectra are very similar, and can be described 
as a power law with spectral index $\alpha_{235}^{1400} \sim$ 1.0.
%

%%%%%%%%%%%%%%%%%%%%%%%% fig. 18 A1775 spectrum tot %%%%%%%%%%%%%%%%%%%%%%
\begin{figure}
\centering
\caption{A\,1775 -- Radio spectrum of the central source between 74 MHz
and 10.7 GHz (the head tail and the double source are considered together), as 
derived from the GMRT 235 MHz and 610 MHz (filled circles) and 
literature data (empty circles). The solid line is the best fit of the 
CI model.}
\label{fig:sp_a1775_2}
\end{figure}
%%%%%%%%%%%%%%%%%%%%%%%%%%%%%%%%%%%%%%%%%%%%%%%%%%%%%%%%%%%%%%%%%%%%%%%

%%%%%%%%%%%%%%%%%%%%%%%% fig. 19 A1775 spectrum D+HT %%%%%%%%%%%%%%%%%%%%%%%
\begin{figure}
\centering
\caption{A\,1775 -- Radio spectrum of the double (D) and head--tail (HT) radio
galaxies. The 1.4 GHz flux density values for both sources are 
from the FIRST survey (empty circles). The filled circles are the GMRT
235 and 610 MHz data.}
\label{fig:sp_a1775}
\end{figure}
%%%%%%%%%%%%%%%%%%%%%%%%%%%%%%%%%%%%%%%%%%%%%%%%%%%%%%%%%%%%%%%%%%%%%%%

In Figure \ref{fig:spix_a1775} (left panel) we show the spectral index image 
of the two sources.
The image is obtained by comparison of the 235 MHz image (contours) and the 
610 MHz image (not shown here).
There is a clear separation in the spectral index distribution between the 
region dominated by the head of the tailed radio galaxy 
($\alpha \sim 0.6-0.8$), and the double source, being this latter steeper, 
with an average value of $\alpha = 1.0 \pm 0.2$, consistent with the spectrum 
in Fig.~\ref{fig:sp_a1775}.
There is no obvious gradient in the spectral index along the tail of the 
head--tail 
source, which appears rather uniform with an average value of 
$<\alpha>\sim 1.5 \pm 0.3$. 

We carried out the spectral index analysis along the tail of A\,1775,
following the approach described above.  
The spectral index (right panel of Fig. \ref{fig:spix_a1775})
steepens from $\sim 0.8$ to $\sim 1.2 - 1.4$ within
the first arcmin from the head, then it remains fairly constant.

In Section \ref{sec:spec_a2480}, we analyzed the spectral profiles in
A2480 assuming that the dominant processes producing the energy losses 
of the relativistic electrons are synchrotron and inverse Compton, neglecting 
both adiabatic expansion and re--acceleration. With the additional assumption 
of a constant advance speed of the synchrotron plasma, we found that 
the fit of the JP model to the data well describes the overall trend of 
the observed spectral index as a function of the distance from the core. 
However, in A\,1775 we observe a saturation of the spectral index along
the tail. This suggests that the break frequency decreases only in the
initial part of the tail, and remains constant going further out.
A possible explanation of this behaviour is that a re--acceleration processes 
are at work, which would partly compensate  
the radiative energy losses and cause a freezing of the break energy at that 
value where the radiative and the acceleration time scales are equal.
The spectral index trend in A\,1775 was therefore fitted with a modified 
JP model.

In particular, a JP model was considered, under the assumption that the radio 
galaxy moves through the intergalactic medium
at constant speed, and that the radiating electrons undergo radiative
losses and are systematically reaccelerated on a Fermi timescale t$_F$
(see Parma et al. \cite{parma99}).
\\
Under such hypothesis, the break frequency scales with time, and hence
with the distance along the tail, according to the following:

\begin{equation}
$$ {\rm \nu_{\rm b} \sim  \frac{B_{\rm eq}}{[(B_{\rm eq}^2 + B_{\rm CMB}^2\times 
t_F\times(1-e^{(-t_{\rm rad}/t_F)})]^2}}.$$
\end{equation}

Note that for t$_{\rm rad} <<$ t$_F$, the above equation provides the ``classical''
scaling

\begin{equation}
$$ {\rm \nu_{\rm b} \sim  \frac{B_{\rm eq}}{[(B_{\rm eq}^2 + B_{\rm CMB}^2)\times t_{\rm rad}]^2}}$$
\end{equation}

while for t$_{\rm rad} >> $ t$_F$ the break frequency is frozen to the value

\begin{equation}
$$ {\rm \nu_{\rm b} \sim  \frac{B_{\rm eq}}{[(B_{\rm eq}^2 + B_{\rm CMB}^2)\times t_F]^2.}}$$
\end{equation}

We further point out that the reacceleration modifies the shape of the 
spectrum in the ``oldest'' part of the tail, where t$_{\rm rad} >$ t$_F$.
\\
The fit provides a break frequency $\nu_{\rm b}=  644^{+195}_{-120}$ MHz, and 
$\alpha_{\rm inj}$=0.6$\pm$0.1. These values which yield an 
equipartition magnetic field B$_{\rm eq} = 8 \mu$G. 
\\
The reacceleration model provides a Fermi scale 
t$_F=6.9^{+0.7}_{-0.8}\times 10^7$ yr, a radiative age 
t$_{\rm rad} = 4.9\pm2.5 \times10^8$ yr and a galaxy velocity 
v$_{gal} = 755^{+299}_{-249}$ km/s.

It is worth mentioning that the fit of the simple JP model without 
re--acceleration (not shown) yields a radiative age of 
t$_{\rm rad} \simeq 5.9\times10^7$ yr. 
This value would underestimate the source age by almost an order of magnitude 
and hence lead to an implausibly high advance speed for the 
galaxy through the ICM, i.e. v$_{gal} \simeq 6300$ km/s.

%%%%%%%%%%%%%%%%%%%%%%% fig. 20 A1775 spectral index map %%%%%%%%%%%%%%%%%%%%
\begin{figure*}
\centering
\caption{A\,1775 -- Left: Colour scale image of the spectral index 
distribution between 235 MHz and 610 MHz over the head--tail radio galaxy and 
the double source, as computed from images with a restoring beam
of 12.7$^{\prime \prime} \times 9.1^{\prime \prime}$. Overlaid
are the GMRT 235 MHz contours at levels 4.5, 9, 18, 36, 72, 144, 288 mJy.
Right: Distribution of the spectral index along the tail.}
\label{fig:spix_a1775}
\end{figure*}
%%%%%%%%%%%%%%%%%%%%%%%%%%%%%%%%%%%%%%%%%%%%%%%%%%%%%%%%%%%%%%%%%%%%%%%

\subsection{A\,2162. Study of the integrated spectrum}\label{sec:spec_a2162}

In order to derive the total synchrotron spectrum of the source 
over a wide range of frequencies, we used the 610 MHz flux densities 
(Table \ref{tab:flux2}), and the literature values.
The spectrum is shown in Fig.~\ref{fig:sp_a2162}, with the best fit 
superposed (solid line).
The GMRT 235 MHz measurement is most likely understimated, due to problems 
with the flux density scale of the secondary calibrator
(see Section \ref{sec:235-610}). For this reason, this value was 
not used for the spectral fit.
The derived integrated spectrum, shown in Fig. \ref{fig:sp_a2162}, has a 
power law shape with an average spectral index $\alpha$=1.0 in the frequency 
range 151 MHz -- 4.7 GHz. 
\\
No strong jets, core or hot spots are present in this source and thus 
its total emission is likely to be dominated by the emission from the lobes. 
We therefore apply the CI model to the study of its integrated spectrum.
The spectral fit of the CI model provided 
$\alpha_{inj} = 0.73^{+0.09}_{-0.08}$ and 
$\nu_{\rm b} = 1.4^{+1.8}_{-0.72}$ GHz. With these values 
the  estimated equipartition magnetic field and radiative age of the electrons 
are respectively B$_{\rm eq} \sim 3 \mu$G and 
t$_{\rm rad} = 1.1 \times10^8$ yr
(see Section \ref{sec:ages} and Table 6).

%%%%%%%%%%%%%%%%%%%%% fig. 21 spectrum of A2162 %%%%%%%%%%%%%%%%%%%%%%%%%%%%%
\begin{figure}
\centering
\caption{A\,2162 -- 74 MHz--4.7 GHz radio spectrum of the central radio 
galaxy, as derived from literature data (empty circles) and the
GMRT 235 MHz and 610 MHz values (filled circles). Note that
the 235 MHz is unreliable due to problems with the secondary
calibrator. The solid line is the best fit of the CI model.}
\label{fig:sp_a2162}
\end{figure}
%%%%%%%%%%%%%%%%%%%%%%%%%%%%%%%%%%%%%%%%%%%%%%%%%%%%%%%%%%%%%%%%%%%%%%%%%%%%%%

\subsection{A\,2372. Spectral analysis along the lobes}\label{sec:spec_a2372}

For this sources we apply both the CI model to the study of the 
integrated spectrum and the JP model to the study of the spectral profiles
along the lobes.

We used our flux density measurements (Tab.~\ref{tab:flux2}), 
the flux density obtained on the 4.9 GHz image from VLA
archival data (Obs. ID AG0293), and collected the 
literature data for this radio galaxy to derive the total radio  
spectrum in the range 74 MHz -- 4.9 GHz, which we report 
in Fig.~\ref{fig:a2372_sp}.  The GMRT flux density values 
nicely fit the gap between 74 MHz and 1.4 GHz.
The spectral index is $\alpha =1.3$ in the 235 MHz-- 4.9  
GHz frequency range, and it flattens to $\alpha=0.5$ below 235 MHz.

The best fit of the CI model to the spectrum is shown in solid line in 
Fig.~\ref{fig:a2372_sp}, and provides a break frequency 
$\nu_{\rm b} = 0.94^{+0.70}_{-0.38}$ GHz and 
$\alpha_{inj} = 0.84^{+0.03}_{0.05}$.
%(see Section \ref{sec:obs} and  Tab. 6). 
The estimated average magnetic field and radiative lifetime of the 
source are respectively B$_{\rm eq} \sim 2\, \mu$ G and 
t$_{\rm rad} \sim 1.5 \times10^8$ yr.
\\
\\
For this source it was possible to image the spectral index distribution 
in the range 235--610 MHz and to fit the spectrum of each lobe. 
The spectral index image and the spectral
trend along the lobes are given in Fig.~\ref{fig:spix_a2372} (left and
right respectively).

%%%%%%%%%%%%%%%%%%%%% fig. 22 spectrum of A2723 %%%%%%%%%%%%%%%%%%%%%%%%%%%%%
\begin{figure}
\centering
\caption{A\,2372 -- Radio spectrum of the central radio galaxy between
74 MHz and 4.9 GHz. The empty circles are literature data, while the
filled circles are the 235 MHz and 610 MHz measurements. The solid line
is the best fit of the CI model.}
\label{fig:a2372_sp}
\end{figure}
%%%%%%%%%%%%%%%%%%%%%%%%%%%%%%%%%%%%%%%%%%%%%%%%%%%%%%%%%%%%%%%%%%%%%%%%%%%%%%

%%%%%%%%%%%%%%%%%%% fig. 23 spectral index map for A2372 %%%%%%%%%%%%%%%%%%% 
\begin{figure*}
\centering
\caption{A\,2372 -- Left: Colour scale image of the spectral index 
distribution between
235 MHz and 610 MHz over the WAT radio galaxy at the cluster centre, derived 
using images with a restoring beam of 
20.0$^{\prime \prime} \times 14.0^{\prime \prime}$. Overlaid are the GMRT 610 
MHz contours at levels 0.6, 1.2, 2.4, 4.8, 9.6 mJy. The error in the spectral 
index at 1$\sigma$ level ranges from $\pm$0.05 and to $\pm$0.1 along the jets, 
and  is $\pm$0.3 at the core and at the end of the tails.
Right: 235 MHz--610 MHz spectral index distribution of the lobes as function 
of the distance from the core. The solid line are the best fit of the
radiative model described in the text.}
\label{fig:spix_a2372}
\end{figure*}
%%%%%%%%%%%%%%%%%%%%%%%%%%%%%%%%%%%%%%%%%%%%%%%%%%%%%%%%%%%%%%%%%%%%%%%%%%%

The synchrotron spectrum of the source between 235 MHz and 610 MHz is 
inverted in the core region ($\alpha \sim -0.5$), while the spectral index 
steepens from 0.7$\pm$0.1 to 1.0$\pm$0.1 along the jets. In the
outer regions of the tails the spectrum further steepens, up
to $\alpha=2.0\pm0.3$. 

We estimated the radiative ages of the radio lobes (B and C in Fig.
\ref{fig:a2372}) by fitting their
spectral index distribution as function of the distance from the core
with a simple JP model as performed for A2480 
(see Section \ref{sec:spec_a2480}}).
The model fit provided $\alpha_{\rm inj} = 0.56\pm 0.04 $ and 
$\nu_{\rm b} = 680^{+260}_{-150}$ MHz,  for both 
lobes. Using these values we estimated
an average equipartition magnetic field B$_{\rm eq} \sim 2 \mu$ G 
and a radiative lifetime  t$_{\rm rad} = 1.5 \pm 0.2 \times10^8$ yr 
for both the lobes. It is noteworthy that 
these estimates are in very good agreement with the
magnetic field and radiative age provided by
the fit of the whole source spectrum (Tab. 6). The CI spectral shape 
should indeed represent the sum of a series of JP spectra with ages
ranging from zero up to the age of the source, while the break frequency 
of the oldest JP spectrum present in the source should correspond 
to the break frequency of the CI model.
\\
Given the total size of the source, we obtained a projected 
expansion velocity of $\sim$ 5000 km/s for each lobe ($\sim 0.02c$).
% 3000$^{+430}_{-310}$ km/s for each lobe ($\sim 0.01c$).

\subsection{A\,2622. Spectral analysis along the lobes}\label{sec:spec_a2622}

The image of the 235 MHz--610 MHz spectral index distribution over the source 
is reported in the left panel of Fig. \ref{fig:spix_a2622}. 
\\
The region with the flattest spectral index ($\alpha \sim0.7$) is not 
coincident with
the peak brightness at either frequencies. Outside this region the
spectrum steepens considerably along the lobes, reaching values  
$\alpha$ \gtsim$~$ 2.
Such offset between the 235 MHz--610 MHz spectral index image and the peak 
in the total intensity is intriguing. One possible
interpretation is that this radio galaxy is old, and the flattest region
reveals the original location of the nucleus.

The hypothesis of an old radio source,
whose nuclear engine might be switched off, 
is supported by the analysis of the integrated spectrum
between 80 MHz and 4.5 GHz, shown in Fig.~\ref{fig:sp_a2622}. 
Open circles are literature data %reported in Tab.~9 
and filled circles are the GMRT data given in Tab.~\ref{tab:flux2}.
Also for this source our observations align very well with the 
flux density measurements known from the literature.
The spectrum can be fitted with a single power law over the whole range  
80 MHz -- 4.5 GHz, with a spectral index of $\alpha=1.7$.
The solid line in the figure represents the best fit of the CI model to the 
integrated spectrum, which provides $\alpha_{inj}=1.1^{+0.0}_{-0.1}$.
%
%%%%%%%%%%%%%%fig. 24 spectral index map for Abell 2622 %%%%%%%%%%%%%%%%%%%%%
\begin{figure*}
\centering
\caption{A\,2622 -- Left: Colour scale image of the spectral index 
distribution between
235 MHz and 610 MHz over central radio galaxy, as 
computed from images with a restoring beam of 
17.1$^{\prime \prime} \times 11.4^{\prime \prime}$. Overlaid are the GMRT 610 
MHz logarithmic contours starting from 0.3 mJy.
Right: 235 MHz--610 MHz spectral index distribution of the lobes
as function of the distance from the core. The solid line are the best fit of 
the radiative model described in the text.}
\label{fig:spix_a2622}
\end{figure*}
%%%%%%%%%%%%%%%%%%%%%%%%%%%%%%%%%%%%%%%%%%%%%%%%%%%%%%%%%%%%%%%%%%%%%%%%%%
%
%%%%%%%%%%%%%%%%%%%%%%%%%%%% fig. 25 spectrum A2622 %%%%%%%%%%%%%%%%%%%%%%%%
\begin{figure}
\centering
\caption{A\,2622 -- Radio spectrum of the central radio galaxy between 80 MHz
and 4.5 GHz. The empty circles are literature data, while the filled circles 
are the 235 MHz and 610 MHz measurements. The solid line is the best fit of 
the CI model.}
\label{fig:sp_a2622}
\end{figure}
%%%%%%%%%%%%%%%%%%%%%%%%%%%%%%%%%%%%%%%%%%%%%%%%%%%%%%%%%%%%%%%%%%%%%%%%%%%%%

For this source we managed to carry out the analysis of the spectral 
distribution along the lobes.
We derived the trend of the spectral index between 235 MHz and 610 MHz 
in the source lobes (B and C in Fig. \ref{fig:a2622} and right panel of 
Fig.~\ref{fig:spix_a2622}) as function of the distance from the peak, 
and fitted the observed steepening.
The spectrum is very steep anywhere along the lobes. The ``flattest'' value
of the spectral index
is $\sim$ 1.1, then it further steepens (up to $\sim$ 2) with 
increasing distance from the core along both lobes.
\\
The overall spectral properties of the source suggest that it is most
likely switched off. For this reason we fitted the spectra along the lobes
assuming an initial phase of electron injection, followed by a switch--off 
of the nuclear activity and then a ``relic'' phase 
(Parma et al. \cite{parma07}). An isotropic distribution of 
the pitch angle (JP model; Jaffe \& Perola \cite{jp74}) is assumed. 

We imposed an initial spectral index $\alpha_{inj}$ = 0.7, which provides a
break frequency $\nu_{\rm b} = 460^{+40}_{-40}$ MHz, B$_{\rm eq} = 4.4 \mu$G,
and a total source age 
t$_{\rm rad} = 1.5 \times 10^8$ yr. Furthermore, we obtained that 
t$_{\rm off}$/t$_{\rm rad}$ = 0.46, where t$_{\rm off}$ is the duration of the
switched--off phase (see also Parma et al. \cite{parma07} for details). 
This means that the source has been active for
79 Myr and switched off 67 Myr ago.

\subsection{MKW\,07. Spectral properties}\label{sec:spec_mkw07}

The total integrated synchrotron spectrum of the source between 235 MHz and 
1.28 GHz is shown in Fig.~\ref{fig:sp_mkw07}.
The source spectrum is steep in this frequency range, with a spectral index
$\alpha_{\rm 235 MHz}^{\rm 1.28 GHz}$=1.2 $\pm$ 0.1. No information was
found in the literature at other frequencies, therefore no detailed study of 
the integrated spectrum was possible.

%%%%%%%%%%%%%%% fig. 26 spectrum of mkw07 %%%%%%%%%%%%%%%%%%%%%%%%%%%%%
\begin{figure}
\centering
\caption{MKW\,07 -- Radio spectrum of the central radio galaxy between
235 MHz and 1.4 GHz. The filled circles are the GMRT data; the empty 
circle is the NVSS value at 1.4 GHz.}
\label{fig:sp_mkw07}
\end{figure}
%%%%%%%%%%%%%%%%%%%%%%%%%%%%%%%%%%%%%%%%%%%%%%%%%%%%%%%%%%%%%%%%%%%%%%%%%%%%%%

We produced an image of the spectral index distribution over the source 
between 235 MHz and 1.28 GHz, which is shown in Figure
\ref{fig:spix_mkw07}. There is no obvious trend of the spectral index in
both lobes. The distribution is patchy, with 
$\alpha_{\rm 235 MHz}^{\rm 1.28 GHz}$ in the range $\sim 1\div$ 1.5, 
suggesting that the lobes are old structures expanding adiabatically. 

%%%%%%%%%%%%%%%%%%%%%% fig. 27 spectral index map MKW07%%%%%%%%%%%%%%%%%%%%
\begin{figure}
\centering
\caption{MKW\,07 -- Colour scale image of the spectral index distribution between
235 MHz and 1.28 GHz over the central radio source, as computed from images with a restoring beam
of 14.0$^{\prime \prime} \times 9.1^{\prime \prime}$. Overlaid
are the GMRT 235 MHz contours at levels 4.5, 9, 18, 36, 72, 144, 288 mJy.}
\label{fig:spix_mkw07}
\end{figure}
%%%%%%%%%%%%%%%%%%%%%%%%%%%%%%%%%%%%%%%%%%%%%%%%%%%%%%%%%%%%%%%%%%%%%%%%%%%%

\subsection{MKW\,03s. Study of the integrated spectrum}\label{sec:spec_mkw03s}

The radio galaxy at the centre of MKW\,03s is a well--known ultra
steep spectrum radio galaxy (3C\,318.1; e.g. Komissarov \& Gubanov 
\cite{komissarov94}, de Breuck et al. \cite{debreuck00}) and 
flux density measurements are available from
16.7 MHz up to 4.9 GHz. We collected these values 
and plotted them in Fig.~\ref{fig:sp_mkw03s}, together with our GMRT 
observations, to derive the source total synchrotron spectrum. 
Also for this radio galaxy our GMRT flux densities align extremely 
well with the literature data. The average spectral index for this source
is $\alpha \sim$1.9. 

The ultra--steep integrated spectrum of this source cannot be
 explained by a simple CI model. It is likely that the injection of 
electrons in the source has stopped. However, the high--frequency cutoff 
of the synchrotron spectrum is less than exponential. A possible 
explanation is that the relic electrons are ageing in a filamentary
magnetic field and that their diffusion is very low. Under these
conditions, different part of the source would have slightly different
break frequencies and this would cause a smoother cutoff of the intregrated 
spectrum. For this source we indeed obtained a very good fit of the 
spectrum between
16.7 MHz-- 4.9 GHz (solid line in Fig.~\ref{fig:sp_mkw03s}) 
using a diffusion MJP model (Slee et al. \cite{slee01}), which accounts 
for the effects of particle diffusion in an inhomogeneous magnetic
field whose intensity is spatially variable. The model includes an initial
phase of electron injection (CI), followed by a switch--off of the nuclear
activity and then a ``relic'' phase. 
An isotropic distribution of the pitch angle 
(JP model; Jaffe \& Perola 1973) is assumed. We fixed the injection 
spectral index of the CI phase to $\alpha_{inj}=0.7$, and 
obtained a best fit value of the break frequency 
$\nu_{\rm b}= 30\pm 1$ MHz (Tab.~6). The model provides a ratio
B$_{\rm rms}$/B$_{\rm CMB}=16^{+18}_{-6}$. 
Since for this source B$_{\rm CMB}$ = 3.5 $\mu$G, we have 
B$_{\rm rms}$=56 $\mu$G, and the  corresponding  radiative age 
(see Formula 2) is  t$_{\rm rad} = 2.1 \times 10^7$ yr.
We were also able to estimate the relative duration of the relic phase 
t$_{\rm off}$ with respect to 
the CI phase t$_{\rm rad}$:
$\rm t_{\rm off}/t_{\rm rad}>$0.46, i.e. the CI phase lasted about
$\sim 1.1\times10^7$ yr and the nuclear engine switched off at least 
$\sim10^7$ yr ago.
\\
For this source the equipartition magnetic field is 
B$_{\rm eq} \sim 16 \mu$G. This value and the overall size of MWK\,03s
are very similar to what is found for the dying sources in Parma
et al. (\cite{parma07}).

%%%%%%%%%%%%%%%%%%%%%%%%%%% fig. 28 spectrum MKW03s %%%%%%%%%%%%%%%%%%%%%%%%
\begin{figure}
\centering
\caption{MKW\,03s -- Radio spectrum of the central radio galaxy between 16.7 
MHz and 1.47 GHz. The empty circles are literature data, 
the filled circle are the GMRT data at 235 MHz, 610 MHz and 1.28 GHz.
The solid line is the best fit of the MJP model.}
\label{fig:sp_mkw03s}
\end{figure}
%%%%%%%%%%%%%%%%%%%%%%%%%%%%%%%%%%%%%%%%%%%%%%%%%%%%%%%%%%%%%%%%%%%%%%%%%%%%%

%%%%%%%%%%%%%%%%% Tab. 6 - campi magnetici ed eta' radiative %%%%%%%%%%%%%%%%%%

\begin{table}[h!]
\label{tab:ages}
\caption[]{Equipartition fields and radiative ages from the analysis of the
radio spectrum}
\begin{center}
\begin{tabular}{lcccccc}
%\noalign{\smallskip}
\hline\noalign{\smallskip}
Source & $\alpha_{obs}$ & $\alpha_{inj}$ & $\nu_{\rm b}$ & B$_{\rm eq}$ &
t$_{\rm rad}$ & v$_{\rm growth}$ \\
       &                &                & GHz & $10^{-6}$ G & Myr & (c)  \\
\hline\noalign{\smallskip}
A\,2480  & 0.6 & 0.55  & 8.4 & $\sim$10 & 15  & 0.8$\times10^{-3}$ \\
AWM\,05  & 0.8 & 0.58  & 1.1 & $\sim$ 7 & 65  & 0.2$\times10^{-3}$ \\
A\,1775  & 1.0 & 0.6   & 0.6 & $\sim$ 8 & 490 & 0.3$\times10^{-2}$ \\
A\,2162  & 1.0 & 0.73  & 1.4 & $\sim$ 3 & 110 & 0.3$\times10^{-2}$ \\
A\,2372  & 1.3 & 0.65  & 0.7 & $\sim$ 2 & 150 & 0.3$\times10^{-1}$ \\
A\,2622  & 1.7 & 0.7$^{*}$  & 0.5 & $\sim$ 4 & 150 & 0.5$\times10^{-2}$ \\
MKW\,03s & 1.9 & 0.7$^{*}$  & 0.03 & $^{\clubsuit}$ & 21  & 0.8$\times10^{-1}$ \\
\hline\noalign{\smallskip}
\end{tabular}
\end{center}
$^{*}$ fixed in the fit.
\\
$^{\clubsuit}$ see Section \ref{sec:spec_mkw03s}.
\end{table}

%%%%%%%%%%%%%%%%%%%%%%%%%%%%%%%%%%%%%%%%%%%%%%%%%%%%%%%%%%%%%%%%%%%%%%%%%%%

\section{Discussion}\label{sec:disc}

\subsection{Radio powers and morphologies}

In this paper we presented new GMRT images at 1.28 GHz, 610 MHz and 235 MHz 
for 13 cD galaxies selected from a statistical sample of rich and poor galaxy 
clusters located at  z$\le 0.1$.
Both the radio power range and the variety of morphologies found are
typical for radio galaxies at the cluster centres. It is well known 
(i.e. Valentjin \& Bijleveld \cite{val83}; Ball et al. \cite{ball93}; 
Burns et al. \cite{burns87}; Bagchi \& Kapahi \cite{bagchi94})
the radio emission associated with cD  galaxies exhibits a large variety 
of morphologies, regardless of the cluster richness: compact radio sources, 
classical doubles with FR\,I and FRI/FRII morphology, wide--angle--tails 
(WAT; e.g. 3C\,465 in A\,2634, Eilek \& Owen \cite{eilek02}), 
core--halo (e.g. 3C\,84 in the Perseus cluster, 
Fabian et al. \cite{fabian00}), and more peculiar structures, as for example 
the complex radio source associated with the cD galaxy in A\,3560 (Bardelli et 
al. \cite{bardelli02}) and that in A\,2199 (3C\,338, Giovannini et al.
\cite{giovannini98}).
\\
Here below we give a brief summary.

\begin{itemize}
\item [i)] All sources are low/intermediate power radio galaxies, 
as typically found in the innermost cluster regions. Their radio 
power at 1.4 GHz (Table 1) ranges from 
logP$_{\rm 1.4~GHz}$ (W Hz$^{-1})$ = 21.86 (MKW\,01s) to 
logP$_{\rm 1.4~GHz}$ (W Hz$^{-1})$ = 24.66 (A\,2480).

\item[ii)] 
We found two sources with galactic size and morphology typical of FR\,Is 
(A\,1663)  
and FR\,I/II (A\,2480); a point--like source associated with the nucleus 
of the cD galaxy in the poor cluster MKW\,01s, whose origin is unclear
and could be due both to a very faint AGN or to nuclear starburst; 
a core--halo source in MKW\,01; a double radio source in AWM\,05, with a
strong nuclear component. Finally, a weak AGN with faint lobes was found in 
MKW\,06.  

\item[iii)] Three sources, A\,2162, MKW\,07 and MKW\,03s, show a fairly
relaxed morphology of the lobes and lack nuclear emission. These features
are consistent with the idea that they are aged radio galaxies, 
as also suggested by the results of the spectral analysis
%suggestive of aged radio galaxies, as confirmed from the spectral analysis 
(see Sections \ref{sec:spec_a2162}, \ref{sec:spec_mkw07}, 
\ref{sec:spec_mkw03s} and  \ref{sec:ages}). 

\item[iv)] The sample includes three very extended radio galaxies. 
In particular, the cluster--size double radio galaxy in A\,2372, 
whose lobes are most likely undergoing strong interaction with the ICM; 
the double source at the centre of MKW\,02, whose images at these two 
frequencies show only the central component and the brightest regions in
the lobes; the dumb--bell system in A\,1775, characterised by a galactic 
size double source and a spectacular head--tail source, extending out to
$\sim$ 380 kpc.
\end{itemize}

In the following subsections we will present a few general considerations on 
the spectral properties and radiative ages for these sources, in the light of 
the extreme environment in which they are hosted.

\subsection{Steepness of the radio spectrum}\label{sec:steep}

It is well known from the literature that the spectra of radio galaxies 
in central cluster regions tend to be steeper than in other environments 
(i.e. Slee et al. \cite{slee83}; Slee \& Reynolds \cite{slee84}; 
Roland et al. \cite{roland85}; Slee et al. \cite{slee01}), with 
$\alpha > 1$ and up to the ultra steep values reported in de Breuck et al.
(\cite{debreuck00}). This has been explained in terms of confinement 
of the ICM of the radio emitting plasma; the confined relativistic 
plasma loses its high energy electrons through synchrotron and Inverse 
Compton losses, resulting in a steepening of the radio spectrum.

The analysis of the spectral index for the sources presented here shows
that also in our sample most of the sources have steep spectrum.
From Tables 4, 5 and 6 we can say that 3/13 sources have a ``normal''
spectrum, with $\alpha$ ranging from 0.7 to 0.8. Among these sources,
we draw the attention to A\,2480. The source is peculiar in many ways,
since it is one of the smallest in the sample, the most powerful,
and on the basis of the spectral analysis it is also the youngest.

Among all the remaining sources, the largest majority (10/13), have observed 
spectra steeper than 1, up to the ultra steep spectrum of MKW\,03s and A\,2622.

\subsection{Source ages. Active and ``dying'' radio galaxies} \label{sec:ages}

Table 6 shows that 4 out of the 7 sources with available estimates of their
radiative age have t$_{\rm rad}~ \gtsim ~ 10^8$ yr. Such ages are 
of the same order of magnitude found for  giant radio galaxies.
There are a few different cases worth to be mentioned.
\\
\\
A\,2622 and MKW\,03s are most likely ``dying'' sources, where the nuclear 
engine has switched off and the spectrum is dominated by synchrotron and 
Inverse Compton losses. In both cases, the radiative model which better 
describes the source spectrum requires a switch--off of the central engine, 
followed by a ``relic'' phase, and this allowed us to estimate the ratio
t$_{\rm off}$/t$_{\rm rad}$.
The radiative age is of the order of few Myr for both sources.
\\
\\
For all the remaining steep spectrum sources, there is evidence of 
nuclear emission with flattish spectrum: the head of the tail 
in A\,1775 (Fig.~\ref{fig:spix_a1775}) has an average spectrum of $\sim$ 0.5; 
the central component in MKW\,02 has a spectral index of 0.5 
(Table 4). 
\\
MKW\,07 (Fig.~\ref{fig:mkw07}) is an intriguing source. Compact emission is
associated with the western optical nucleus in the galaxy, but the bulk of the 
radio emission comes from two lobes with steep spectrum. The measurement of 
the flux density of the compact component is difficult at 610 MHz and
235 MHz, since it is blended with the emission from the south--western lobe,
and it is most likely overstimated. For this reason the spectral index given 
in Tab. 4 for this component is very uncertain.
The radio--optical overlay in the image is reminiscent of 3C\,338, which is 
thought to be the superposition of old emission and restarted nuclear 
activity. 
Here at 1.28 GHz and at 610 MHz the two lobes are clearly connected 
with a ridge of emission superposed on the eastern nucleus. A possible 
interpretation of this peculiar source is that the two lobes are associated 
with a past cycle of radio activity associated with the eastern nucleus.

Table 6 reports also a crude estimate of the source growth velocity, based on
the radiative ages estimated from the analysis of the integrated spectrum
and from the total source size (end to end). These values are considerably
smaller than the typical expansion velocities estimated for the large radio 
galaxies, due to the combination of the high radiative ages found and
the relatively small/average sizes of the radio galaxies (except A\,2372
and the tail in A\,1775).
It is tempting to relate these values with
the dense environment at the centres of galaxy clusters, causing a slow
growth rate. The growth 
velocity estimated for A\,2480 would fit this idea, since it is a young 
source still embedded in the optical host.
Unfortunately, high resolution X--ray data are unavailable for all sources
presented here, and this does not allow us to derive the properties of the
external medium and check whether they are overpressured, and/or
highly confined.
\\
We underline that the radiative ages estimated here are all based on
the assumption that the spectral steepening is due to particle ageing in a
constant magnetic field. Alternative models have been developed for classical
double radio galaxies, which
consider the role of a decreasing magnetic field along the strucure of
extended radio galaxies (e.g. Blundell \& Rawlings \cite{br00} and
references therein). Such configuration of the magnetic field would also
reflect into a spectral steepening, with no direct implication on the source
ageing. However, it should be noted that
the radio sources in our sample are significantly weaker than the classical
FRII type sources. In particular, the magnetic field strenght
in the sources of our sample is of the same order of the equivalent
magnetic field associated to the inverse Compton losses $B\sim B_{\rm IC}$.
This implies a scaling for the break frequency of
$\nu_{\rm b} \propto B t_{rad}^{-2}$ (see Eq.\ref{trad1}) rather than
$\nu_{\rm b} \propto B^{-3} t_{rad}^{-2}$, as expected in the case of
powerful radio sources where $B\gg B_{\rm IC}$. Consequently, in the low
power sources considered in this work, the observed spectral steepening
should be less sensitive to systematic variations of magnetic field
strenght. Although these variations should have a low impact
on the spectral steepening, we cannot completely rule out their effects
and thus our age estimates should be considered upper limits, and the
computed velocity growth as lower limits.

\subsection{A\,2372: a restarted radio galaxy?}\label{sec:restarted}

The WAT in A\,2372 is unique in many ways. Beyond the very large 
angular extent, the gap of radio emission visible between the inner 
jets of A\,2372 (coincident with the optical galaxy) and the large scale lobes
(see Fig.~\ref{fig:a2372}) may contain very relevant pieces of information.
This feature appears also in higher frequency images available in the
literature (1.4 GHz, Owen \& Ledlow \cite{owen97}; 4.9 GHz, Gregorini et al. 
\cite{gregorini94}). Such large gap is uncommon in low power radio galaxies, 
and
suggests that this galaxy might be undergoing intermittent activity of the 
radio nucleus. The double morphology of the nuclear component (insert in the 
upper panel of Fig. \ref{fig:a2372}) is a clear indication that the radio 
nucleus is currently active.
\\
The emission gap is $\sim$ 40 kpc (from the galaxy to the edge of the lobes).
Using the source LLS (Tab.~\ref{tab:flux2}) and its radiative age (Tab. 6),
we made a first order estimate of the growth velocity of this source,
and obtained $\rm{v_{growth} \sim LLS/t_{rad}} \sim$ 0.03c.
These numbers allow us to make a rough estimate of the time elapsed since
the last refurnishment of the radio lobes, i.e. t$_{\rm gap} \sim 10^6$ yr.
\\
We suggest therefore that this radio galaxy might be characterised by 
intermittent activity of the radio nucleus, on a time scale which is 
$\sim$ 1/100 of the total lifetime of the source. We point out that,
if confirmed, this would be the first case of a direct measurement
of a cycle of activity in a radio galaxy. 

One possible way to check for further hints of recurrent activity is to
study the properties of the intracluster medium at high sensitivity and
resolution, in search for bubbles and/or temperature gradients. A similar
study was carried out for the cluster 2A\,0335+096 by 
Mazzotta et al. (\cite{mazzotta03}) who estimated
expected bursts of radio activity with cycles of the order of 10$^7$ years.
However, such study is not possible for A\,2372 with the current generation 
of X--ray satellites. 
Ledlow et al. (\cite{ledlow03}) report a ROSAT (0.2 -- 2 keV) 
luminosity L$_{\rm X} = 2.2 \times 10^{42}$ erg/s, with an observed flux 
$f_{\rm X} = 0.22\times10^{-12} \rm erg~cm^{-2}s{-1}$.
This cluster is therefore too faint to detect cavities with some
significance.

\section{Summary and conclusions}\label{sec:summary}

In this paper we presented high resolution and high sensitivity GMRT 
radio images for 13 radio galaxies located at the centres of rich galaxy 
clusters and poor groups, and carried out a morphological and spectral
analysis.
\\
Our deep images revealed a variety of morphologies, i.e. compact, double 
and tailed sources, and linear sizes ranging from the sub--kpc scale to 
almost the Mpc scale. No difference in the morphology and size was
found among cluster and group cD galaxies.
\\
On the basis of our observations and literature data, we derived steep
spectra ($\alpha \ge 1$ in the range $\sim 0.2 - 1.4$ GHz) 
for 10/13 radio galaxies, confirming the
known result that the central regions of clusters tend to host steep radio
sources.
\\
For 7 sources in the sample we could carry out a detailed spectral analysis,
which allowed us to estimate the equipartition parameters and source 
radiative ages. We found radiative ages in the range $\sim 1.5\times10^7 \div 
\sim 5\times10^8$ yr.
These ages, coupled with the linear sizes measured from our images, provide 
growth velocities considerably lower than what is found 
in the literature for the velocity advance speed of double radio galaxies.
There seems to be no connection between the growth velocity and the 
large scale environment (i.e. no difference between rich and poor clusters).
\\
Our analysis suggests that A\,2622 and MKW\,03s might be ``dying'' radio 
galaxies, where the nucleus has switched off and the lobes are not currently 
fed by an active nucleus. The overall properties of MKW\,03s (age, size,
magnetic field strength) are very similar to those reported by Parma et al.
(\cite{parma07}) for the dying source in their sample.
\\
Finally, A\,2372 has aged lobes and an active nucleus, and might represent a 
clear example of a restarted radio galaxy. For this source we 
estimated a life cycle of the order of 10$^6$ yr.
\\ 
\\
{\it Acknowledgements.}
We thank the staff of the GMRT for their help during the observations.
GMRT is run by the National Centre for Radio Astrophysics of the Tata 
Institute of Fundamental Research.
S.G. and T.V. acknowledge partial support from the Italian Ministry
of Foreing Affairs.  
This research has made use of the NASA/IPAC Extragalactic Database 
(NED) which is operated by the Jet Propulsion Laboratory, 
California Institute of Technology. The authors made use of the database
CATS of the Special Astrophysical Observatory (Russia).
Funding for the SDSS and SDSS-II has been provided by the Alfred P. Sloan 
Foundation, the Participating Institutions, the National Science Foundation, 
the U.S. Department of Energy, the National Aeronautics and Space 
Administration, the Japanese Monbukagakusho, the Max Planck Society, and the
Higher Education Funding Council for England. 
The SDSS Web Site is http://www.sdss.org/.
The SDSS is managed by the Astrophysical Research Consortium for the 
Participating Institutions. 

\appendix
\section{Abell 2634 (3C\,465)}
In this Appendix we present the GMRT 235 MHz image of the
central cD galaxy in A\,2634 (z=0.029). This cluster
belongs to our complete sample of cD galaxies, selected as
described in Sec. \ref{sec:sample}. Due to the amount of radio
information available in the literature, A\,2634
was not selected for the GMRT observations presented in this paper.
However the cluster is located within the field of view of the
GMRT 235 MHz observation of A\,2622, presented in Sec. \ref{sec:235-610}.
In particular A\,2634 is $\sim 50^{\prime}$ South--East of A\,2622,
and thus within the 108$^{\prime}$ radius primary beam of the 
GMRT antenna at 235 MHz. This is the first radio image of 3C\,465 
ever published at this radio frequency.

The radio source 3C\,465, associated with the cD galaxy n A\,2634 
is considered the prototype of WAT sources.  It has been studied in detail
both in the radio (e.g. Eilek \& Owen \cite{eilek02}; Hardcastle \& Sakelliou 
\cite{hardcastle04} and references therein) and X--ray band 
(Hardcastle et al. \cite{hardcastle05}). 

The full resolution 235 MHz image of 3C\,465
(corrected for the primary beam of the GMRT antenna
at 235 MHz) is shown in Fig. \ref{fig:a2634}.  
Its total flux density is S$_{\rm 235\,MHz}$=32.9 Jy,
and the derived radio power is logP$_{\rm 235\,MHz}$(W Hz$^{-1}$)=25.84.
At this frequency and resolution 3C\,465 exhibits some interesting
features. A ridge is visible North of the western jet, with no
obvious associated optical counterparts (at the magnitude limit of the
POSS--2 image). Hints of this feature are visible also at 74 MHz on
the VLSS, and for this reason we tend to rule out the possibility that it
is an artifact of our image. The radio jets are very well confined all the 
way to the outermost edges, then they spread out in two converging tails,
again in agreement with the VLSS.

In Fig. \ref{fig:sp_a2634} we show the integrated synchrotron  
spectrum of the source, determined using the flux density
measured from our 235 MHz image and the literature data.
%reported in Tab. \ref{tab:a2634_flux}. Once again we not that the
GMRT 235 MHZ flux density perfectly aligns with all the spectral
data.

%%%%%%%%%%%%%%%%%%%%%%%%%%%%% fig. A.1 A2634 %%%%%%%%%%%%%%%%%%%%%%%%%%%%%%%%%%%%%%%%%%
\begin{figure}
\centering
\caption{GMRT 235 MHz radio contours of the central WAT radio galaxy
in A\,2634, overlaid on the POSS--2 optical image. 
The 1$\sigma$ level in the image is 1 mJy b$^{-1}$. 
Logarithmic contours are reported, starting 
from $\pm$5 mJy b$^{-1}$. The contour peak flux is 808 mJy b$^{-1}$.
The HPWB is $17.1^{\prime\prime} \times 11.4^{\prime\prime}$, p.a. $53^{\circ}$. The image
 is corrected for the primary beam of the GMRT antenna at 235 MHz.}
\label{fig:a2634}
\end{figure}
%%%%%%%%%%%%%%%%%%%%%%%%%%%%%%%%%%%%%%%%%%%%%%%%%%%%%%%%%%%%%%%%%%%%%%%%%%%%%

%%%%%%%%%%%%%%%%%%%%%%%%%%%%% fig. A.2 spectrum A2634 %%%%%%%%%%%%%%%%%%%%%%%%
\begin{figure}
\centering
\caption{Radio spectrum of the central radio galaxy in A\,2634 between 10 MHz
and 8.4 GHz. The empty circles are literature data, the filled circle is the 
GMRT flux density of the source at 235 MHz.}
\label{fig:sp_a2634}
\end{figure}
%%%%%%%%%%%%%%%%%%%%%%%%%%%%%%%%%%%%%%%%%%%%%%%%%%%%%%%%%%%%%%%%%%%%%%%%%%%%%

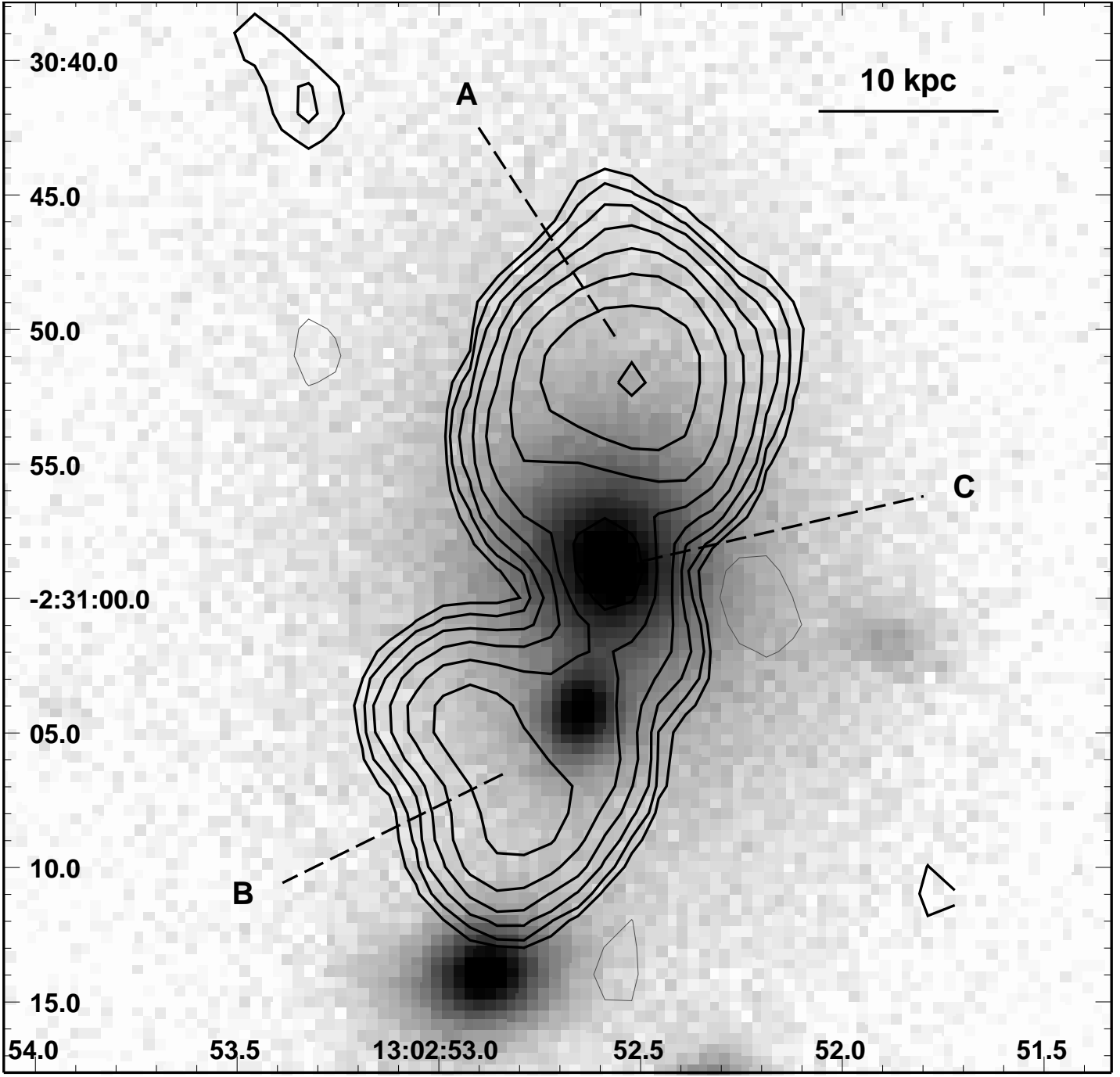
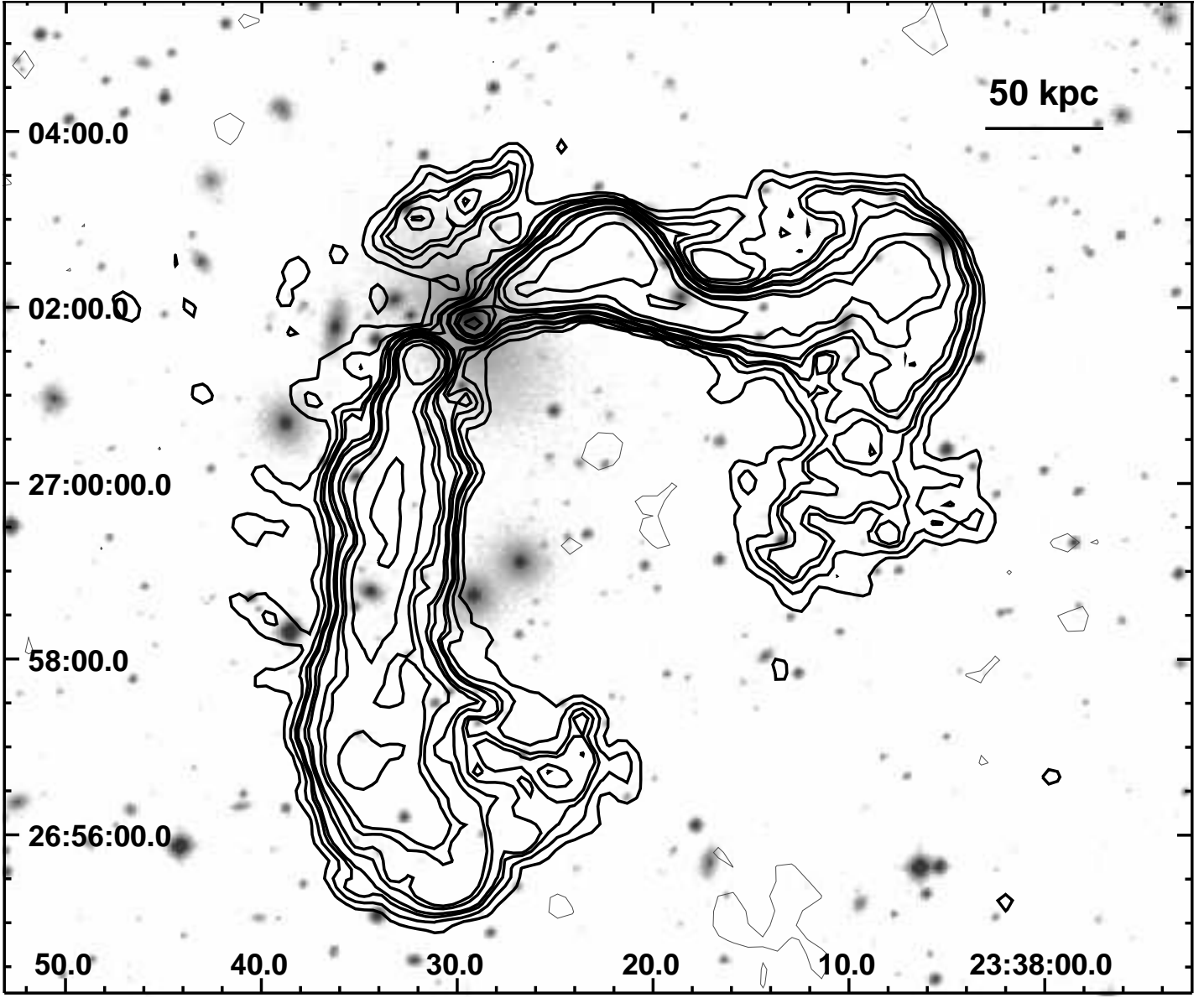
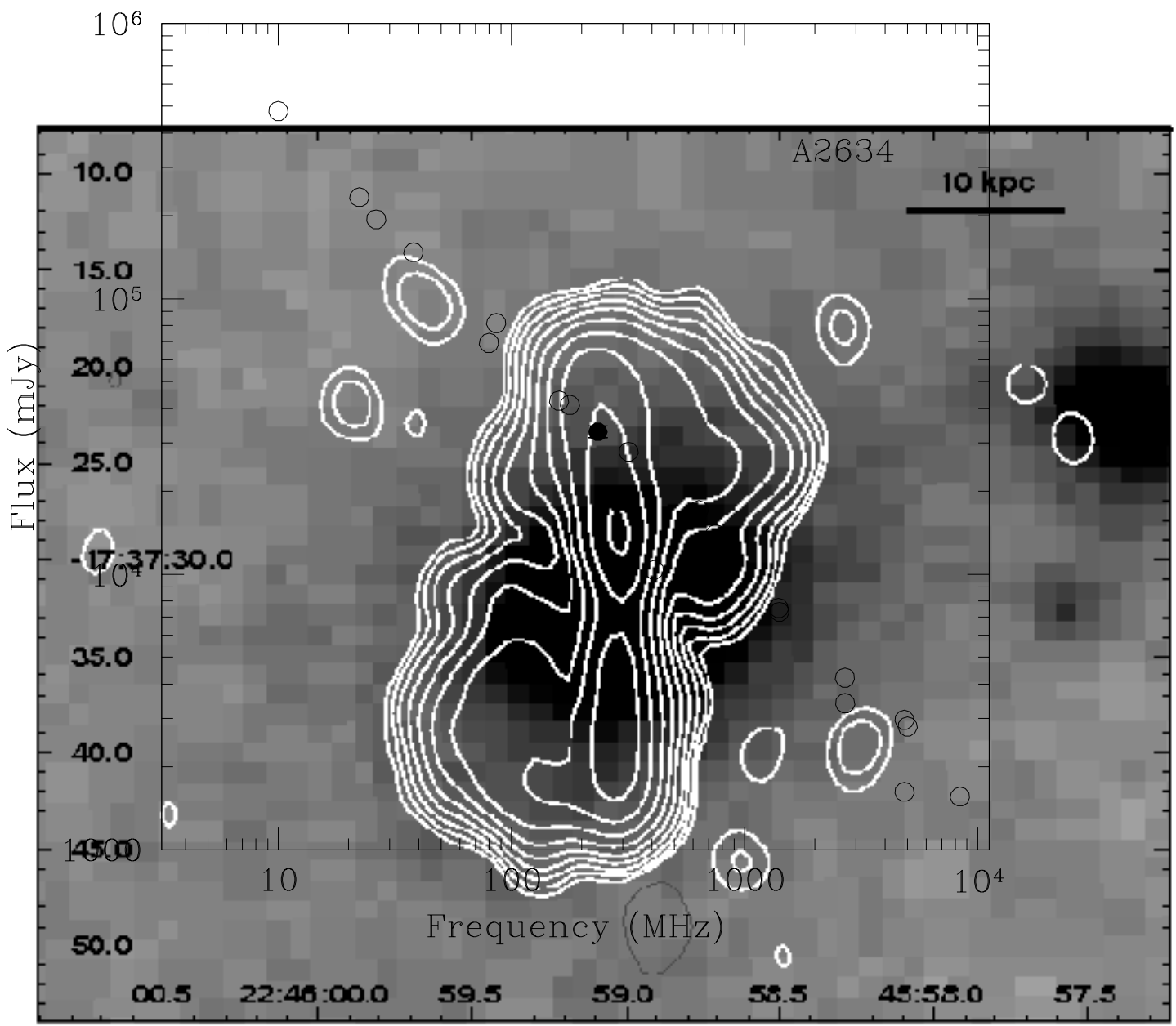
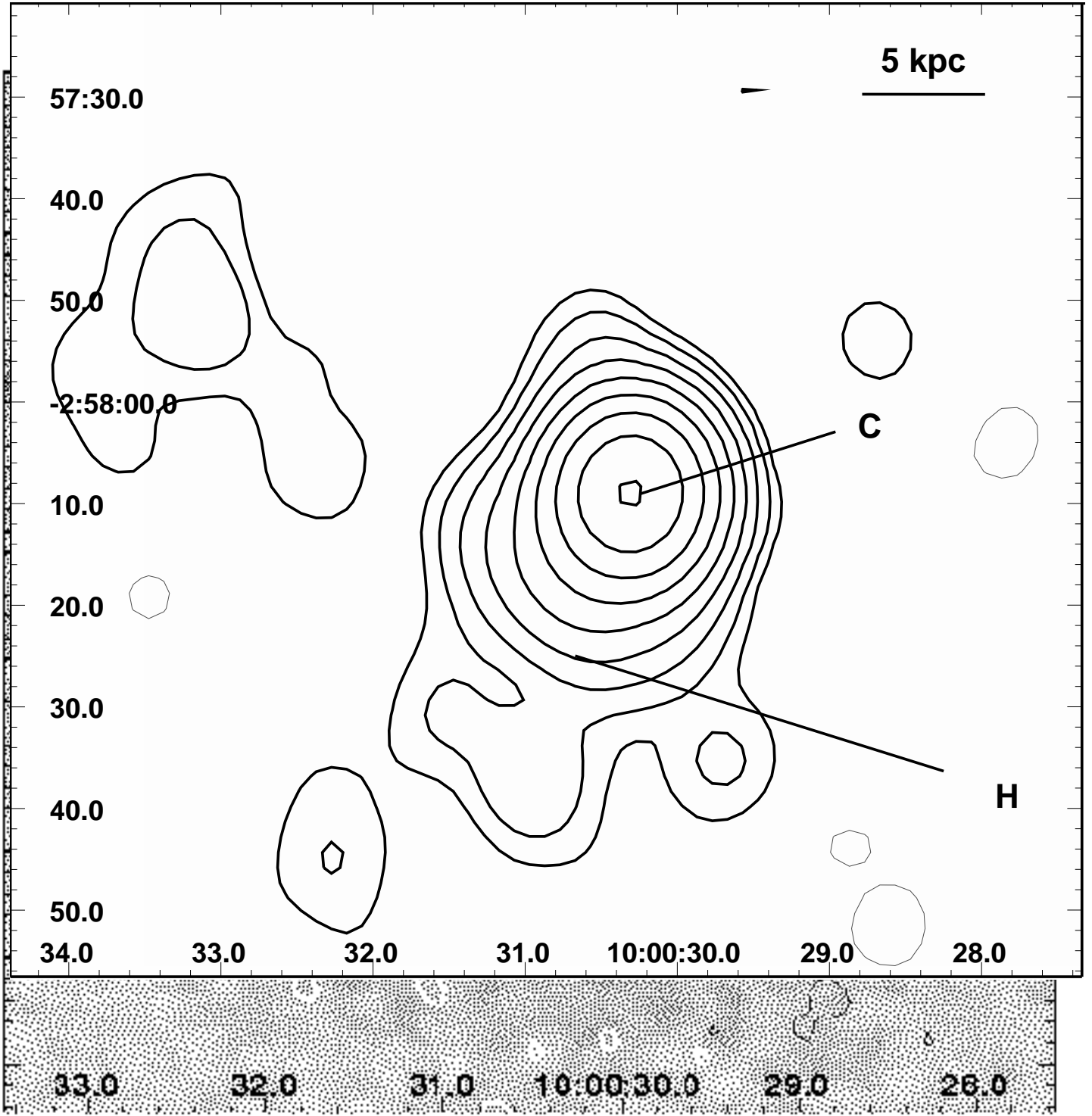
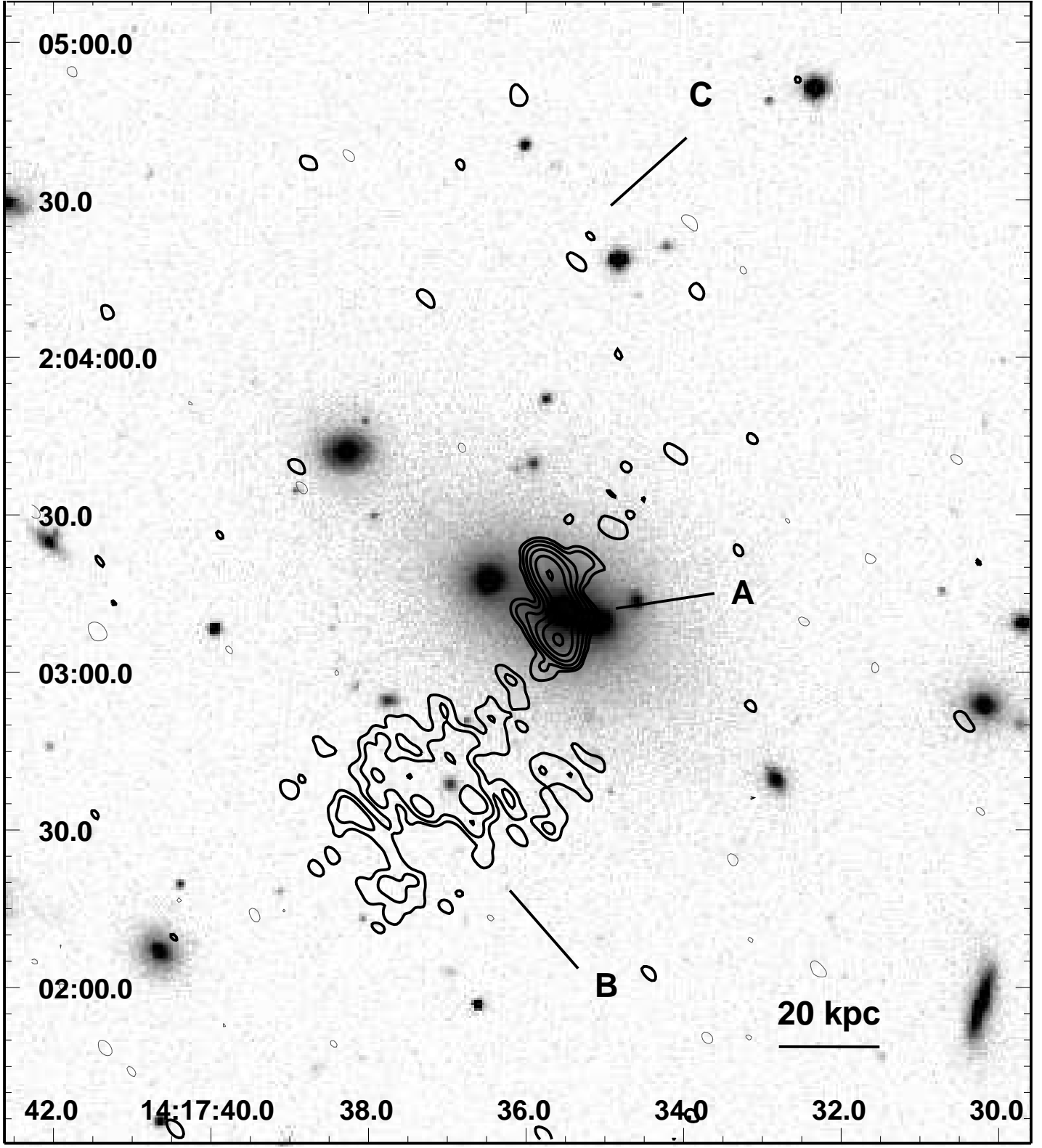
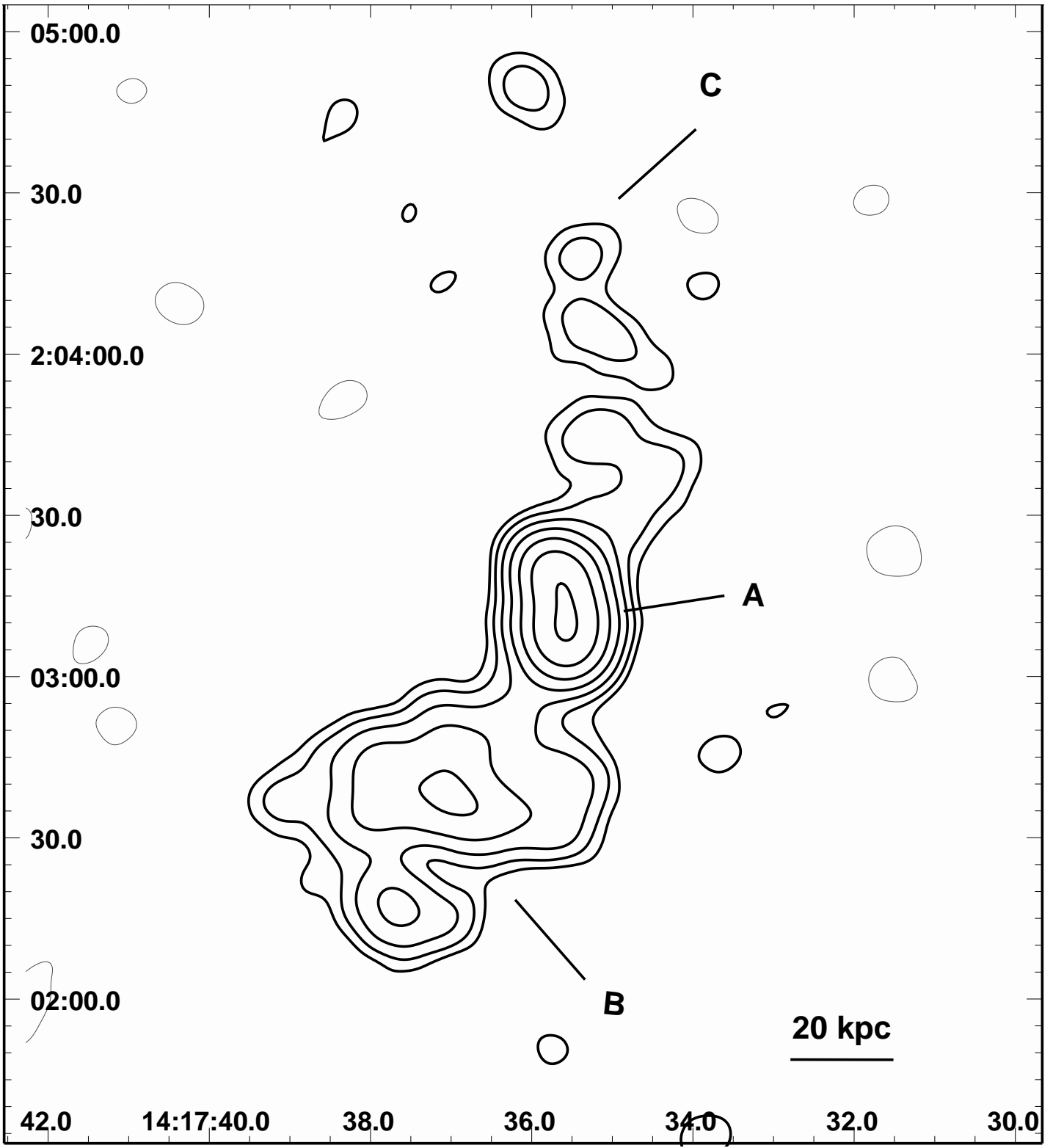
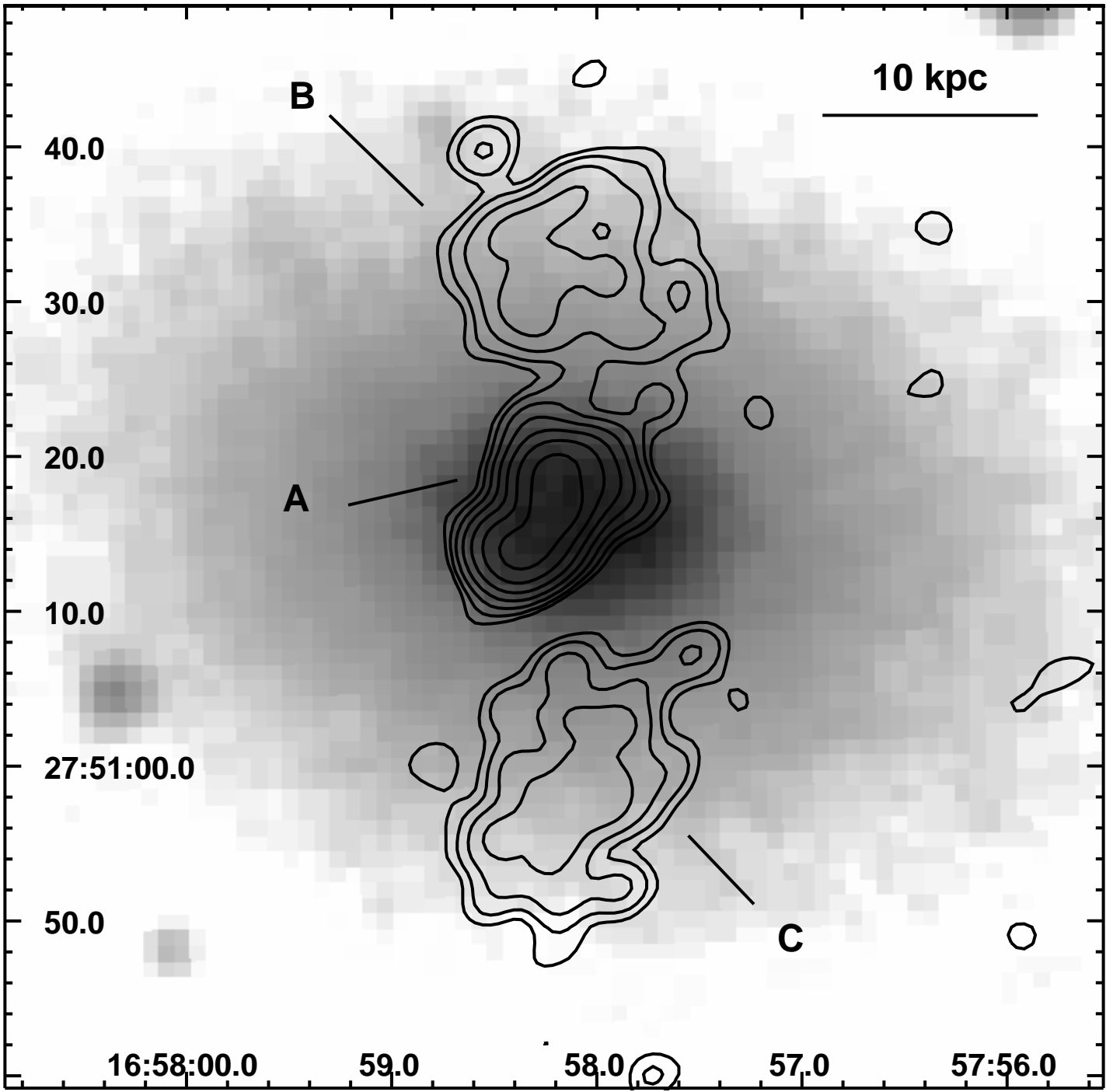
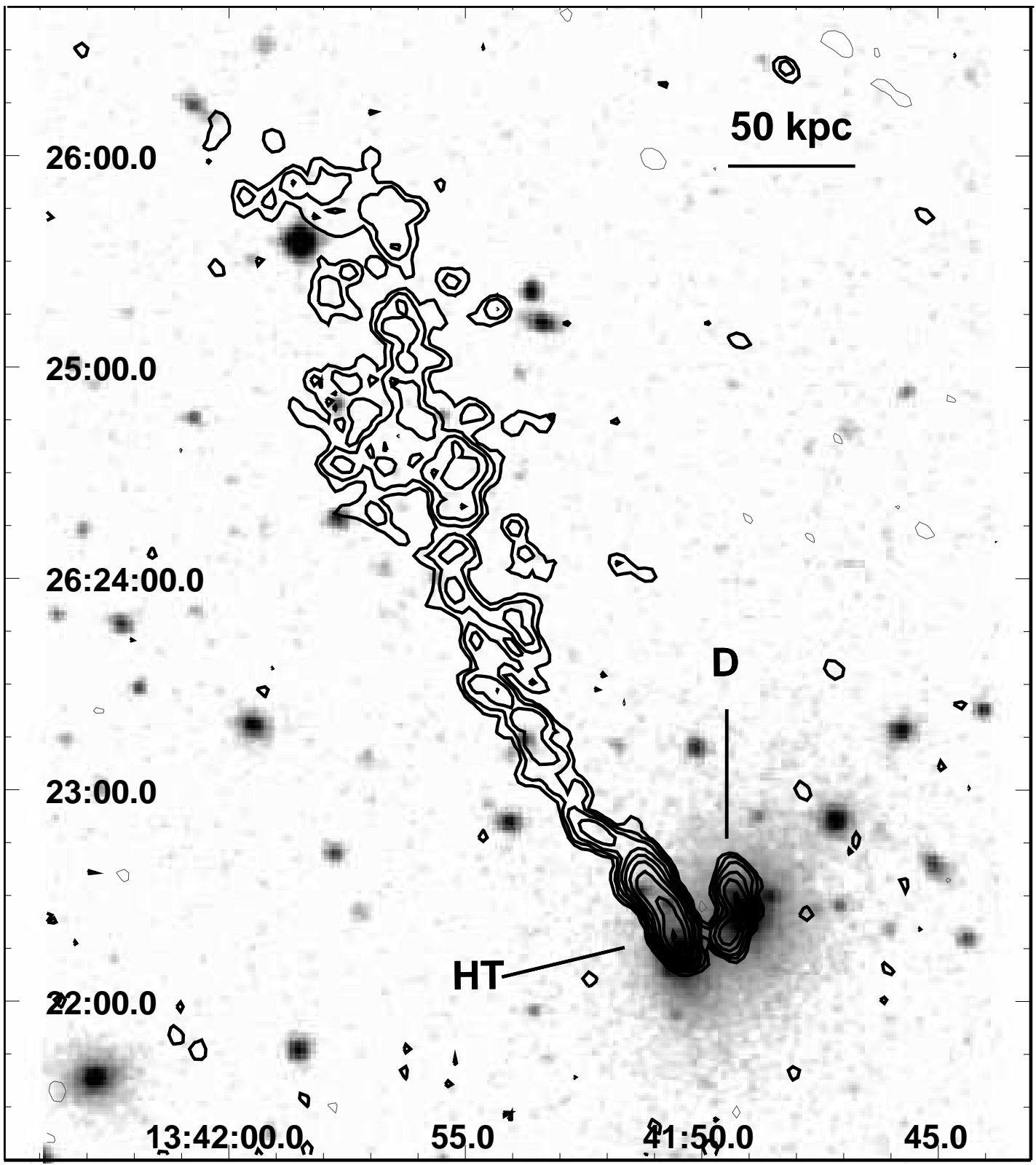
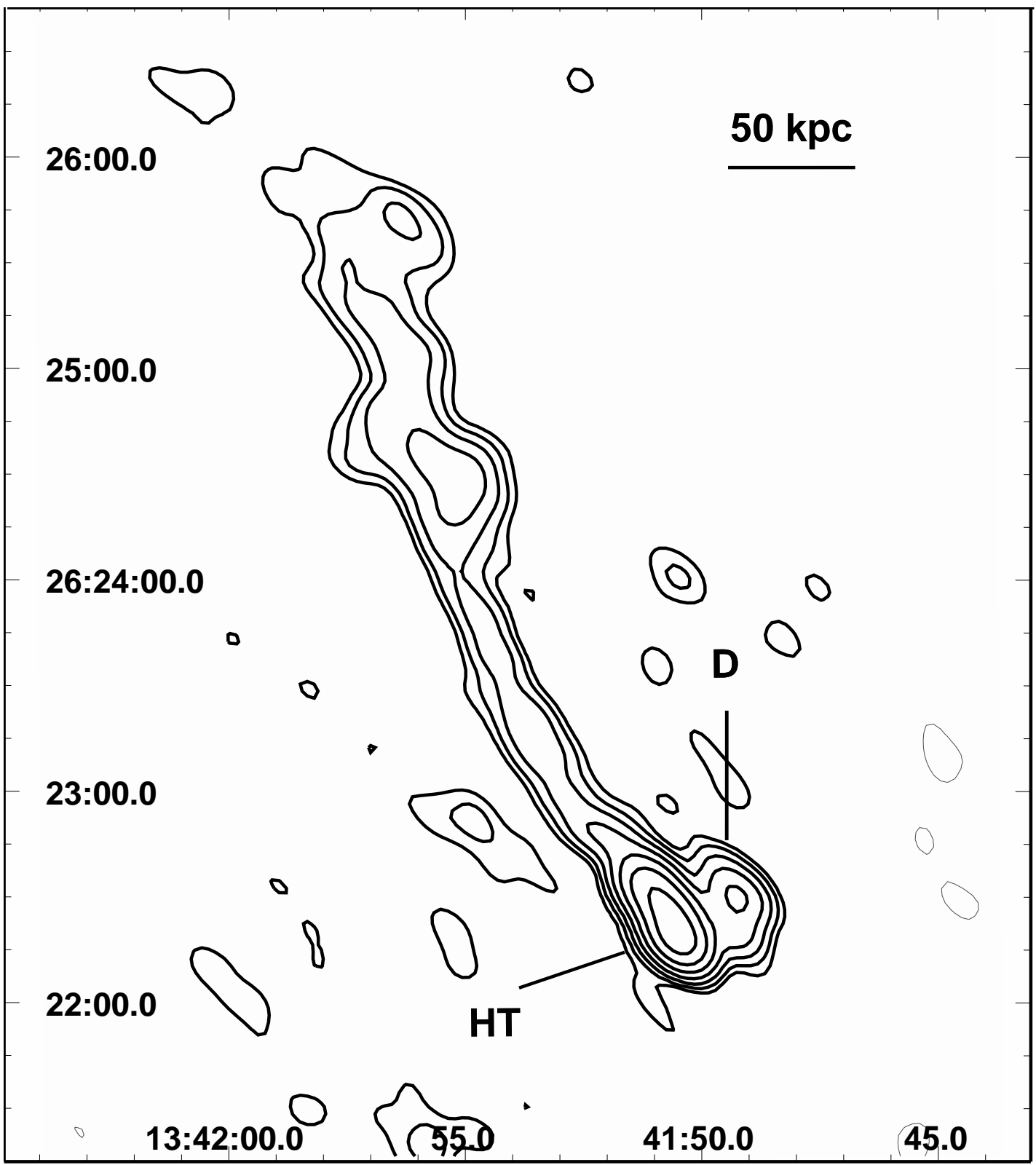
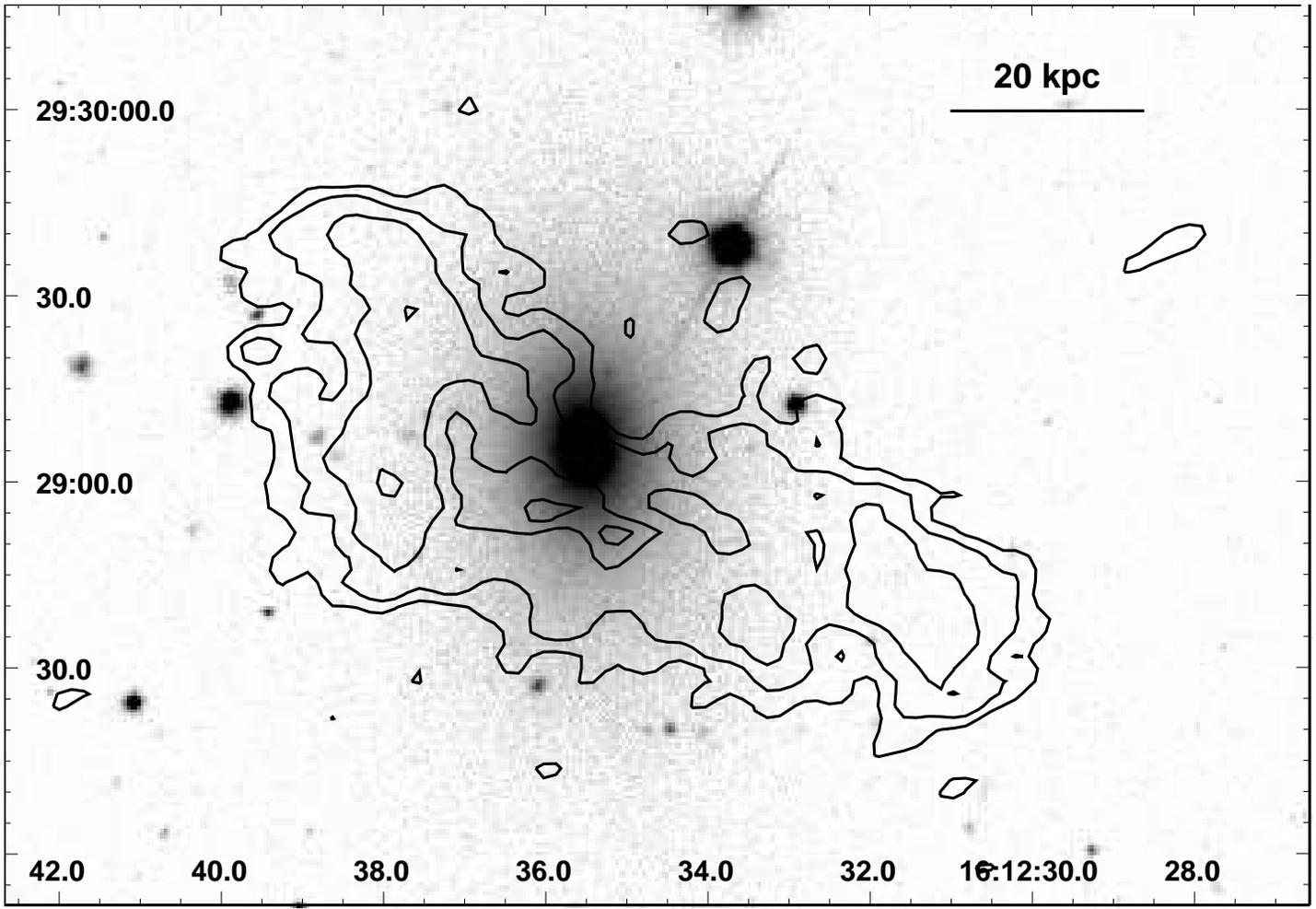
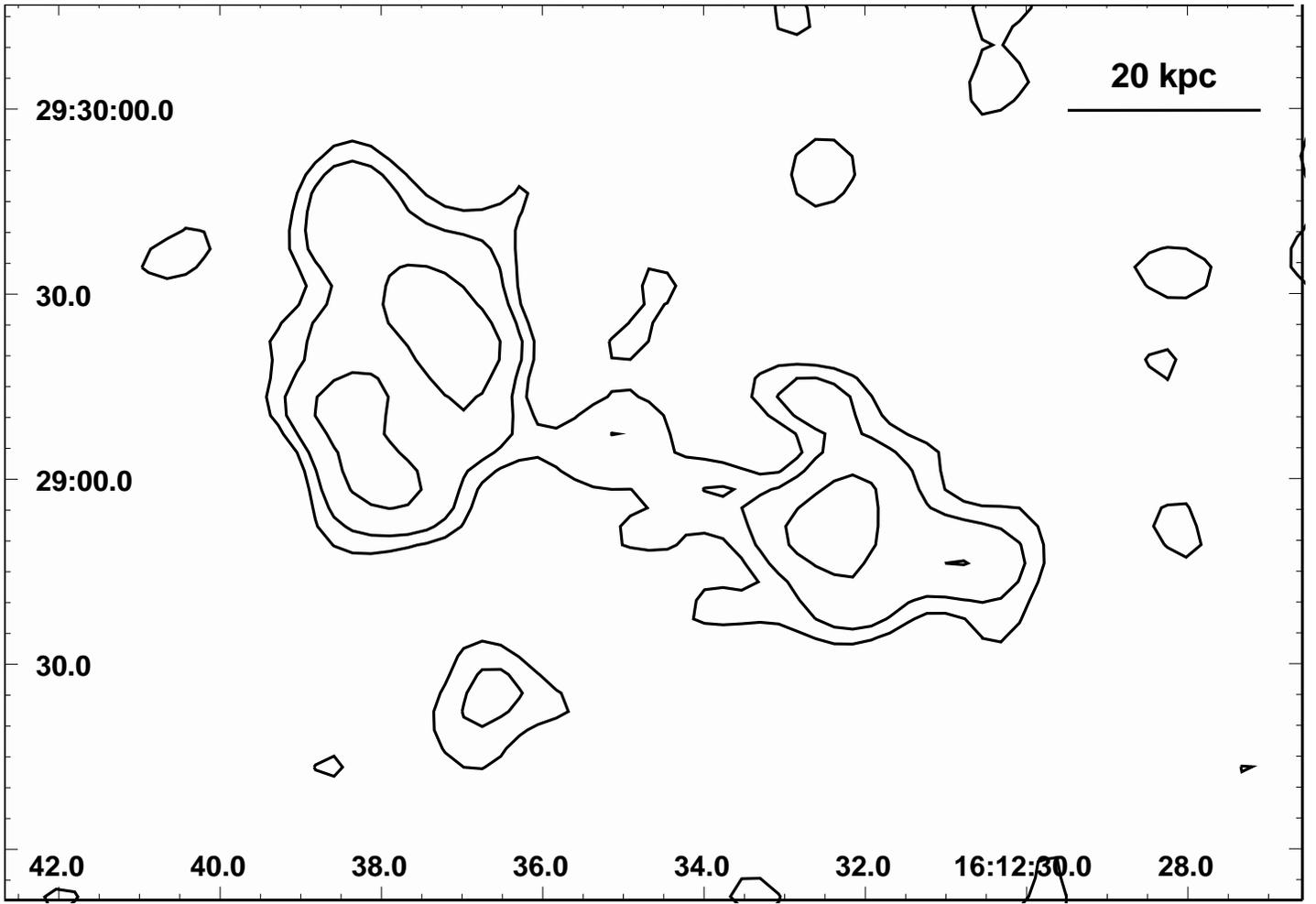
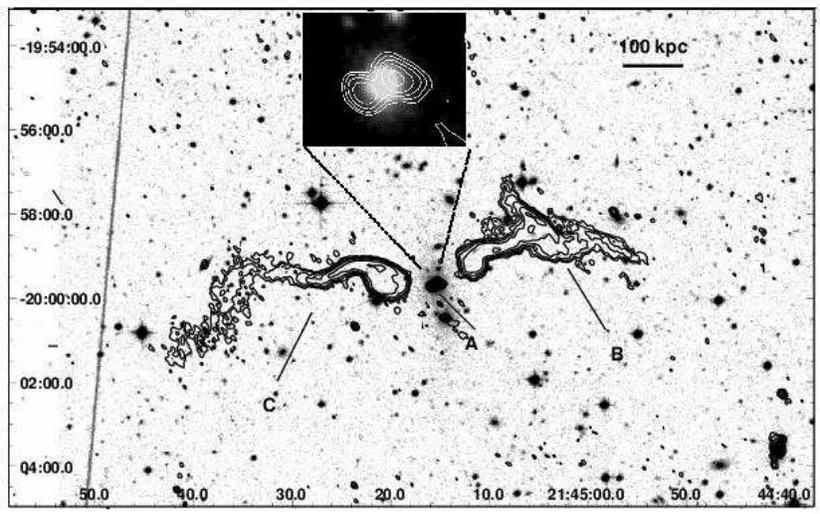
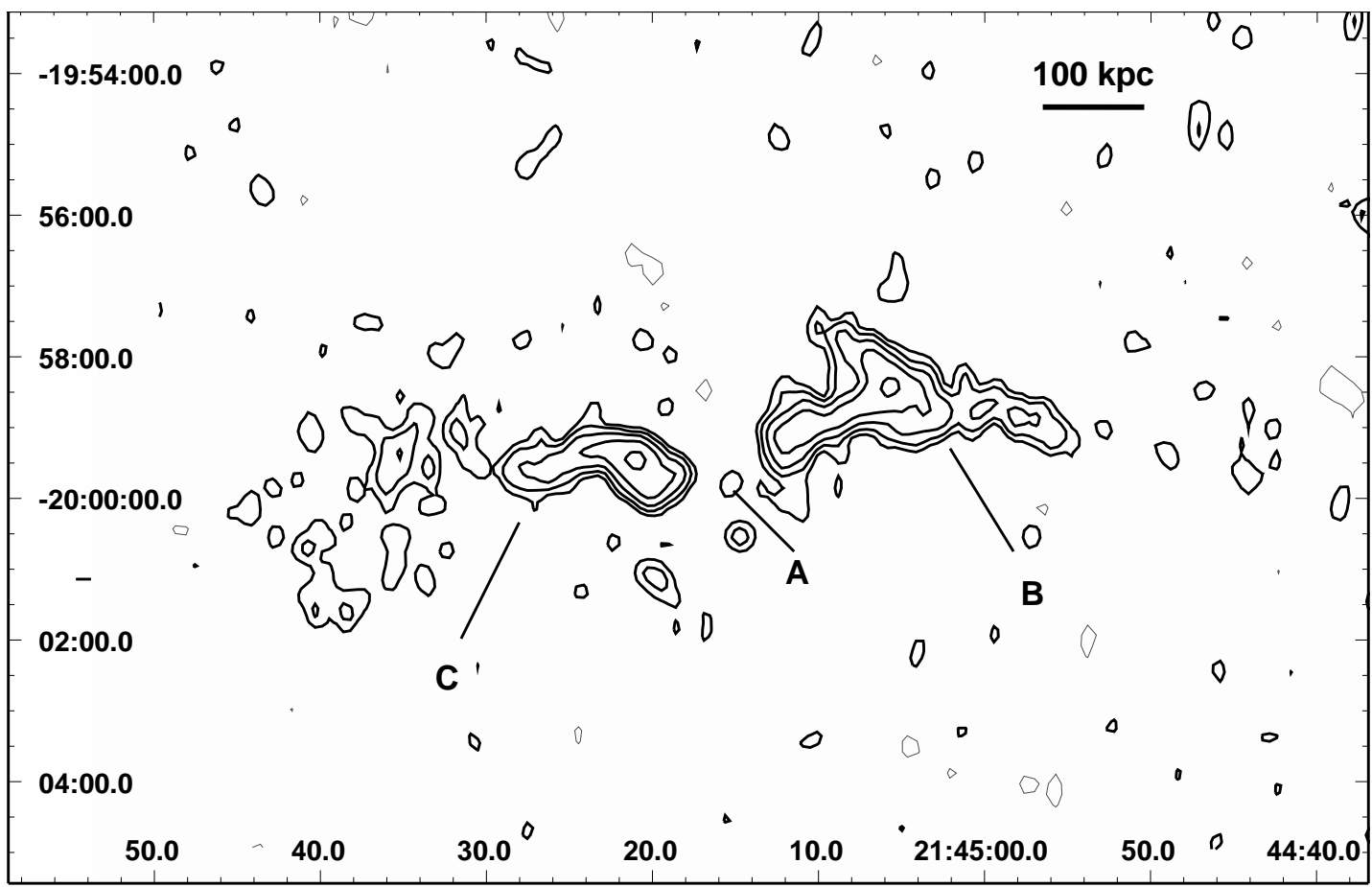
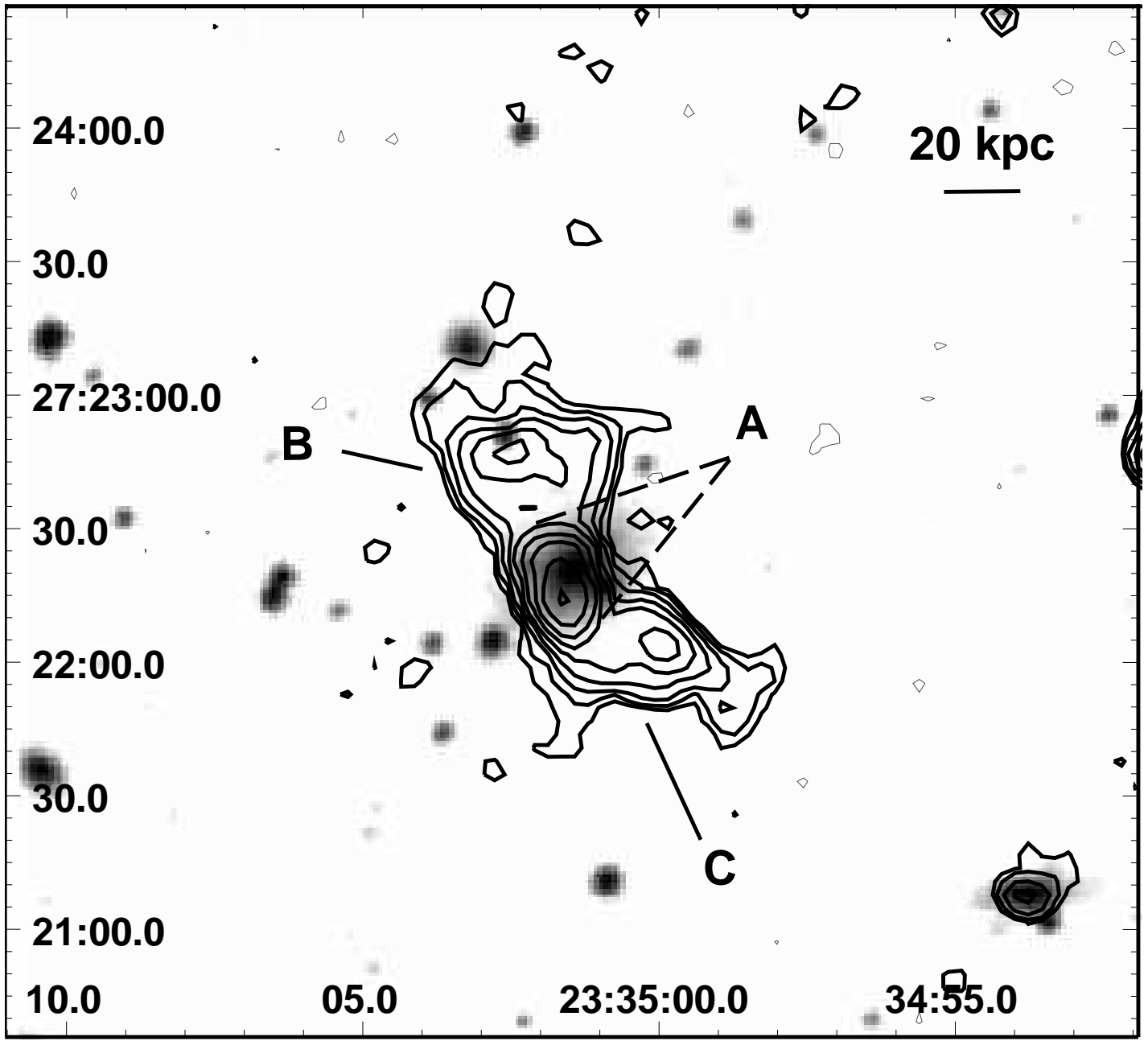
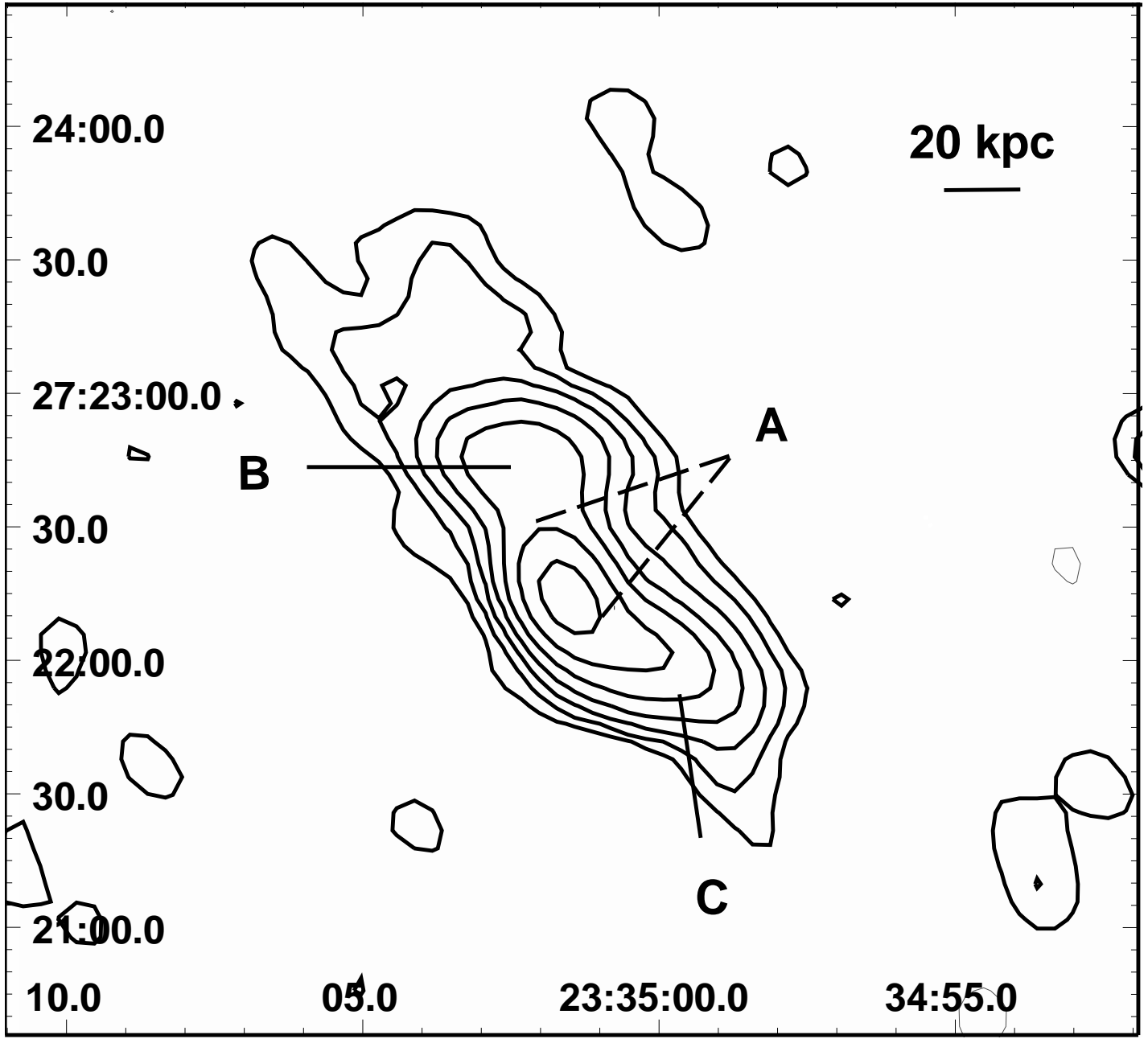
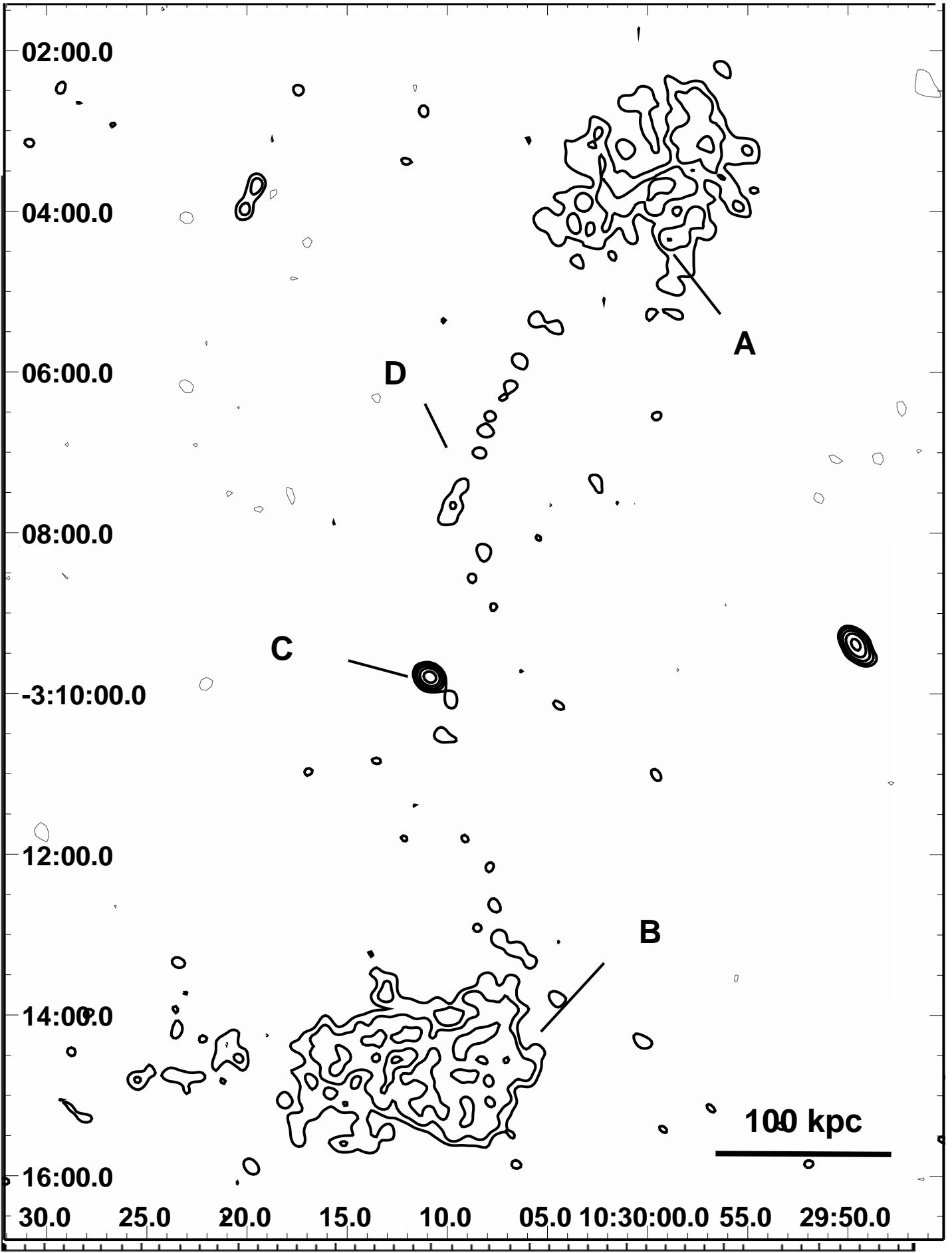
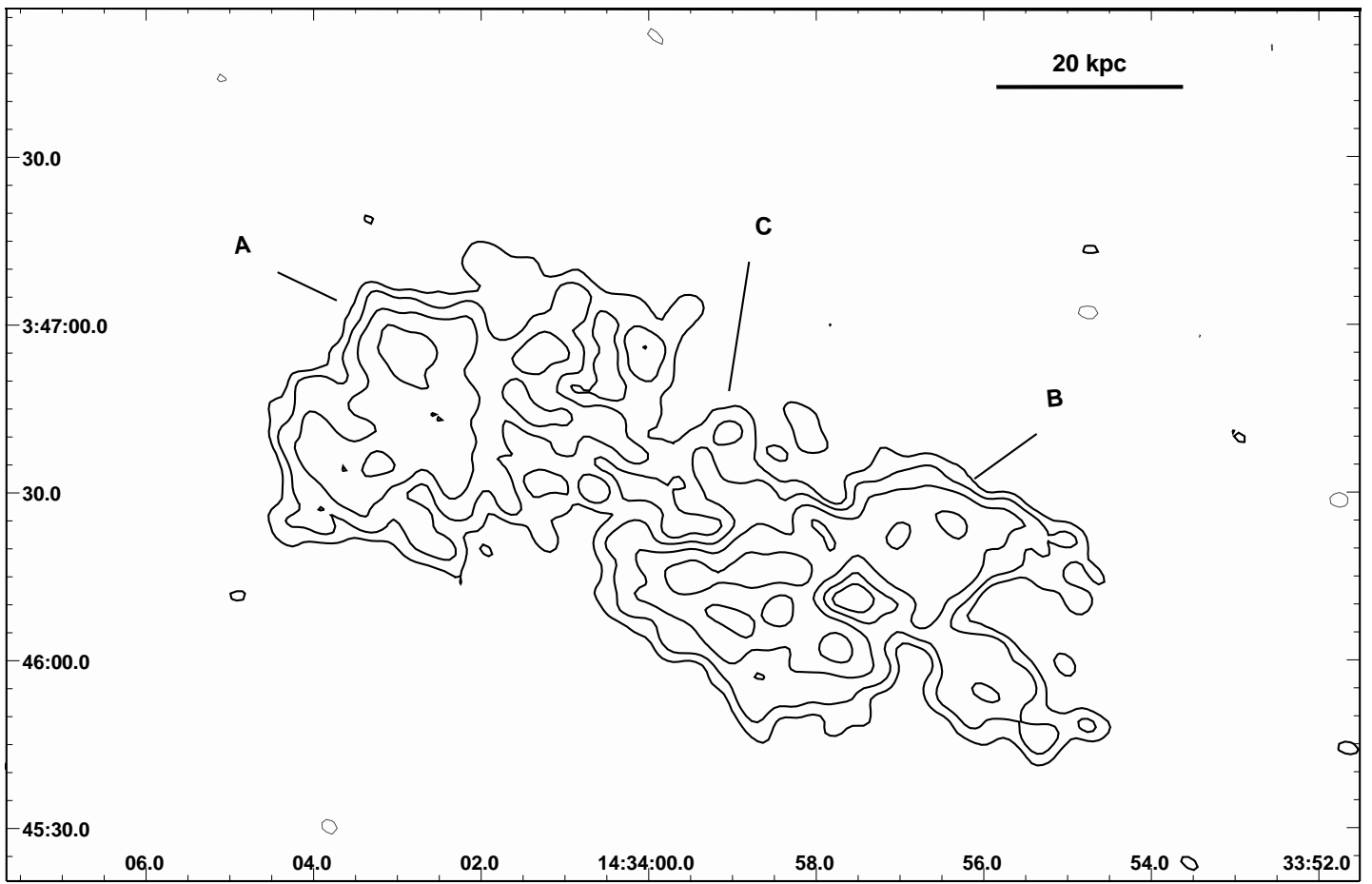
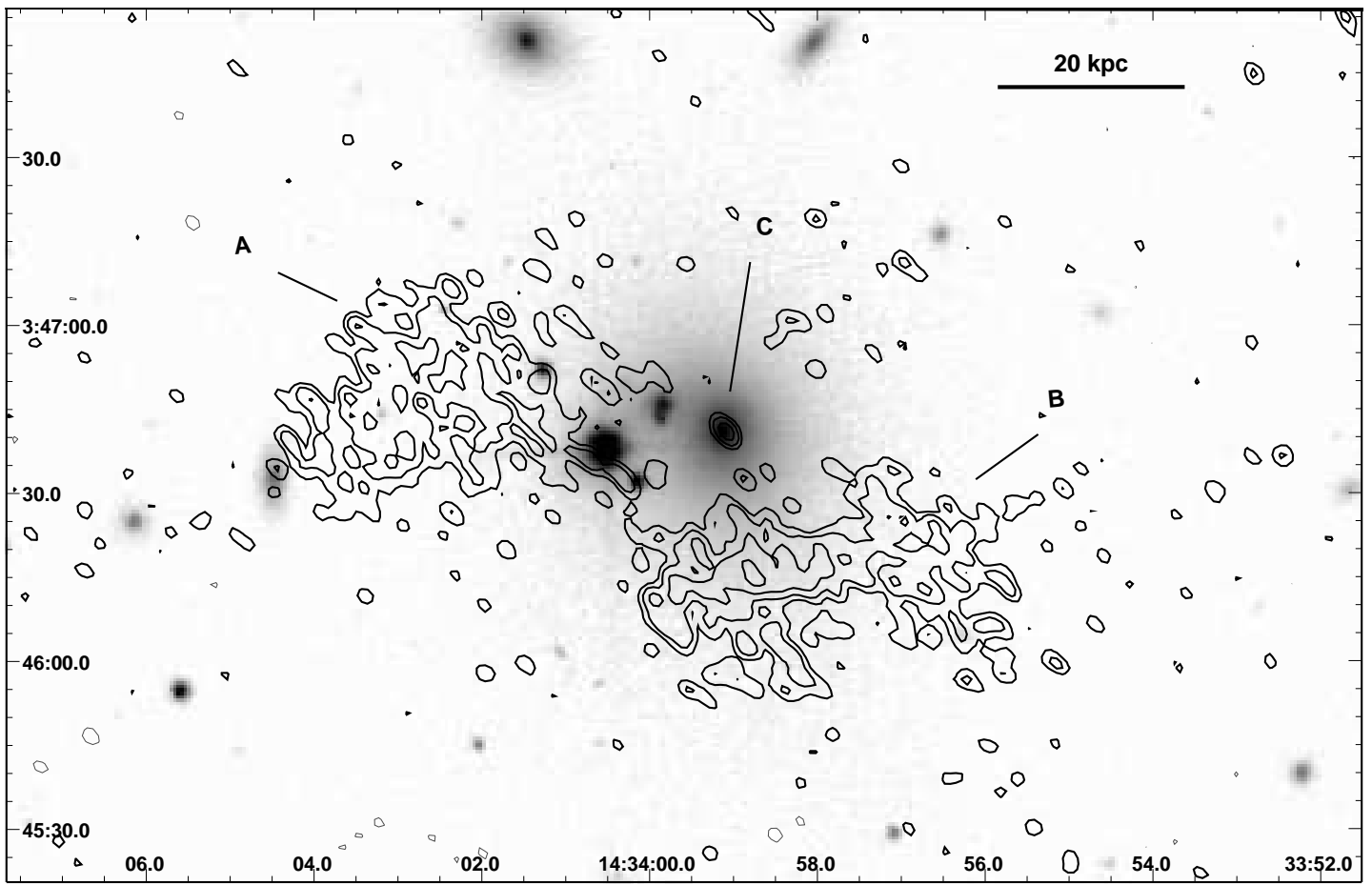
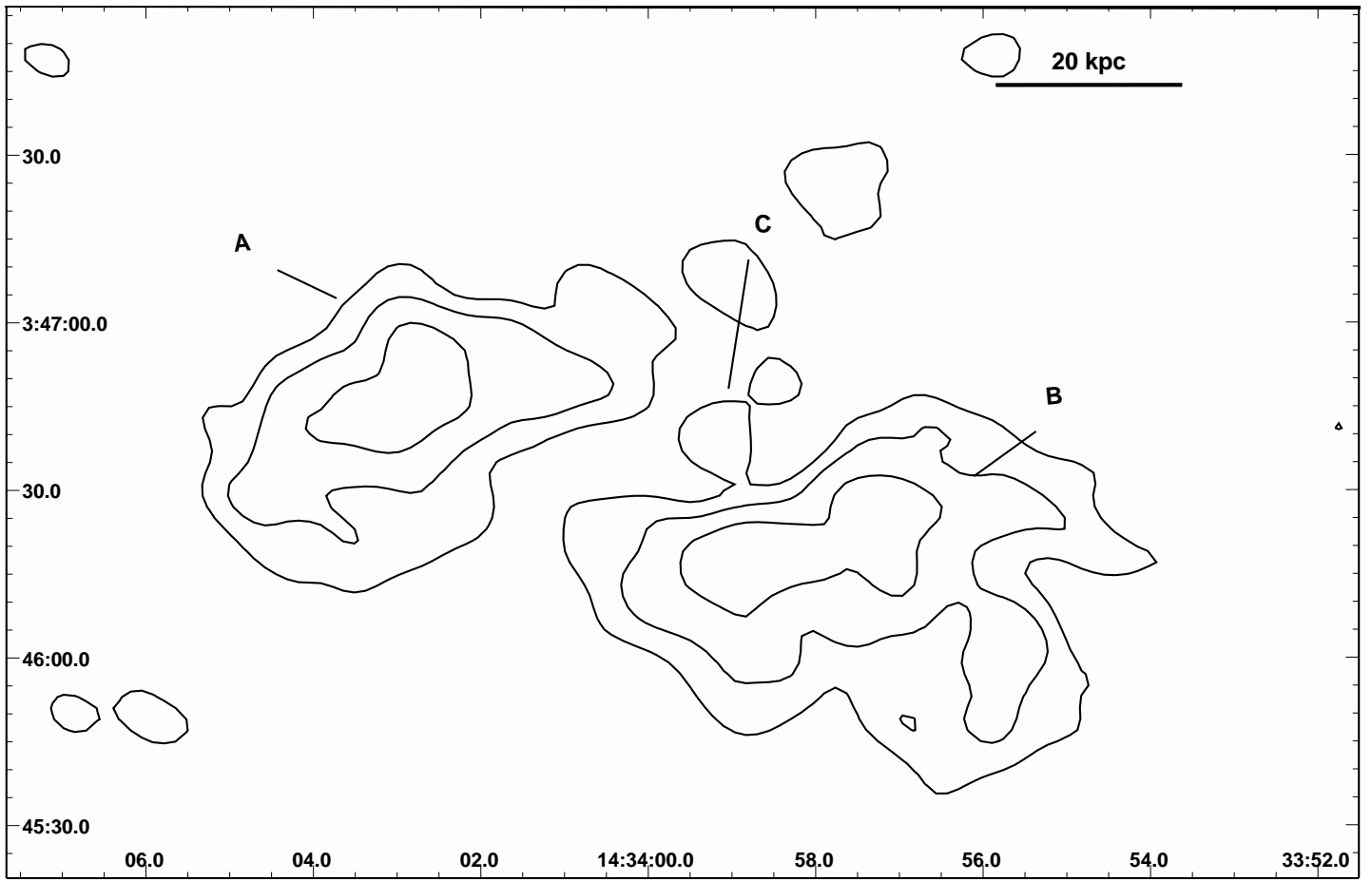
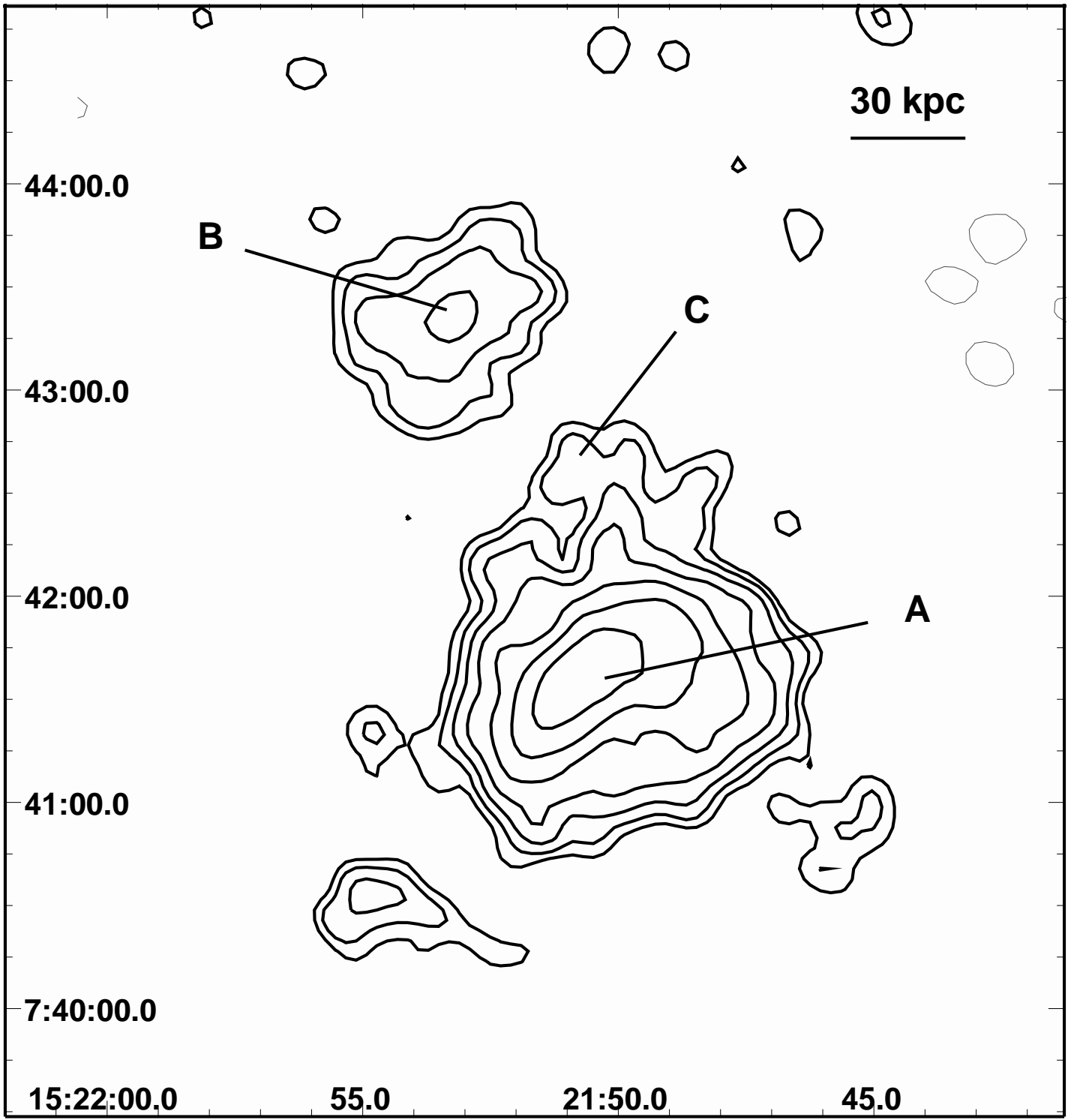
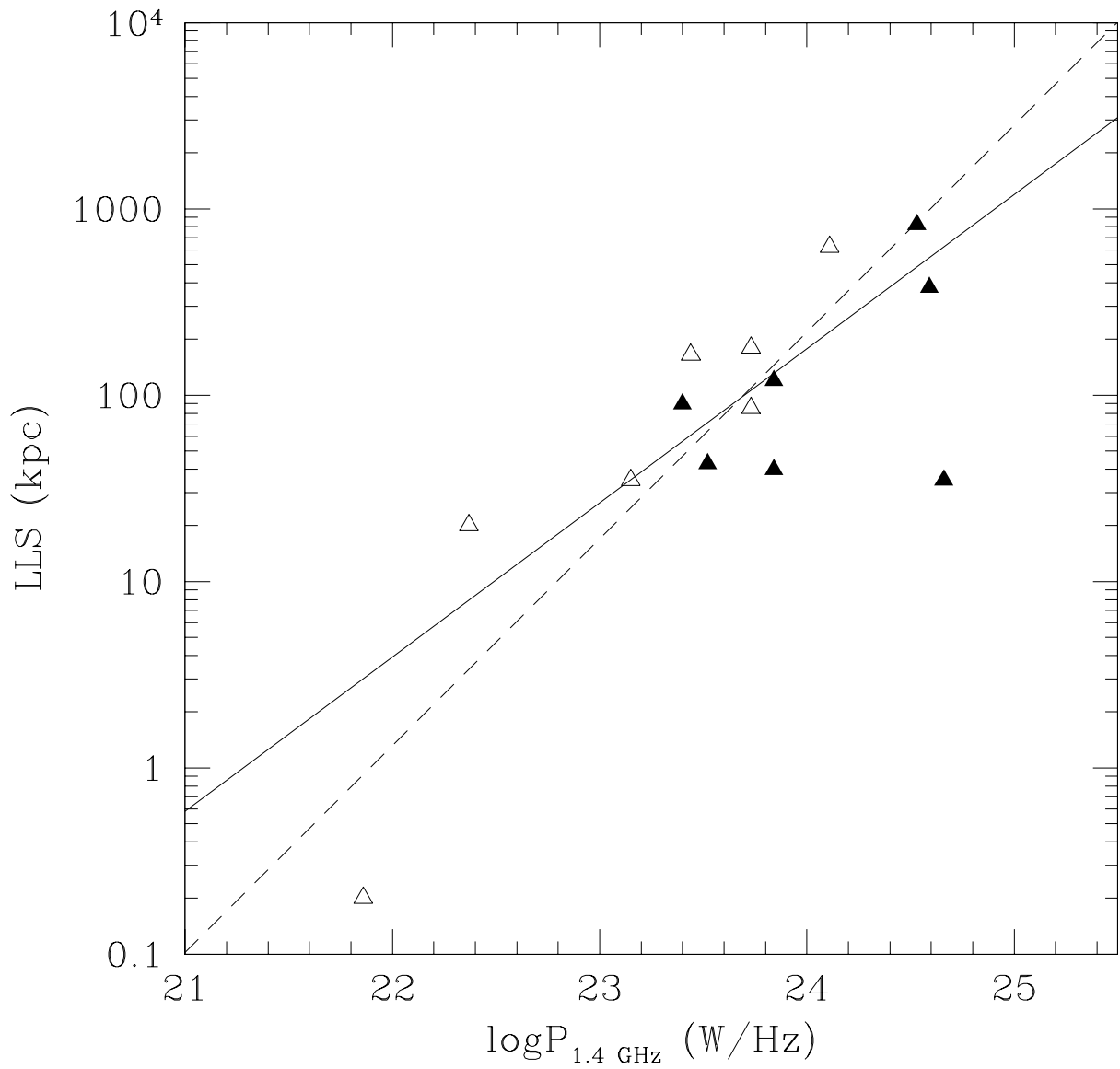
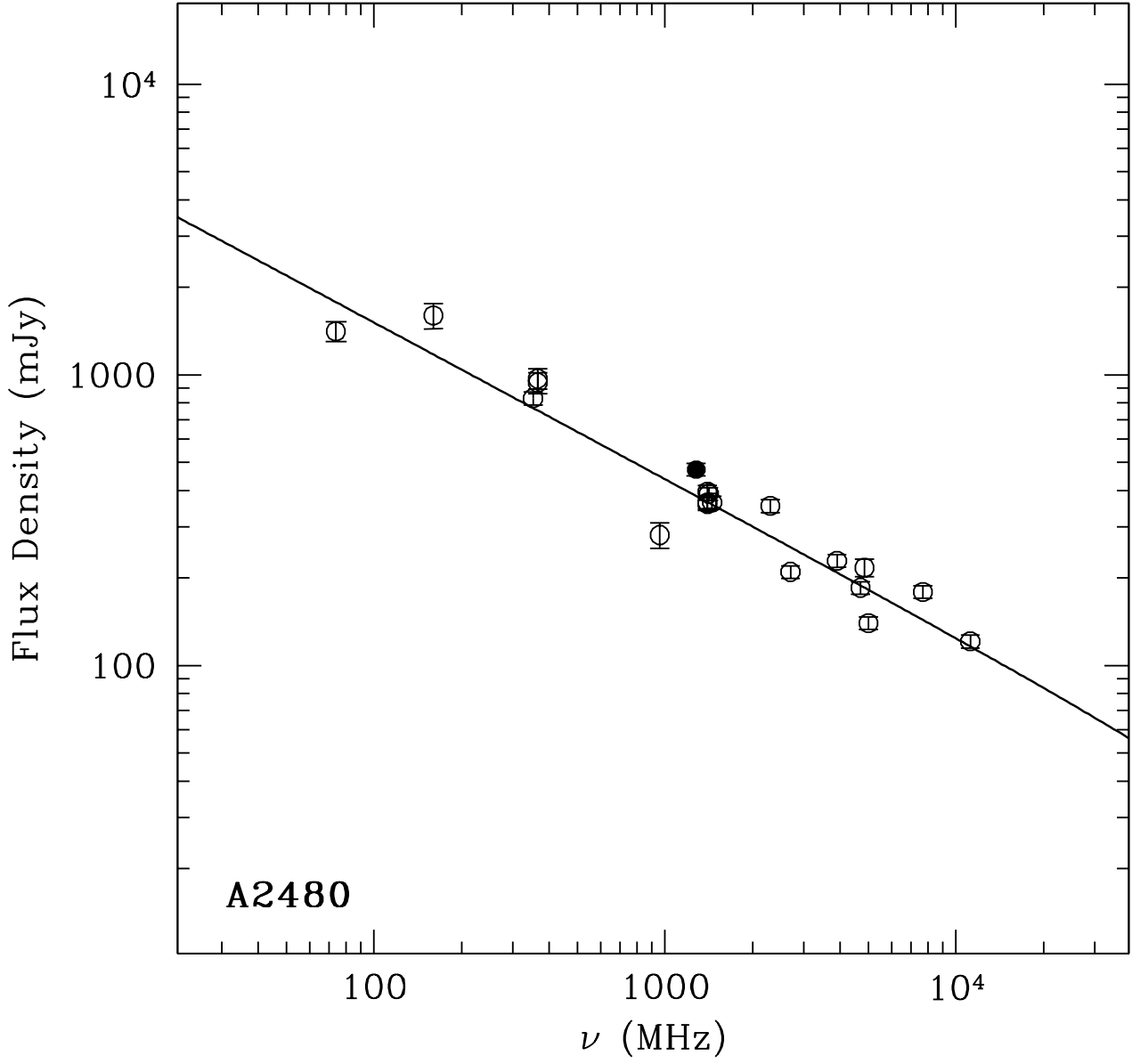
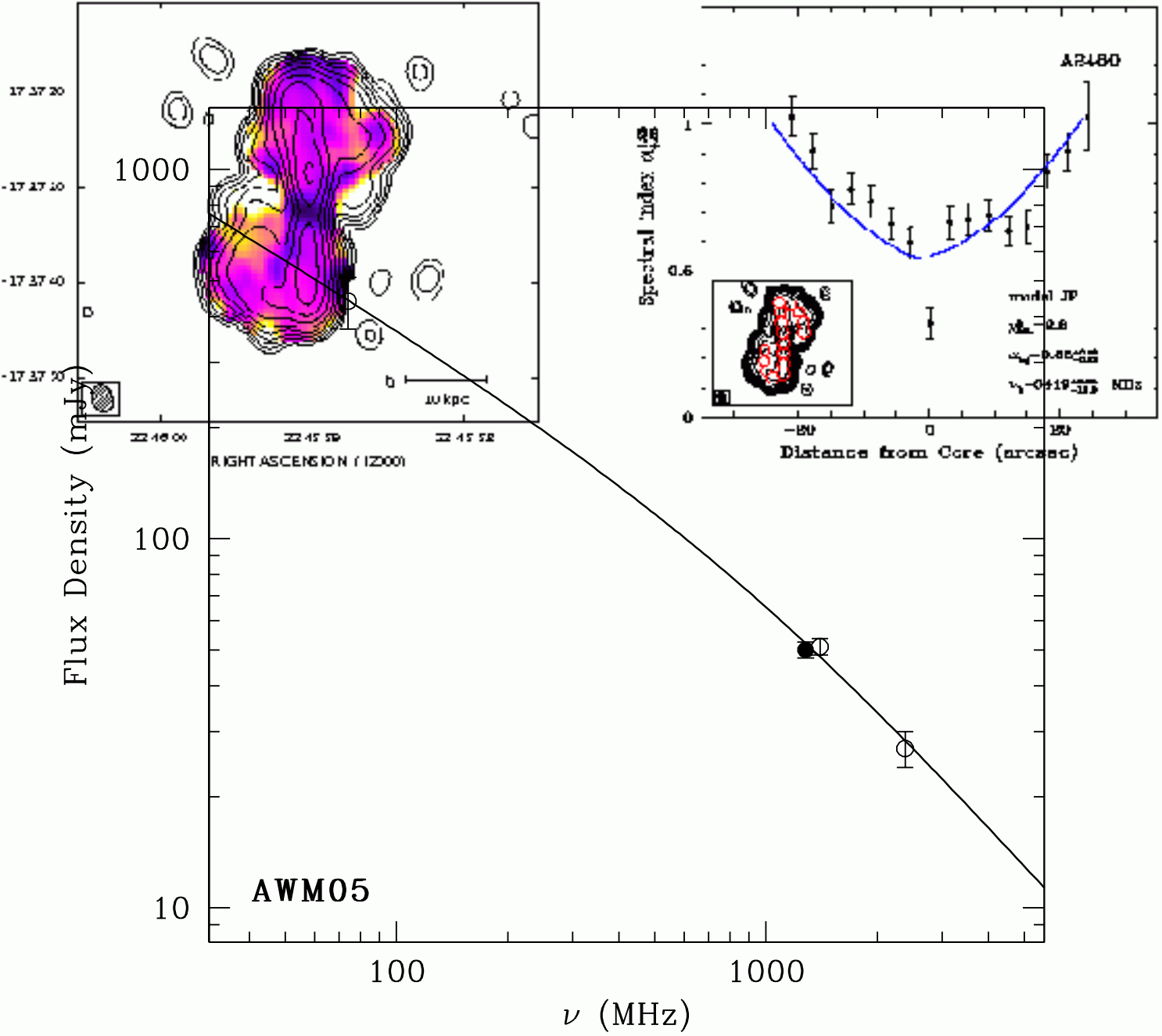
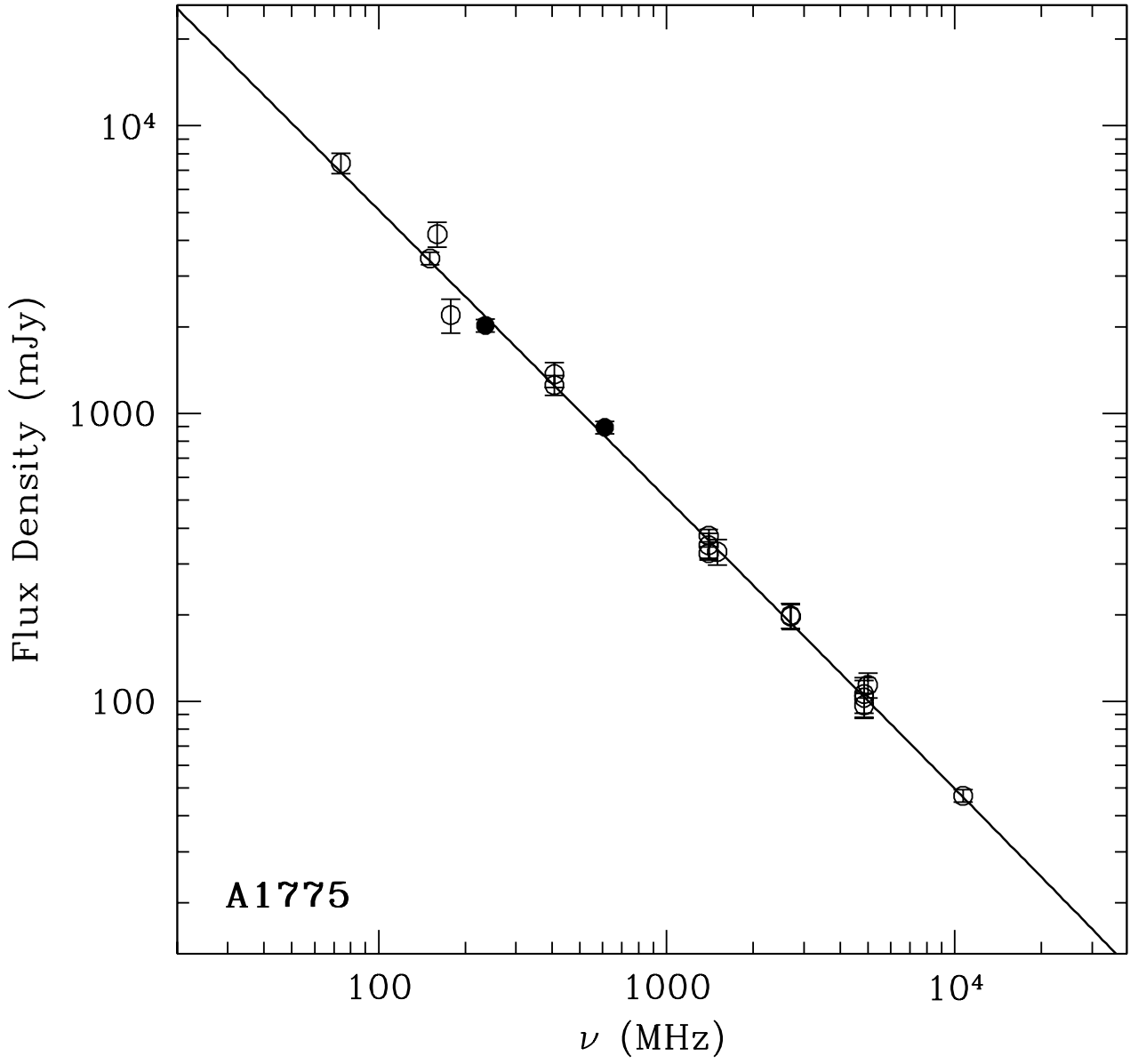
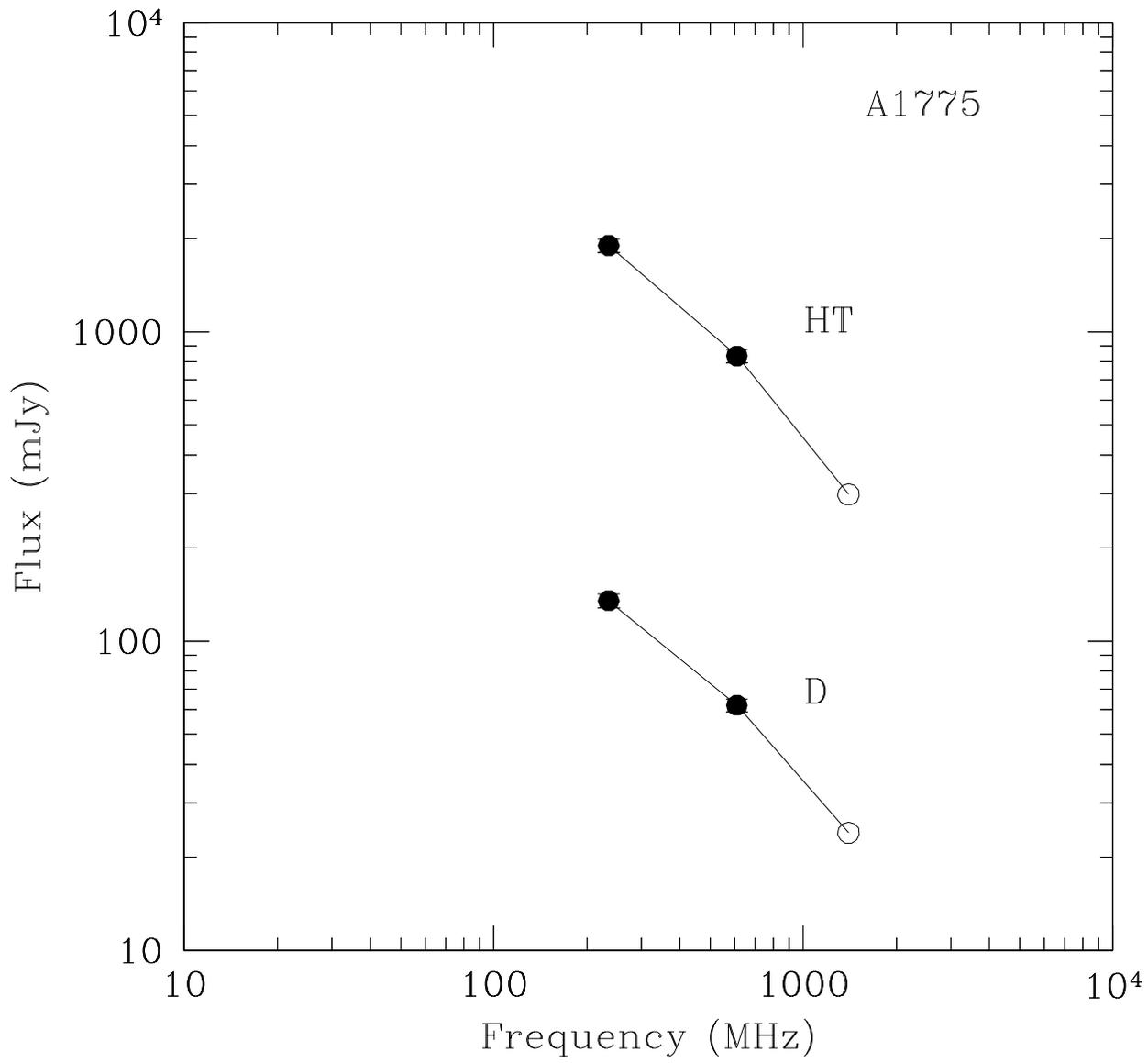
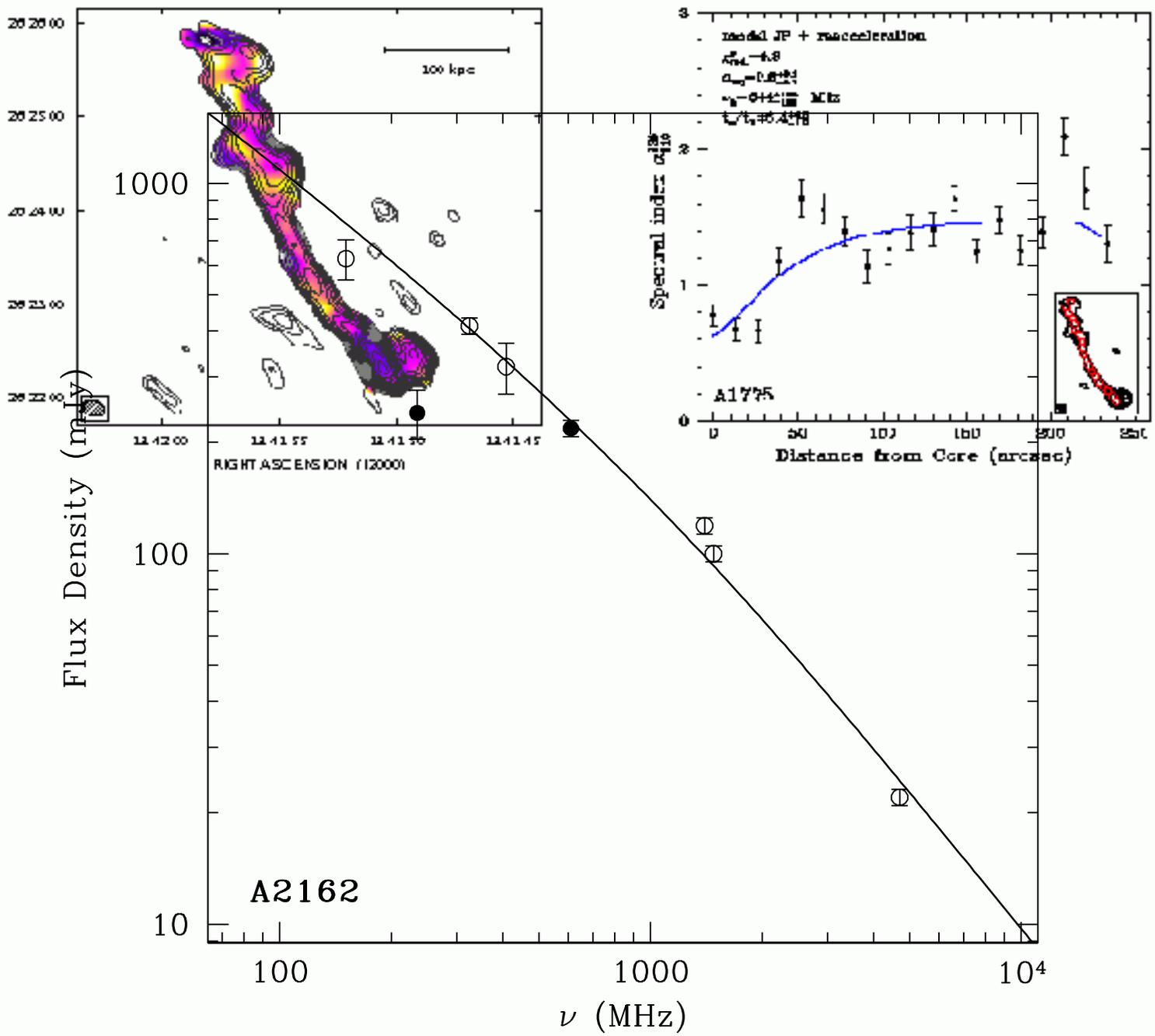
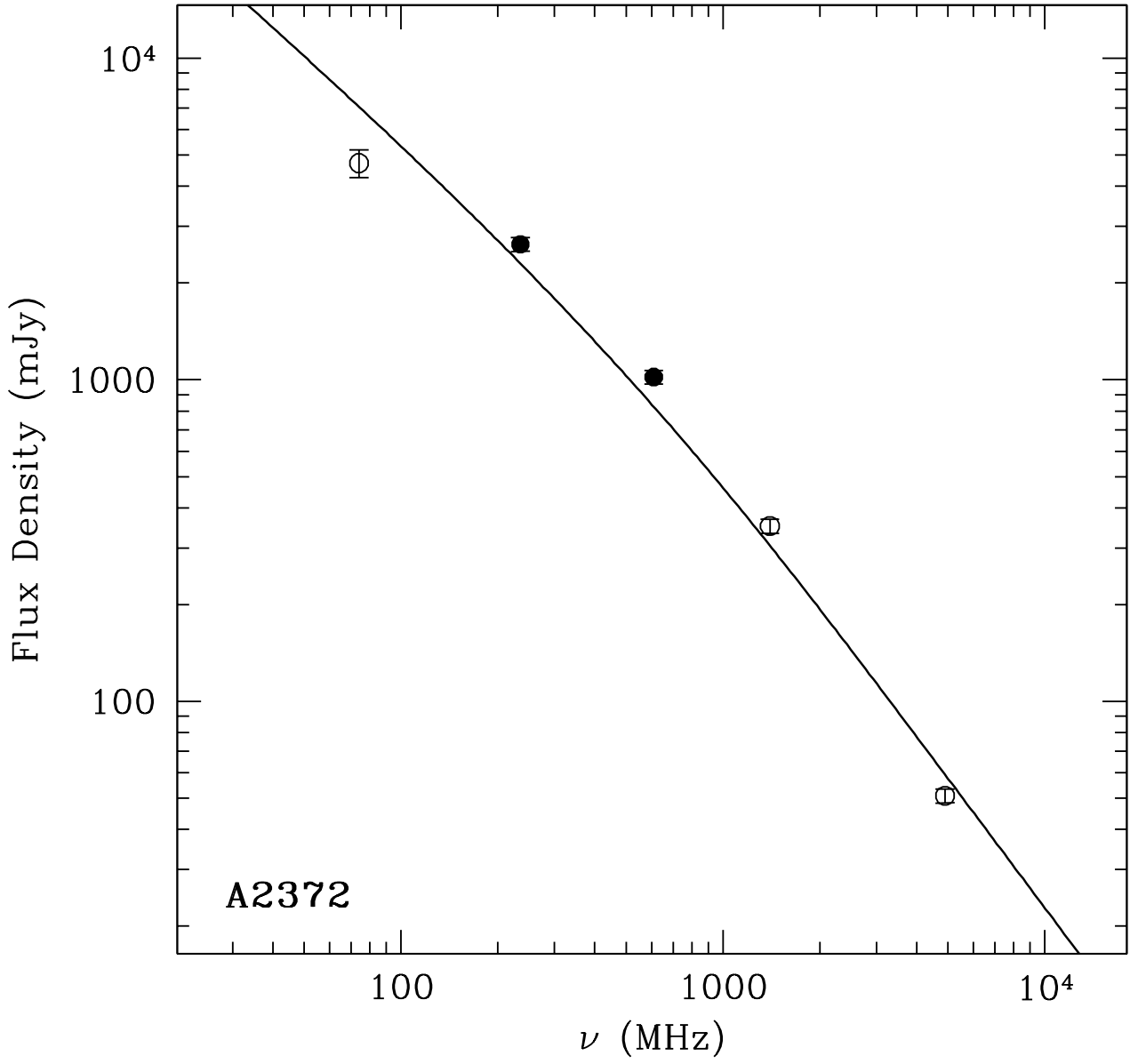
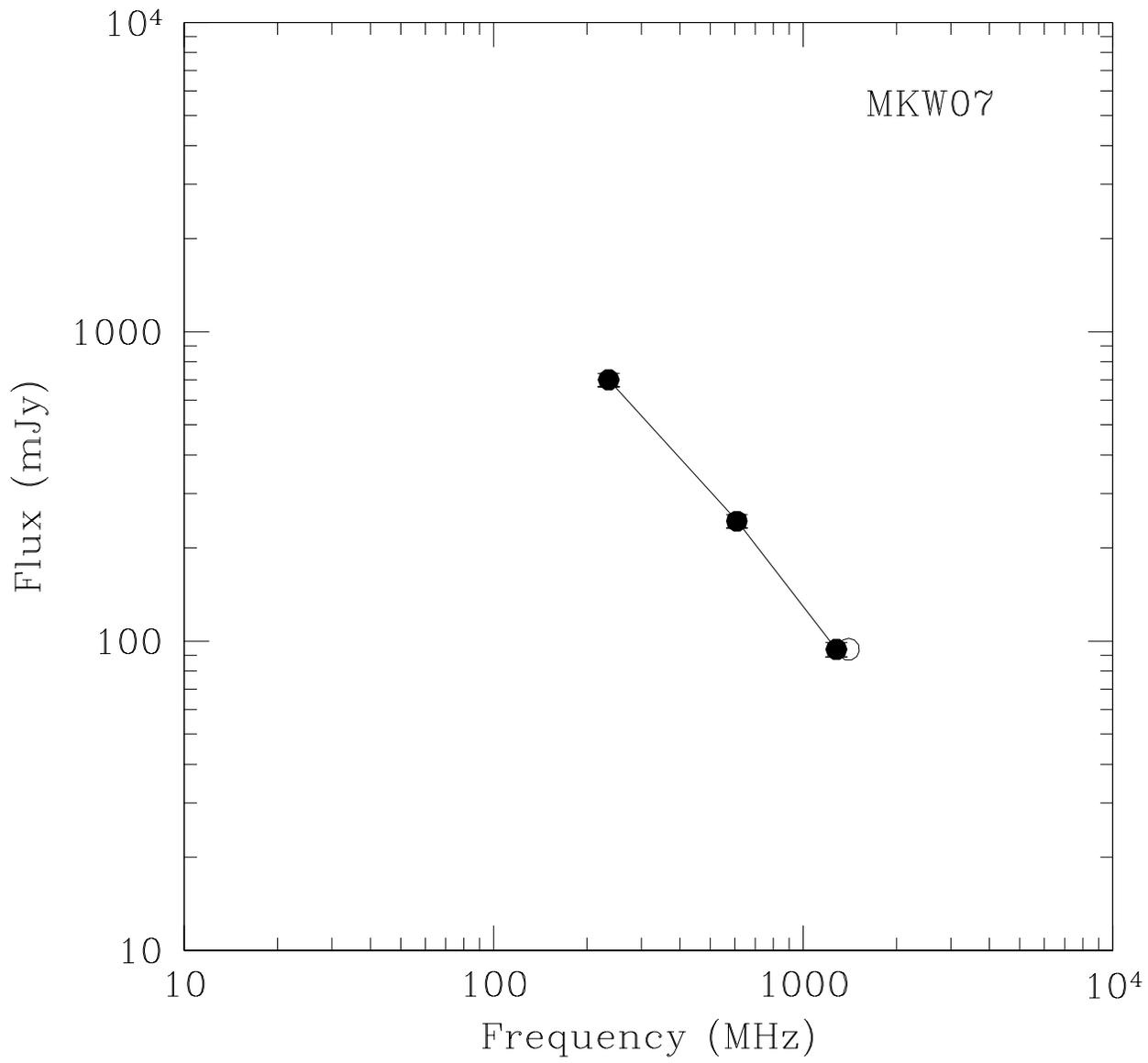
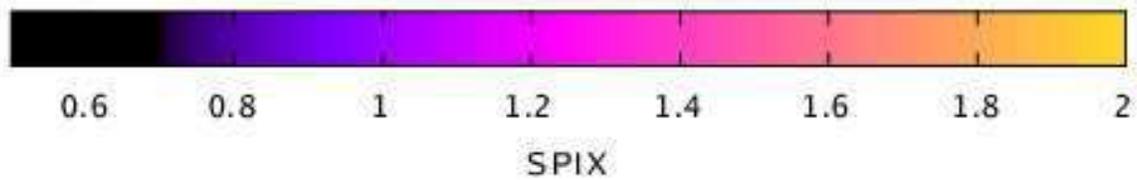
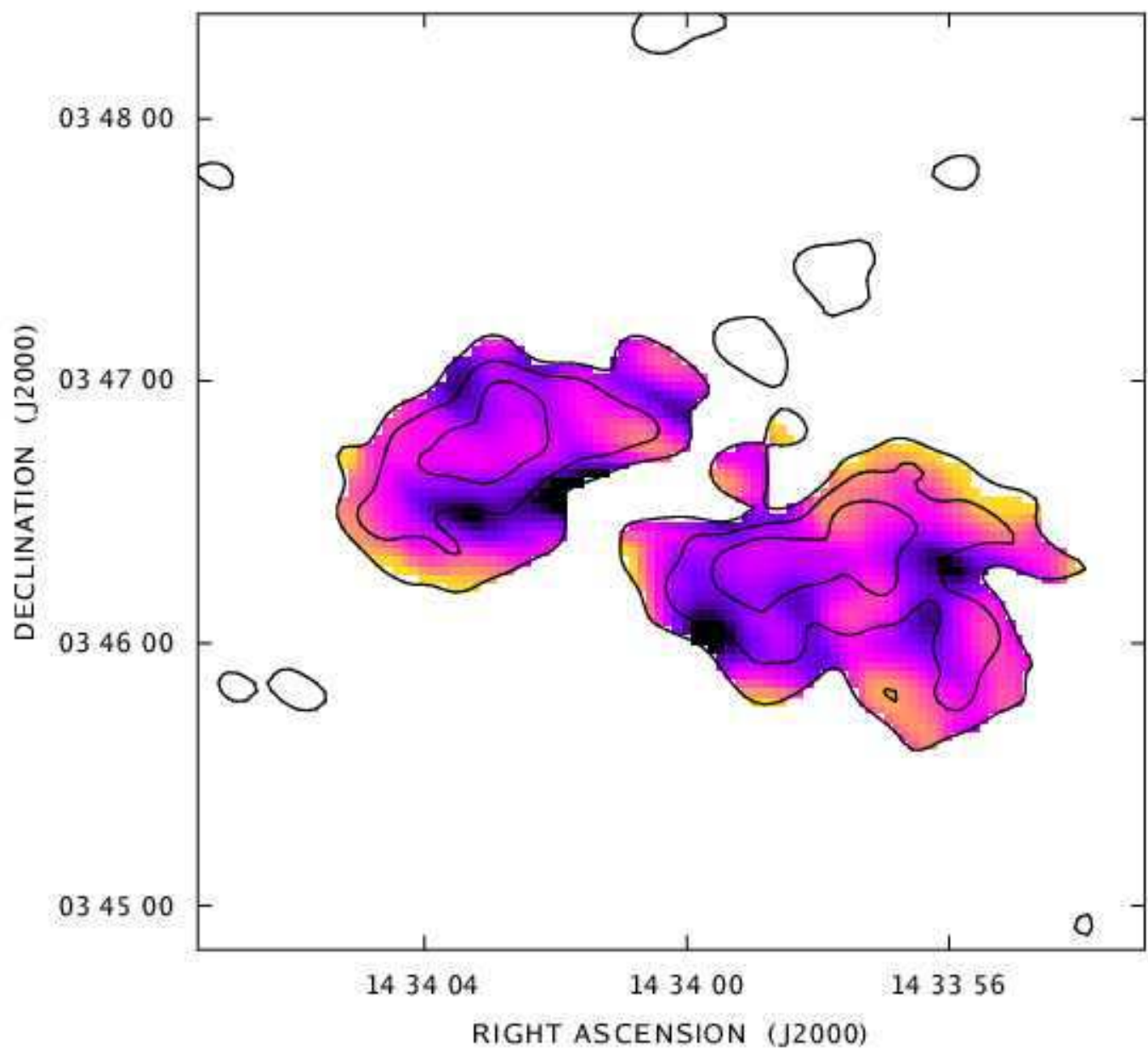
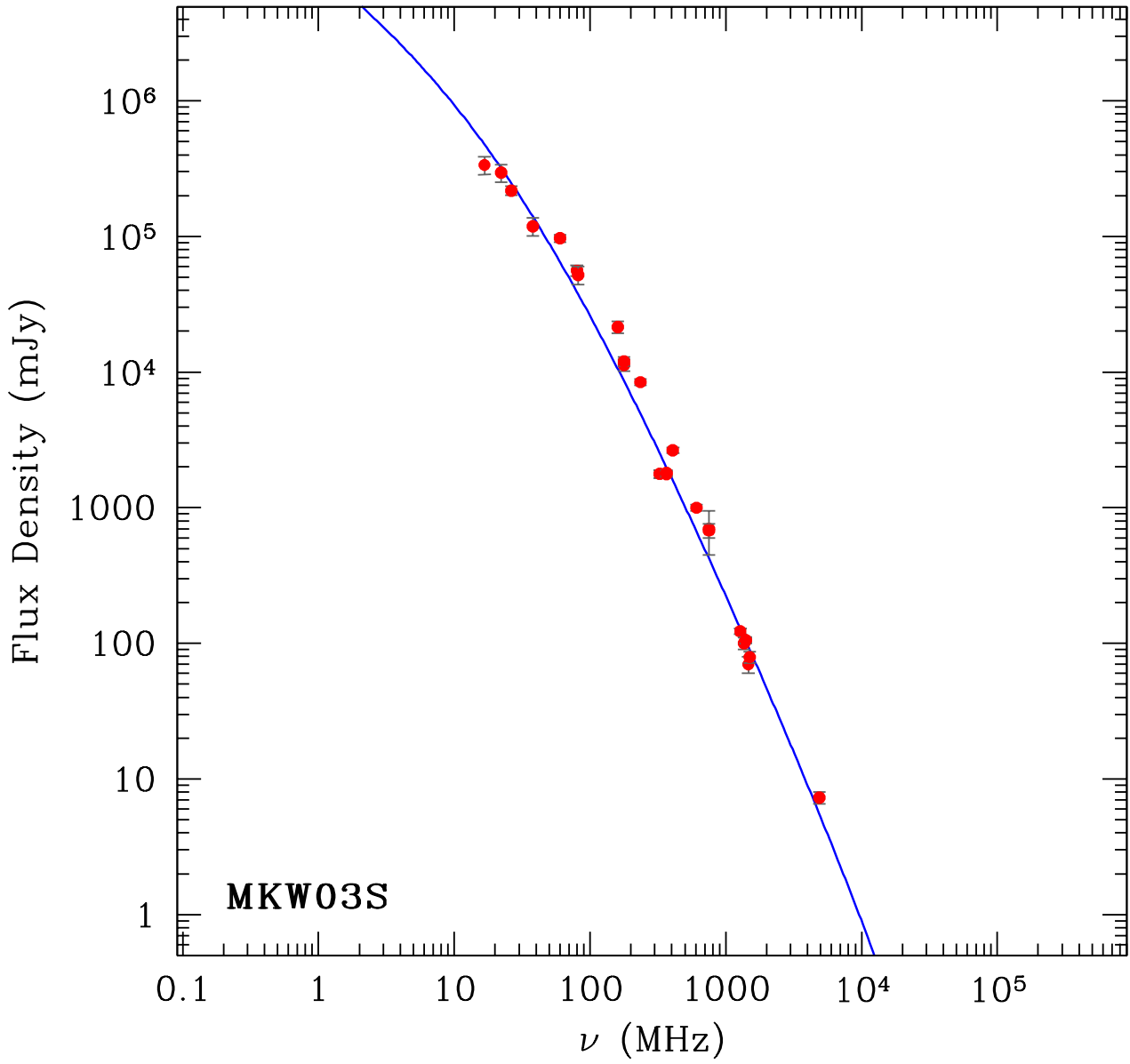
\end{document}